\UseRawInputEncoding
\documentclass[aps,prx,twocolumn,
superscriptaddress,nofootinbib]{revtex4-2}
\usepackage{amsmath,amssymb}
\usepackage{graphicx}
\usepackage{bm}
\usepackage{multirow}
\usepackage{hyperref}
\usepackage{braket}

\usepackage{color}

\begin{document}

\title{
Effective spin model in momentum space: Toward a systematic understanding of multiple-$Q$ instability by momentum-resolved anisotropic exchange interactions
}
\author{Ryota Yambe}
\email{yambe@jphys.t.u-tokyo.ac.jp}
\author{Satoru Hayami}
\email{hayami@ap.t.u-tokyo.ac.jp}
\affiliation{Department of Applied Physics, The University of Tokyo, Tokyo 113-8656, Japan }

\begin{abstract}
Multiple-$Q$ magnetic states, such as a skyrmion crystal, become a source of unusual transport phenomena and dynamics. 
Recent theoretical and experimental studies clarify that such multiple-$Q$ states ubiquitously appear under different crystal structures in metals and insulators.  
Toward a systematic understanding of the formation of the multiple-$Q$ states in various crystal systems, in this theoretical study, we present a low-energy effective spin model with anisotropic exchange interactions in momentum space. 
We summarize specific six symmetry rules for nonzero symmetric and antisymmetric anisotropic exchange interactions in momentum space, which are regarded as an extension of Moriya's rule.
According to the rules, we construct the effective spin model for tetragonal, hexagonal, and trigonal magnets with crystal- and momentum-dependent anisotropic exchange interactions based on magnetic representation analysis. 
Furthermore, we describe the origin of the effective anisotropic exchange interactions in itinerant magnets by perturbatively analyzing a multi-band periodic Anderson model with the spin-orbit coupling. 
We apply the effective spin model to an itinerant magnet in a $P6/mmm$ crystal and find various multiple-$Q$ states with a spin scalar chirality in the ground state.  
Our results provide a foundation of constructing effective phenomenological spin models for any crystal systems hosting the multiple-$Q$ states, which will stimulate further exploration of exotic multiple-$Q$ states in materials with the spin-orbit coupling. 
\end{abstract}

\maketitle

\section{Introduction}

Topological spin textures have attracted much attention as a source of unconventional physical phenomena and a candidate for robust information carriers against external stimuli~\cite{nagaosa2013topological,fert2013skyrmions,zhang2020skyrmion,PhysRevLett.127.067201,gobel2021beyond}. 
Since the first discovery of a magnetic skyrmion~\cite{skyrme1962unified,Bogdanov89,Bogdanov94,rossler2006spontaneous} in a chiral magnet MnSi~\cite{Muhlbauer_2009skyrmion}, the active searches have revealed its existence in a variety of crystal systems irrespective of spatial inversion symmetry~\cite{Tokura_doi:10.1021/acs.chemrev.0c00297}: cubic~\cite{Muhlbauer_2009skyrmion,tokunaga2015new}, hexagonal~\cite{kezsmarki_neel-type_2015,kurumaji2019skyrmion,hirschberger2019skyrmion}, and tetragonal~\cite{nayak2017discovery,Kurumaji_PhysRevLett.119.237201,khanh2020nanometric,Karube2021Room} crystal systems.
The skyrmion spin structure is characterized by an integer topological number (skyrmion number), which gives rise to intriguing transport phenomena, such as the topological Hall and Nernst effects~\cite{Neubauer_PhysRevLett.102.186602,Hamamoto_PhysRevB.92.115417,Gobel_PhysRevB.95.094413,Saha_PhysRevB.60.12162,kurumaji2019skyrmion,Hirschberger_PhysRevLett.125.076602,Shiomi_PhysRevB.88.064409}.  
In addition, a variety of new topological spin textures beyond the skyrmions have also been extensively investigated~\cite{gobel2021beyond}, some of which have been observed in experiments, such as a hedgehog~\cite{Kanazawa_PhysRevB.96.220414,fujishiro2019topological,Ishiwata_PhysRevB.101.134406},  biskyrmion~\cite{yu2014biskyrmion,wang2016centrosymmetric}, skyrmionium~\cite{Zhang2018Skyrmionium}, ferrimagnetic skyrmion~\cite{woo2018current}, and antiferromagnetic skyrmion~\cite{dohi2019formation}.

In the crystal systems, such topological spin textures often appear in a periodic form, which are expressed as a superposition of multiple spin density waves termed as a multiple-$Q$ state~\cite{Bak_PhysRevLett.40.800,Shapiro_PhysRevLett.43.1748,bak1980theory,Forgan_PhysRevLett.62.470,batista2016frustration,hayami2021topological}.
The multiple-$Q$ spin configuration consisting of $n$-tuple spin density waves with the ordering wave vectors $\{\bm{Q}_1,\bm{Q}_2,\cdots,\bm{Q}_n\}$ is given by
\begin{align}
\label{eq:multipleQ}
\bm{S}_i = \sum_{\nu=1}^n\left( e^{i\bm{Q}_\nu\cdot\bm{R}_i}\bm{S}_{\bm{Q}_\nu} + e^{-i\bm{Q}_\nu\cdot\bm{R}_i}\bm{S}_{-\bm{Q}_\nu} \right),
\end{align}
where $\bm{S}_{\bm{q}}$ is the Fourier transform of the spin at the wave vector $\bm{q}$.
A variety of the multiple-$Q$ magnetic structures are realized depending on the type of the constituent waves characterized by $\bm{S}_{\bm{Q}_\nu}$. 
For example, a triple-$Q$ superposition of the proper-screw (cycloidal) spiral waves leads to a Bloch-type (N\'{e}el-type) skyrmion crystal (SkX) with a skyrmion number of one, while that of the sinusoidal waves induces different types of the SkX with a skyrmion number of two.
In addition, the multiple-$Q$ spin configuration in Eq.~(\ref{eq:multipleQ}) can describe other periodic topological spin textures, such as a hedgehog lattice~\cite{Binz_PhysRevB.74.214408,Park_PhysRevB.83.184406,Okumura_PhysRevB.101.144416,grytsiuk2020topological,Shimizu_PhysRevB.103.054427,Aoyama_PhysRevB.103.014406,kato2021spin}, meron-antimeron crystal~\cite{Lin_PhysRevB.91.224407,yu2018transformation,Hayami_PhysRevLett.121.137202,Bera_PhysRevResearch.1.033109,Hayami_PhysRevB.103.024439,Wang_PhysRevB.103.104408,Utesov_PhysRevB.103.064414,Hayami_PhysRevB.103.054422,Hayami_PhysRevB.104.094425}, and vortex crystal~\cite{Momoi_PhysRevLett.79.2081,Kamiya_PhysRevX.4.011023,Marmorini2014,Wang_PhysRevLett.115.107201,Hayami_PhysRevB.94.174420,yambe2021skyrmion,hayami2021phase}.

The emergence of multiple-$Q$ states largely depends on the microscopic mechanisms, which have been extensively studied for various systems.
The typical mechanisms are dipolar interactions~\cite{Garel_PhysRevB.26.325,Utesov_PhysRevB.103.064414,utesov2021mean}, competing exchange interactions in frustrated magnets~\cite{Okubo_PhysRevLett.108.017206,Shimokawa_PhysRevB.100.224404,Aoyama_PhysRevB.103.014406,mitsumoto2021replica,mitsumoto2021skyrmion}, a biquadratic spin interaction in itinerant magnets~\cite{Akagi_PhysRevLett.108.096401,Hayami_PhysRevB.90.060402,Ozawa_doi:10.7566/JPSJ.85.103703,Ozawa_PhysRevLett.118.147205,Hayami_PhysRevB.95.224424}, and symmetric and antisymmetric magnetic anisotropy in systems with the spin-orbit coupling (SOC)~\cite{rossler2006spontaneous,Butenko_PhysRevB.82.052403,Wilson_PhysRevB.89.094411,leonov2015multiply,Lin_PhysRevB.93.064430,Hayami_PhysRevB.93.184413,leonov2016properties,Leonov_PhysRevB.96.014423,Hayami_PhysRevB.99.094420,brinker2019chiral,Wang_PhysRevLett.124.207201,Hayami_PhysRevLett.121.137202,Wang_PhysRevB.103.104408,Kathyat_PhysRevB.102.075106,amoroso2020spontaneous}.          

Among the various stabilization mechanisms for the multiple-$Q$ states, we focus on the role of symmetric and antisymmetric anisotropic exchange interactions including the antisymmetric Dzyaloshinskii-Moriya (DM) interaction. 
The DM interaction is the most familiar anisotropic exchange interaction arising in noncentrosymmetric materials~\cite{dzyaloshinsky1958thermodynamic,moriya1960anisotropic}.
It favors the spiral spin density wave with a spiral plane perpendicular to the DM vector in the combination of the ferromagnetic exchange interaction, which results in the multiple-$Q$ spiral states in an external magnetic field~\cite{rossler2006spontaneous,Park_PhysRevB.83.184406}.
As the DM vector is determined by the crystal symmetry, which is so-called Moriya's rule~\cite{moriya1960anisotropic}, one can expect what types of multiple-$Q$ states appear in the DM-based systems~\cite{Bogdanov89}.
For example, the DM interaction in $P422$, $P4mm$, and $P\bar{4}m2$ crystals tends to favor the Bloch-, N\'{e}el-, and anti-type SkXs, respectively. 

In contrast to the early studies based on the antisymmetric DM interaction, recent discoveries of SkX and hedgehog lattices in centrosymmetric magnets~\cite{kurumaji2019skyrmion,hirschberger2019skyrmion,khanh2020nanometric,Ishiwata_PhysRevB.101.134406} open up the possibility that various multiple-$Q$ states can emerge by symmetric anisotropic exchange interactions.
In fact, some model calculations have clarified that such a symmetric anisotropic exchange interaction stabilizes the multiple-$Q$ states including the SkXs in centrosymmetric crystals with the space group $P4/mmm$~\cite{Hayami_PhysRevB.103.024439,Wang_PhysRevB.103.104408,Hayami_doi:10.7566/JPSJ.89.103702,doi:10.7566/JPSJ.91.023705}  $P6/mmm$~\cite{Hayami_PhysRevB.103.054422,hayami2020multiple,takagi2018multiple,Hirschberger_10.1088/1367-2630/abdef9}, and $P\bar{3}m1$~\cite{yambe2021skyrmion,amoroso2020spontaneous,amoroso2021tuning}. 
However, there have been few studies focusing on the symmetric anisotropic exchange interactions in spite of various types of them depending on the crystal symmetry classified by the space group.
 To understand which types of anisotropic exchange interactions play an important role in inducing the multiple-$Q$ states, it is highly desired to perform a systematic investigation for various space groups irrespective of the centrosymmetric and noncentrosymmetric lattice structures. 
Furthermore, it is important to clarify relevant microscopic model parameters for the emergence of the anisotropic exchange interactions beyond the symmetry argument.

To systematically investigate the multiple-$Q$ instability induced by the anisotropic exchange interactions, we present how to construct an effective spin model in discrete lattice systems based on the magnetic representation~\cite{Bertaut:a05871} and perturbation analyses~\cite{Schrieffer_PhysRev.149.491,Akagi_PhysRevLett.108.096401,Hayami_PhysRevB.95.224424}. 
Our effective spin model can be applied to magnetic systems with both a short-range exchange interaction in insulators and a long-range one in metals.
First, we find important six symmetry rules to specify both symmetric and antisymmetric exchange interactions in momentum space, the latter of which is a counterpart of Moriya's rule in real space. 
The obtained rules give a foundation of constructing the effective low-energy spin model with the momentum-resolved anisotropic exchange interactions in any crystals.
As an example, we demonstrate how to construct the effective spin model in tetragonal, hexagonal, and trigonal crystal systems by applying the rules to 24 gray space groups.
Next, we show one of the microscopic origins of the long-range anisotropic exchange interactions by starting from the periodic Anderson model (PAM) incorporating the effect of the SOC~\cite{Anderson_PhysRev.124.41,PhysRevB.55.12561,yambe2021skyrmion}. 
We present important microscopic model parameters for the anisotropic exchange interactions based on the perturbation analysis. 
The perturbation analysis beyond the symmetry argument gives a way to quantitatively evaluate the anisotropic exchange interactions. 
Finally, we demonstrate that the anisotropic exchange interactions stabilize various multiple-$Q$ states with a spin scalar chirality by considering a specific example in a $P6/mmm$ crystal and by performing the simulated annealing for the effective spin model. 
The present results to construct the effective spin model with the momentum-resolved interactions provide both symmetric and microscopic ways of investigating a plethora of multiple-$Q$ instabilities in various crystal systems.
Especially, the present effective spin model is useful to identify complicated spin configurations including the SkX in materials, such as Gd$_2$PdSi$_3$~\cite{kurumaji2019skyrmion}, Gd$_3$Ru$_4$Al$_{12}$~\cite{hirschberger2019skyrmion,Hirschberger_10.1088/1367-2630/abdef9}, GdRu$_2$Si$_2${~\cite{khanh2020nanometric,Yasui2020imaging,khanh2022zoology}, EuPtSi~\cite{kakihana2018giant,kaneko2019unique,kakihana2019unique,tabata2019magnetic,hayami2021field}, and EuAl$_4$~\cite{Shang_PhysRevB.103.L020405,kaneko2021charge}. 
        
This paper is organized as follows.
In Sec.~\ref{sec:Symmetry}, we show a way of obtaining the momentum-resolved anisotropic exchange interactions under the crystal symmetry based on the magnetic representation analysis.
In Secs.~\ref{sec:Origin1} and \ref{sec:Origin2}, we discuss the origin of momentum-resolved anisotropic exchange interactions in itinerant electron models and localized spin models, respectively. 
In particular, we show the relationship between the long-range anisotropic exchange interaction and microscopic model parameters by performing the perturbation calculation in the PAM in Sec.~\ref{sec:Origin1}.
In Sec.~\ref{sec:SpecificExample}, we present how to construct and analyze the effective spin model by taking an example of the system belonging to the $P6/mmm$ space group.
We summarize the obtained results and discuss a perspective in Sec.~\ref{sec:Summary}.

\section{Symmetry analysis of anisotropic exchange interactions}
\label{sec:Symmetry}

In this section, we show a complete classification of anisotropic exchange interaction in momentum space in 
crystal systems based on the symmetry argument.
In Sec.~\ref{sec:Rule}, we present general six rules to give nonzero momentum-resolved anisotropic exchange interactions.
Then, we explicitly show the effective spin model in tetragonal, hexagonal, and trigonal crystal systems in Sec.~\ref{sec:Model1}.
We also discuss a tendency of modulations from the single-$Q$ spin configuration and a possible multiple-$Q$ spin configuration in the presence of
the anisotropic exchange interactions in Sec.~\ref{sec:Modulation}.

\subsection{General six symmetry rules}
\label{sec:Rule} 

\begin{figure*}[tp!]
\begin{center}
\includegraphics[width=1.0\hsize]{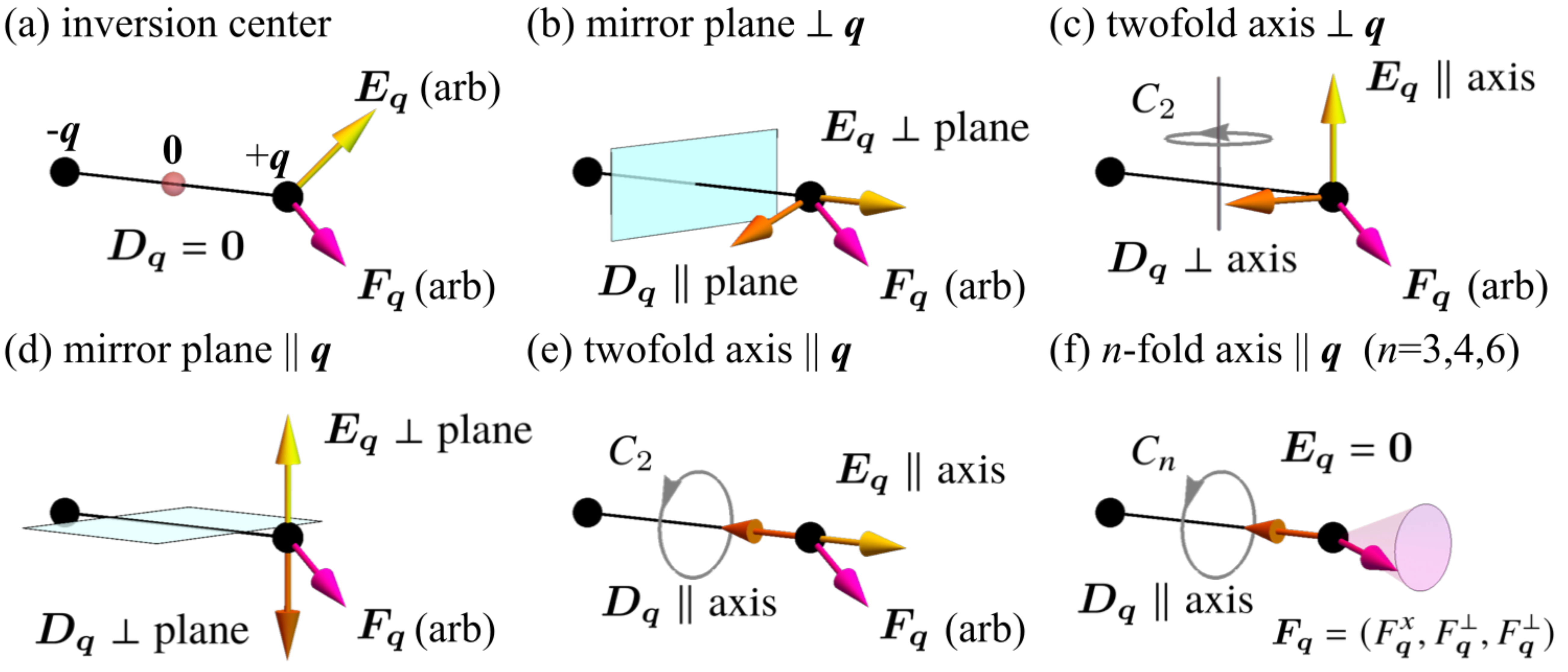} 
\caption{
\label{fig:Rule}
Symmetry operations for wave vectors $\bm{q}$ and $-\bm{q}$ in momentum space:
(a) space inversion at the inversion center denoted as $\bm{0}$, 
(b) mirror perpendicular to $\bm{q}$,
(c) twofold rotation perpendicular to $\bm{q}$, 
(d) mirror parallel to $\bm{q}$,
(e) twofold rotation around $\bm{q}$, and (f) $n$-fold ($n=3,4,6$) rotation around $\bm{q}$.
The nonzero components of  $\bm{D}_{\bm{q}}$, $\bm{E}_{\bm{q}}$, and $\bm{F}_{\bm{q}}$ 
are represented by the orange, yellow, and magenta arrows, respectively.
$\bm{E}_{\bm{q}}$ in (a) and $\bm{F}_{\bm{q}}$ in (a)-(e) are arbitrary (arb).
}
\end{center}
\end{figure*}

Let us start by considering a general form of the momentum-resolved exchange interaction with wave vector $\bm{q}$ in the presence of the time-reversal symmetry. 
It is given by
\begin{align}
\label{eq:Interaction}
\bm{S}_{\bm{q}}^T X_{\bm{q}}  \bm{S}_{-\bm{q}},
\end{align}
with
\begin{align}
\label{eq:CouplingMatrix}
X_{\bm{q}}=
\begin{pmatrix}
F_{\bm{q}}^{x_\mathrm{s}} & E_{\bm{q}}^{z_\mathrm{s}}+iD_{\bm{q}}^{z_\mathrm{s}} & E_{\bm{q}}^{y_\mathrm{s}}-iD_{\bm{q}}^{y_\mathrm{s}} \\
E_{\bm{q}}^{z_\mathrm{s}}-iD_{\bm{q}}^{z_\mathrm{s}} & F_{\bm{q}}^{y_\mathrm{s}}  & E_{\bm{q}}^{x_\mathrm{s}}+iD_{\bm{q}}^{x_\mathrm{s}} \\
E_{\bm{q}}^{y_\mathrm{s}}+iD_{\bm{q}}^{y_\mathrm{s}} & E_{\bm{q}}^{x_\mathrm{s}}-iD_{\bm{q}}^{x_\mathrm{s}}  & F_{\bm{q}}^{z_\mathrm{s}}  
\end{pmatrix}.
\end{align}
In Eq.~(\ref{eq:Interaction}), $\bm{S}_{\bm{q}}^T=(S^{x_\mathrm{s}}_{\bm{q}},S^{y_\mathrm{s}}_{\bm{q}},S^{z_\mathrm{s}}_{\bm{q}})$ is the Fourier transform of the classical spin, $(x_\mathrm{s}, y_\mathrm{s}, z_\mathrm{s})$ are cartesian spin coordinates, and $T$ denotes the transpose of the vector. 
In Eq.~(\ref{eq:CouplingMatrix}), $X_{\bm{q}}$ stands for the general form of the interaction matrix with the real coupling constants $\bm{D}_{\bm{q}}=(D^{x_\mathrm{s}}_{\bm{q}}, D^{y_\mathrm{s}}_{\bm{q}}, D^{z_\mathrm{s}}_{\bm{q}})$, $\bm{E}_{\bm{q}}=(E^{x_\mathrm{s}}_{\bm{q}}, E^{y_\mathrm{s}}_{\bm{q}}, E^{z_\mathrm{s}}_{\bm{q}})$, and $\bm{F}_{\bm{q}}=(F^{x_\mathrm{s}}_{\bm{q}}, F^{y_\mathrm{s}}_{\bm{q}}, F^{z_\mathrm{s}}_{\bm{q}})$; $\bm{D}_{\bm{q}}$ corresponds to the antisymmetric interaction in spin space, while $\bm{E}_{\bm{q}}$ and $\bm{F}_{\bm{q}}$ correspond to the symmetric off-diagonal and diagonal ones, respectively.
For instance, the $x_\mathrm{s}$ components of $\bm{D}_{\bm{q}}$, $\bm{E}_{\bm{q}}$, and $\bm{F}_{\bm{q}}$ are expressed as
\begin{align}
&iD_{\bm{q}}^{x_\mathrm{s}}\left( S^{y_\mathrm{s}}_{\bm{q}}S^{z_\mathrm{s}}_{-\bm{q}} -S^{z_\mathrm{s}}_{\bm{q}}S^{y_\mathrm{s}}_{-\bm{q}} \right),\\
&E_{\bm{q}}^{x_\mathrm{s}}\left( S^{y_\mathrm{s}}_{\bm{q}}S^{z_\mathrm{s}}_{-\bm{q}} + S^{z_\mathrm{s}}_{\bm{q}}S^{y_\mathrm{s}}_{-\bm{q}} \right),\\
&F_{\bm{q}}^{x_\mathrm{s}}\left( S^{x_\mathrm{s}}_{\bm{q}}S^{x_\mathrm{s}}_{-\bm{q}}\right).
\end{align}
It is noted that $\bm{D}_{\bm{q}}$, $\bm{E}_{\bm{q}}$, and $\bm{F}_{\bm{q}}$ show a different transformation by reversing $\bm{q} \to -\bm{q}$ due to the time-reversal symmetry; $\bm{D}_{\bm{q}}=-\bm{D}_{-\bm{q}}$, $\bm{E}_{\bm{q}}=\bm{E}_{-\bm{q}}$, and $\bm{F}_{\bm{q}}=\bm{F}_{-\bm{q}}$.
The nonzero components in $X_{\bm{q}}$ depend on the crystal symmetry.

We find six rules to determine nonzero $\bm{q}$-resolved anisotropic exchange interactions.
In the following, we consider the wave vector $\bm{q}$, which lies inside the Brillouin zone for simplicity; $\bm{q}$ does not lie on the Brillouin zone boundary.
We ignore the sublattice structure in the lattice system, although the following result can be extended to a multi-sublattice case. 
In these assumptions, the six rules for $\bm{D}_{\bm{q}}$, $\bm{E}_{\bm{q}}$, and $\bm{F}_{\bm{q}}$ under the specific crystal (point group) symmetry are given by using the magnetic representation theory:
\begin{itemize}
\item[(a)] The space inversion symmetry imposes $\bm{D}_{\bm{q}}=\bm{0}$, while there is no constraint on $\bm{E}_{\bm{q}}$ and $\bm{F}_{\bm{q}}$ [Fig.~\ref{fig:Rule}(a)].
\item[(b)] The mirror symmetry with respect to the plane perpendicular to $\bm{q}$ imposes $\bm{D}_{\bm{q}}\parallel$ plane and $\bm{E}_{\bm{q}}\perp$ plane, while there is no constraint on $\bm{F}_{\bm{q}}$ [Fig.~\ref{fig:Rule}(b)].
\item[(c)] The twofold rotational symmetry around the axis perpendicular to $\bm{q}$ imposes $\bm{D}_{\bm{q}}\perp$ axis and $\bm{E}_{\bm{q}}\parallel$ axis, while there is no constraint on $\bm{F}_{\bm{q}}$ [Fig.~\ref{fig:Rule}(c)].
\item[(d)] The mirror symmetry with respect to the plane parallel to $\bm{q}$ imposes $\bm{D}_{\bm{q}}\perp$ plane and $\bm{E}_{\bm{q}}\perp$ plane, while there is no constraint on $\bm{F}_{\bm{q}}$ [Fig.~\ref{fig:Rule}(d)].
\item[(e)] The twofold rotational symmetry around the axis parallel to $\bm{q}$ imposes $\bm{D}_{\bm{q}}\parallel$ axis and $\bm{E}_{\bm{q}}\parallel$ axis, while there is no constraint on $\bm{F}_{\bm{q}}$ [Fig.~\ref{fig:Rule}(e)].
\item[(f)] The $n$-fold ($n=3,4,6$) rotational symmetries around the axis parallel to $\bm{q}$ imposes $\bm{D}_{\bm{q}}\parallel$ axis, $\bm{E}_{\bm{q}}=\bm{0}$, and $\bm{F}_{\bm{q}} =(F_{\bm{q}}^{x_\mathrm{s}},F_{\bm{q}}^\perp,F_{\bm{q}}^\perp)$ [Fig.~\ref{fig:Rule}(f)].
\end{itemize}
Here, $x_\mathrm{s}$ is taken along the $\bm{q}$ direction and each operation leaves the origin $\bm{q}=(0,0,0)$ invariant.
The detailed discussion is given in Appendix~\ref{ap:MagRep}. 

The above rules indicate that nonzero components of $\bm{D}_{\bm{q}}$ and $\bm{E}_{\bm{q}}$ largely depend on the point group symmetry, while there is only one constraint for $\bm{F}_{\bm{q}}$.
In particular, the rules for $\bm{D}_{\bm{q}}$ are the counterpart in momentum space of Moriya's rule in real space~\cite{moriya1960anisotropic}. 
In addition, the condition for the symmetric off-diagonal interaction $\bm{E}_{\bm{q}}$ is also obtained, where the nonzero component of $\bm{E}_{\bm{q}}$ is different from (the same as) that of $\bm{D}_{\bm{q}}$ for the rules (a), (b), (c), and (f) [(d) and (e)].

\subsection{Effective spin model under space groups}
\label{sec:Model1}

\begin{figure}[t!]
\begin{center}
\includegraphics[width=1.0\hsize]{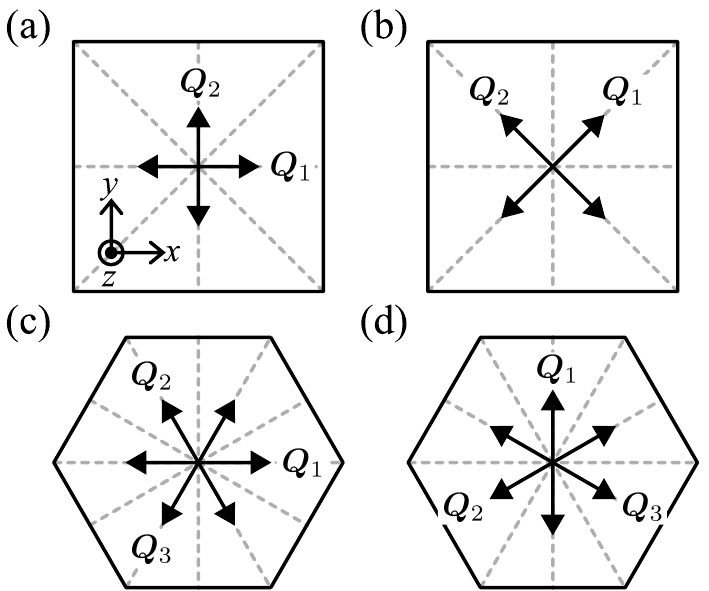} 
\caption{
\label{fig:Qset} 
A set of $\{\bm{Q}\}$ along the different high symmetric lines inside the first Brillouin zone in (a, b) tetragonal crystal systems and (c, d) hexagonal and trigonal crystal systems. 
In (a) and (b), $\bm{Q}_1$ and $\bm{Q}_2$ are connected by the fourfold rotation around the $z$ axis, while in (c) and (d), $\bm{Q}_1$, $\bm{Q}_2$, and $\bm{Q}_3$ are connected by the threefold rotation. 
The wave vectors in $\{\bm{Q}\}$ lie on the $xy$ plane.   
}
\end{center}
\end{figure}

The above symmetry argument gives a general form of the spin Hamiltonian in any primitive lattice systems, which is obtained by summing the contribution from entire $\bm{q}$ in the first Brillouin zone as
\begin{align}
\label{eq:ABS}
\mathcal{H}= -\sum_{\bm{q}}\bm{S}_{\bm{q}}^T X_{\bm{q}}  \bm{S}_{-\bm{q}}, 
\end{align}
where the minus sign is added for later convenience (see Eq.~(\ref{eq:Omega}) in Sec.~\ref{sec:PAM}).
As a demonstration, we here discuss a specific form of $X_{\bm{q}}$ for 24 gray symmorphic space groups belonging to the tetragonal, hexagonal, and trigonal crystal systems shown in Tables~\ref{tab:tetragonal}-\ref{tab:trigonal}.
The gray symmorphic space group $\mathbf{M}$ defined as $\mathbf{M}=\mathbf{H}+\theta\mathbf{H}$ with the symmorphic space group $\mathbf{H}$ and the time-reversal operation $\theta$~\cite{bradley2009mathematical}.
In the model in Eq.~(\ref{eq:ABS}), we take the spin coordinates $x_\mathrm{s}$, $y_\mathrm{s}$, and $z_\mathrm{s}$ along the $x$, $y$, and $z$ directions shown in Fig.~(\ref{fig:Qset})(a), respectively.

Although one can obtain the optimal spin configurations by performing unbiased numerical simulations, such as the Monte Carlo simulations, for the model in Eq.~(\ref{eq:ABS}),
one notices that a part of $X_{\bm{q}}$ is enough to discuss the magnetic instability at low temperatures in most cases. 
The unbiased numerical simulations for the model in Eq.~(\ref{eq:ABS}) needs tremendous computational cost, since the model roughly has $9N/2$ independent interaction parameters in $X_{\bm{q}}$ at most, where $N$ is the system size and the factor $1/2$ is owing to the constraint under the time-reversal symmetry, $X_{\bm{q}}=X^*_{-\bm{q}}$.
However, only a few $\bm{q}$ components of the interactions contribute to the ground-state energy in most cases, which enables us to reduce the computational cost. 
For example, in the case of the isotropic spin model, which corresponds to $F_{\bm{q}}^{x}=F_{\bm{q}}^{y}=F_{\bm{q}}^{z}$ and $\bm{D}_{\bm{q}}=\bm{E}_{\bm{q}}=0$, the ground state becomes the spiral ordering with the wave vector $\bm{q}^*$ that gives the largest value of $X_{\bm{q}}$. 
In this case, the interactions with other $\bm{q} (\neq \bm{q}^*)$ do not contribute to the energy, which can be neglected.  
Meanwhile, in the case of the anisotropic spin model, the instability toward a multiple-$Q$ state, which is a superposition of spin density waves at different $\bm{q}$, can occur. 
In such a situation, the superposition between the wave vectors connected by the point group operations tends to be favored, since they give the same largest eigenvalue of $X_{\bm{q}}$ so as to satisfy the lattice symmetry. 
When considering the tetragonal crystal system, there are at least two equivalent wave vectors connected by the fourfold rotation, $\bm{q}^*=\bm{Q}_1$ and $\bm{Q}_2$, whereas there are at least three equivalent wave vectors connected by the threefold rotation, $\bm{q}^*=\bm{Q}_1$, $\bm{Q}_2$, and $\bm{Q}_3$, in the hexagonal and trigonal crystal systems. 
The schematic pictures of the symmetry-related wave vectors are shown in Fig.~\ref{fig:Qset}. 
Thus, it is natural to take into account their contributions in Eq.~(\ref{eq:ABS}) to examine the low-temperature spin configuration. 

From the above argument, the effective model at low temperatures is simplified as
\begin{align}
\label{eq:ABS2}
\mathcal{H}^{\rm eff}
= -\sum_{\bm{q}\in\{\bm{Q}\}
}\bm{S}_{\bm{q}}^T X_{\bm{q}}  \bm{S}_{-\bm{q}},
\end{align}
where $\{\bm{Q}\}$ represents a symmetry-related wave vectors: For example, $\{\bm{Q}\}=\pm\bm{Q}_1, \pm\bm{Q}_2$ for the tetragonal crystal system with the fourfold rotational symmetry and $\{\bm{Q}\}=\pm\bm{Q}_1, \pm\bm{Q}_2, \pm\bm{Q}_3$ for the hexagonal and trigonal crystal systems with the threefold rotational symmetry. 
This type of the effective bilinear spin model has been studied to examine the multiple-$Q$ instabilities including the SkX and hedgehog lattice in both insulating~\cite{leonov2015multiply,Hayami_PhysRevB.103.224418,Hayami_PhysRevB.105.014408} and itinerant magnetic systems~\cite{Hayami_PhysRevB.103.054422,kato2021spin,Hayami_PhysRevLett.121.137202,yambe2021skyrmion}. 
It is noted that one might additionally consider the effect of the interaction from the higher-harmonics wave vectors, such as $2\bm{Q}_1$ and $\bm{Q}_1+\bm{Q}_2$, in Eq.~(\ref{eq:ABS2}) when discussing the stability of the multiple-$Q$ states~\cite{Hayami_doi:10.7566/JPSJ.89.103702,doi:10.7566/JPSJ.91.023705}. 
We demonstrate such a case in Sec~\ref{sec:SpecificExample}.

\begin{table*}
\caption{\label{tab:tetragonal}
Interaction matrix $X_{\bm{Q}_1}$ and the number of independent components $N_\mathrm{c}$ in the tetragonal crystal systems for the high-symmetric wave vector $\bm{Q}_1$ shown in Figs.~\ref{fig:Qset}(a) and \ref{fig:Qset}(b).
The spin coordinates $x_\mathrm{s}$, $y_\mathrm{s}$, and $z_\mathrm{s}$ are taken along the $x$, $y$, and $z$ directions in Fig.~(\ref{fig:Qset})(a), respectively. 
}
\begin{ruledtabular}
\begin{tabular}{ccccc}
   &\multicolumn{2}{c}{$\bm{Q}_1\parallel \hat{\bm{x}}$}  & \multicolumn{2}{c}{$\bm{Q}_1\parallel[110]$}  \\
 \cline{2-3} \cline{4-5}
space group $\mathbf{H}$&  $X_{\bm{Q}_1}$ & $N_\mathrm{c}$ & $X_{\bm{Q}_1}$ & $N_\mathrm{c}$ \\ \hline
$P4/mmm$ & $\begin{pmatrix}
F_{\bm{Q}_1}^{x} & 0 & 0 \\
0 & F_{\bm{Q}_1}^{y}  & 0 \\
0& 0  & F_{\bm{Q}_1}^{z}  
\end{pmatrix}$ 
& 3& $\begin{pmatrix}
F_{\bm{Q}_1}^{x} & E_{\bm{Q}_1}^{z} & 0 \\
E_{\bm{Q}_1}^{z} & F_{\bm{Q}_1}^{x}  &0  \\
0& 0 & F_{\bm{Q}_1}^{z}  
\end{pmatrix}$ & 3
 \\
$P422$ & $\begin{pmatrix}
F_{\bm{Q}_1}^{x} & 0 & 0 \\
0 & F_{\bm{Q}_1}^{y}  & iD_{\bm{Q}_1}^{x} \\
0 &  -iD_{\bm{Q}_1}^{x}  & F_{\bm{Q}_1}^{z}  
\end{pmatrix}$ & 4& $\begin{pmatrix}
F_{\bm{Q}_1}^{x} & E_{\bm{Q}_1}^{z} & -iD_{\bm{Q}_1}^{x} \\
E_{\bm{Q}_1}^{z} & F_{\bm{Q}_1}^{x}  & iD_{\bm{Q}_1}^{x} \\
iD_{\bm{Q}_1}^{x} &-iD_{\bm{Q}_1}^{x}  & F_{\bm{Q}_1}^{z}  
\end{pmatrix}$ & 4
\\
$P\bar{4}2m$ &  $\begin{pmatrix}
F_{\bm{Q}_1}^{x} & 0 & 0 \\
0 & F_{\bm{Q}_1}^{y}  & iD_{\bm{Q}_1}^{x} \\
0 &  -iD_{\bm{Q}_1}^{x}  & F_{\bm{Q}_1}^{z}  
\end{pmatrix}$ 
& 4& $\begin{pmatrix}
F_{\bm{Q}_1}^{x} & E_{\bm{Q}_1}^{z} & iD_{\bm{Q}_1}^{x} \\
E_{\bm{Q}_1}^{z} & F_{\bm{Q}_1}^{x}  & iD_{\bm{Q}_1}^{x} \\
-iD_{\bm{Q}_1}^{x} &-iD_{\bm{Q}_1}^{x}  & F_{\bm{Q}_1}^{z}  
\end{pmatrix}$ & 4
 \\
 $P\bar{4}m2$ & $\begin{pmatrix}
F_{\bm{Q}_1}^{x} & 0 & -iD_{\bm{Q}_1}^{y} \\
0 & F_{\bm{Q}_1}^{y}  & 0 \\
iD_{\bm{Q}_1}^{y} & 0 & F_{\bm{Q}_1}^{z}  
\end{pmatrix}$ 
& 4& $\begin{pmatrix}
F_{\bm{Q}_1}^{x} & E_{\bm{Q}_1}^{z} & -iD_{\bm{Q}_1}^{x} \\
E_{\bm{Q}_1}^{z} & F_{\bm{Q}_1}^{x}  & iD_{\bm{Q}_1}^{x} \\
iD_{\bm{Q}_1}^{x} &-iD_{\bm{Q}_1}^{x}  & F_{\bm{Q}_1}^{z}  
\end{pmatrix}$ & 4
\\
$P4mm$ & $\begin{pmatrix}
F_{\bm{Q}_1}^{x} & 0 & -iD_{\bm{Q}_1}^{y} \\
0 & F_{\bm{Q}_1}^{y}  & 0 \\
iD_{\bm{Q}_1}^{y} & 0 & F_{\bm{Q}_1}^{z}  
\end{pmatrix}$ & 4 & $\begin{pmatrix}
F_{\bm{Q}_1}^{x} & E_{\bm{Q}_1}^{z} & iD_{\bm{Q}_1}^{x} \\
E_{\bm{Q}_1}^{z} & F_{\bm{Q}_1}^{x}  & iD_{\bm{Q}_1}^{x} \\
-iD_{\bm{Q}_1}^{x} &-iD_{\bm{Q}_1}^{x}  & F_{\bm{Q}_1}^{z}  
\end{pmatrix}$ & 4
\\
$P4/m$ & $\begin{pmatrix}
F_{\bm{Q}_1}^{x} & E_{\bm{Q}_1}^{z} & 0 \\
E_{\bm{Q}_1}^{z} & F_{\bm{Q}_1}^{y}  &0  \\
0& 0 & F_{\bm{Q}_1}^{z}  
\end{pmatrix}$  
& 4& $\begin{pmatrix}
F_{\bm{Q}_1}^{x} & E_{\bm{Q}_1}^{z} & 0 \\
E_{\bm{Q}_1}^{z} & F_{\bm{Q}_1}^{y}  &0  \\
0& 0 & F_{\bm{Q}_1}^{z}  
\end{pmatrix}$ & 4
\\
$P4$ &  $\begin{pmatrix}
F_{\bm{Q}_1}^{x} & E_{\bm{Q}_1}^{z} & -iD_{\bm{Q}_1}^{y} \\
E_{\bm{Q}_1}^{z} & F_{\bm{Q}_1}^{y}  & iD_{\bm{Q}_1}^{x} \\
iD_{\bm{Q}_1}^{y} &-iD_{\bm{Q}_1}^{x}  & F_{\bm{Q}_1}^{z}  
\end{pmatrix}$ & 6& $\begin{pmatrix}
F_{\bm{Q}_1}^{x} & E_{\bm{Q}_1}^{z} & -iD_{\bm{Q}_1}^{y} \\
E_{\bm{Q}_1}^{z} & F_{\bm{Q}_1}^{y}  & iD_{\bm{Q}_1}^{x} \\
iD_{\bm{Q}_1}^{y} &-iD_{\bm{Q}_1}^{x}  & F_{\bm{Q}_1}^{z}  
\end{pmatrix}$ & 6
\\ 
$P\bar{4}$ & $\begin{pmatrix}
F_{\bm{Q}_1}^{x} & E_{\bm{Q}_1}^{z} & -iD_{\bm{Q}_1}^{y} \\
E_{\bm{Q}_1}^{z} & F_{\bm{Q}_1}^{y}  & iD_{\bm{Q}_1}^{x} \\
iD_{\bm{Q}_1}^{y} &-iD_{\bm{Q}_1}^{x}  & F_{\bm{Q}_1}^{z}  
\end{pmatrix}$ 
& 6& $\begin{pmatrix}
F_{\bm{Q}_1}^{x} & E_{\bm{Q}_1}^{z} & -iD_{\bm{Q}_1}^{y} \\
E_{\bm{Q}_1}^{z} & F_{\bm{Q}_1}^{y}  & iD_{\bm{Q}_1}^{x} \\
iD_{\bm{Q}_1}^{y} &-iD_{\bm{Q}_1}^{x}  & F_{\bm{Q}_1}^{z}  
\end{pmatrix}$ & 6
\end{tabular}
\end{ruledtabular}
\end{table*}

\begin{table*}
\caption{\label{tab:hexagonal}
Interaction matrix $X_{\bm{Q}_1}$ and the number of independent components $N_\mathrm{c}$ in the hexagonal crystal systems for the high-symmetric wave vector $\bm{Q}_1$ shown in Figs.~\ref{fig:Qset}(c) and \ref{fig:Qset}(d).
The spin coordinates $x_\mathrm{s}$, $y_\mathrm{s}$, and $z_\mathrm{s}$ are taken along the $x$, $y$, and $z$ directions in Fig.~(\ref{fig:Qset})(a), respectively.
}
\begin{ruledtabular}
\begin{tabular}{ccccc}
 & \multicolumn{2}{c}{$\bm{Q}_1\parallel  \hat{\bm{x}}$}  & \multicolumn{2}{c}{$\bm{Q}_1\parallel  \hat{\bm{y}}$}  \\ \cline{2-3}\cline{4-5} 
space group $\mathbf{H}$& $X_{\bm{Q}_1}$ & $N_\mathrm{c}$ & $X_{\bm{Q}_1}$ & $N_\mathrm{c}$ \\ \hline
$P6/mmm$ & $\begin{pmatrix}
F_{\bm{Q}_1}^{x} & 0 & 0 \\
0 & F_{\bm{Q}_1}^{y}  & 0 \\
0& 0  & F_{\bm{Q}_1}^{z}  
\end{pmatrix}$ & 3&$\begin{pmatrix}
F_{\bm{Q}_1}^{x} & 0 & 0 \\
0 & F_{\bm{Q}_1}^{y}  & 0 \\
0& 0  & F_{\bm{Q}_1}^{z}  
\end{pmatrix}$ & 3
\\
$P622$ & $\begin{pmatrix}
F_{\bm{Q}_1}^{x} & 0 & 0 \\
0 & F_{\bm{Q}_1}^{y}  & iD_{\bm{Q}_1}^{x} \\
0 &  -iD_{\bm{Q}_1}^{x}  & F_{\bm{Q}_1}^{z}  
\end{pmatrix}$ & 4& $\begin{pmatrix}
F_{\bm{Q}_1}^{x} & 0 & -iD_{\bm{Q}_1}^{y} \\
0 & F_{\bm{Q}_1}^{y}  & 0 \\
iD_{\bm{Q}_1}^{y} & 0 & F_{\bm{Q}_1}^{z}  
\end{pmatrix}$ & 4 
\\
$P\bar{6}m2$ &$\begin{pmatrix}
F_{\bm{Q}_1}^{x} & iD_{\bm{Q}_1}^{z} &0  \\
-iD_{\bm{Q}_1}^{z}  & F_{\bm{Q}_1}^{y}  & 0 \\
0 & 0& F_{\bm{Q}_1}^{z}  
\end{pmatrix}$ & 4&$\begin{pmatrix}
F_{\bm{Q}_1}^{x} &0 &0  \\
0 & F_{\bm{Q}_1}^{y}  & 0 \\
0 & 0& F_{\bm{Q}_1}^{z}  
\end{pmatrix}$ & 3
 \\
$P\bar{6}2m$ &$\begin{pmatrix}
F_{\bm{Q}_1}^{x} &0 &0  \\
0 & F_{\bm{Q}_1}^{y}  & 0 \\
0 & 0& F_{\bm{Q}_1}^{z}  
\end{pmatrix}$  & 3& $\begin{pmatrix}
F_{\bm{Q}_1}^{x} & iD_{\bm{Q}_1}^{z} &0  \\
-iD_{\bm{Q}_1}^{z}  & F_{\bm{Q}_1}^{y}  & 0 \\
0 & 0& F_{\bm{Q}_1}^{z}  
\end{pmatrix}$ & 4
\\
 $P6mm$ & $\begin{pmatrix}
F_{\bm{Q}_1}^{x} & 0 & -iD_{\bm{Q}_1}^{y} \\
0 & F_{\bm{Q}_1}^{y}  & 0 \\
iD_{\bm{Q}_1}^{y} & 0 & F_{\bm{Q}_1}^{z}  
\end{pmatrix}$ & 4& $\begin{pmatrix}
F_{\bm{Q}_1}^{x} & 0 & 0 \\
0 & F_{\bm{Q}_1}^{y}  & iD_{\bm{Q}_1}^{x} \\
0 &  -iD_{\bm{Q}_1}^{x}  & F_{\bm{Q}_1}^{z}  
\end{pmatrix}$ & 4
\\
$P6/m$ &$\begin{pmatrix}
F_{\bm{Q}_1}^{x} & E_{\bm{Q}_1}^{z} & 0 \\
E_{\bm{Q}_1}^{z} & F_{\bm{Q}_1}^{y}  &0  \\
0& 0 & F_{\bm{Q}_1}^{z}  
\end{pmatrix}$ & 4&$\begin{pmatrix}
F_{\bm{Q}_1}^{x} & E_{\bm{Q}_1}^{z} & 0 \\
E_{\bm{Q}_1}^{z} & F_{\bm{Q}_1}^{y}  &0  \\
0& 0 & F_{\bm{Q}_1}^{z}  
\end{pmatrix}$ & 4
\\
$P\bar{6}$ & $ \begin{pmatrix}
F_{\bm{Q}_1}^{x} & E_{\bm{Q}_1}^{z}+iD_{\bm{Q}_1}^{z} &0  \\
E_{\bm{Q}_1}^{z}-iD_{\bm{Q}_1}^{z}  & F_{\bm{Q}_1}^{y}  & 0 \\
0 & 0& F_{\bm{Q}_1}^{z}  
\end{pmatrix}$ & 4& $ \begin{pmatrix}
F_{\bm{Q}_1}^{x} &  E_{\bm{Q}_1}^{z}+iD_{\bm{Q}_1}^{z} &0  \\
E_{\bm{Q}_1}^{z}-iD_{\bm{Q}_1}^{z}  & F_{\bm{Q}_1}^{y}  & 0 \\
0 & 0& F_{\bm{Q}_1}^{z}  
\end{pmatrix}$ & 4\\
$P6$ &$\begin{pmatrix}
F_{\bm{Q}_1}^{x} & E_{\bm{Q}_1}^{z} & -iD_{\bm{Q}_1}^{y} \\
E_{\bm{Q}_1}^{z} & F_{\bm{Q}_1}^{y}  & iD_{\bm{Q}_1}^{x} \\
iD_{\bm{Q}_1}^{y} &-iD_{\bm{Q}_1}^{x}  & F_{\bm{Q}_1}^{z}  
\end{pmatrix}$ & 6& $\begin{pmatrix}
F_{\bm{Q}_1}^{x} & E_{\bm{Q}_1}^{z} & -iD_{\bm{Q}_1}^{y} \\
E_{\bm{Q}_1}^{z} & F_{\bm{Q}_1}^{y}  & iD_{\bm{Q}_1}^{x} \\
iD_{\bm{Q}_1}^{y} &-iD_{\bm{Q}_1}^{x}  & F_{\bm{Q}_1}^{z}  
\end{pmatrix}$ & 6 
\end{tabular}
\end{ruledtabular}
\end{table*}

\begin{table*}
\caption{\label{tab:trigonal}
Interaction matrix $X_{\bm{Q}_1}$ and the number of independent components $N_\mathrm{c}$ in the trigonal crystal systems for the high-symmetric wave vector $\bm{Q}_1$ shown in Figs.~\ref{fig:Qset}(c) and \ref{fig:Qset}(d).
The spin coordinates $x_\mathrm{s}$, $y_\mathrm{s}$, and $z_\mathrm{s}$ are taken along the $x$, $y$, and $z$ directions in Fig.~(\ref{fig:Qset})(a), respectively.
}
\begin{ruledtabular}
\begin{tabular}{ccccc}
 &\multicolumn{2}{c}{$\bm{Q}_1\parallel \hat{\bm{x}}$}  & \multicolumn{2}{c}{$\bm{Q}_1\parallel \hat{\bm{y}}$}  \\ 
\cline{2-3} \cline{4-5}
space group $\mathbf{H}$&  $X_{\bm{Q}_1}$ & $N_\mathrm{c}$ & $X_{\bm{Q}_1}$ & $N_\mathrm{c}$ \\  \hline
$P\bar{3}m1$ & $\begin{pmatrix}
F_{\bm{Q}_1}^{x} & 0 & 0 \\
0 & F_{\bm{Q}_1}^{y}  & E_{\bm{Q}_1}^{x} \\
0 & E_{\bm{Q}_1}^{x}  & F_{\bm{Q}_1}^{z}  
\end{pmatrix}$ & 4  &  $ \begin{pmatrix}
F_{\bm{Q}_1}^{x} & 0 & 0 \\
0 & F_{\bm{Q}_1}^{y}  & E_{\bm{Q}_1}^{x} \\
0 & E_{\bm{Q}_1}^{x}  & F_{\bm{Q}_1}^{z}   
\end{pmatrix}$ & 4 
 \\
$P\bar{3}1m$ & $\begin{pmatrix}
F_{\bm{Q}_1}^{x} & 0 & E_{\bm{Q}_1}^{y} \\
0 & F_{\bm{Q}_1}^{y}  & 0 \\
 E_{\bm{Q}_1}^{y} & 0 & F_{\bm{Q}_1}^{z}  
\end{pmatrix}$ & 4  &  $ \begin{pmatrix}
F_{\bm{Q}_1}^{x} & 0 & E_{\bm{Q}_1}^{y} \\
0 & F_{\bm{Q}_1}^{y}  & 0 \\
 E_{\bm{Q}_1}^{y} & 0 & F_{\bm{Q}_1}^{z}   
\end{pmatrix}$ & 4
\\
$P321$ & $ \begin{pmatrix}
F_{\bm{Q}_1}^{x} & 0 & 0  \\
0 & F_{\bm{Q}_1}^{y}  & E_{\bm{Q}_1}^{x}+iD_{\bm{Q}_1}^{x} \\
0 & E_{\bm{Q}_1}^{x}-iD_{\bm{Q}_1}^{x}  & F_{\bm{Q}_1}^{z}  
\end{pmatrix}$ & 5  &  $ \begin{pmatrix}
F_{\bm{Q}_1}^{x} & iD_{\bm{Q}_1}^{z} & -iD_{\bm{Q}_1}^{y} \\
-iD_{\bm{Q}_1}^{z} & F_{\bm{Q}_1}^{y}  & E_{\bm{Q}_1}^{x} \\
iD_{\bm{Q}_1}^{y} & E_{\bm{Q}_1}^{x}  & F_{\bm{Q}_1}^{z}  
\end{pmatrix}$ & 6 
\\ 
$P312$ & $\begin{pmatrix}
F_{\bm{Q}_1}^{x} & iD_{\bm{Q}_1}^{z} & E_{\bm{Q}_1}^{y} \\
-iD_{\bm{Q}_1}^{z} & F_{\bm{Q}_1}^{y}  & iD_{\bm{Q}_1}^{x} \\
E_{\bm{Q}_1}^{y} & -iD_{\bm{Q}_1}^{x}  & F_{\bm{Q}_1}^{z}  
\end{pmatrix}$ & 6  &  $ \begin{pmatrix}
F_{\bm{Q}_1}^{x} & 0  & E_{\bm{Q}_1}^{y}-iD_{\bm{Q}_1}^{y} \\
0 & F_{\bm{Q}_1}^{y}  & 0 \\
E_{\bm{Q}_1}^{y}+iD_{\bm{Q}_1}^{y} & 0  & F_{\bm{Q}_1}^{z}  
\end{pmatrix}$ & 5
\\
$P3m1$ & $\begin{pmatrix}
F_{\bm{Q}_1}^{x} & iD_{\bm{Q}_1}^{z} & -iD_{\bm{Q}_1}^{y} \\
-iD_{\bm{Q}_1}^{z} & F_{\bm{Q}_1}^{y}  & E_{\bm{Q}_1}^{x} \\
iD_{\bm{Q}_1}^{y} & E_{\bm{Q}_1}^{x}  & F_{\bm{Q}_1}^{z}  
\end{pmatrix}$ & 6  &  $ \begin{pmatrix}
F_{\bm{Q}_1}^{x} & 0 & 0  \\
0 & F_{\bm{Q}_1}^{y}  & E_{\bm{Q}_1}^{x}+iD_{\bm{Q}_1}^{x} \\
0 & E_{\bm{Q}_1}^{x}-iD_{\bm{Q}_1}^{x}  & F_{\bm{Q}_1}^{z}  
\end{pmatrix}$ & 5  \\
$P31m$ & $\begin{pmatrix}
F_{\bm{Q}_1}^{x} & 0 & E_{\bm{Q}_1}^{y}-iD_{\bm{Q}_1}^{y} \\
0 & F_{\bm{Q}_1}^{y}  & 0\\
E_{\bm{Q}_1}^{y}+iD_{\bm{Q}_1}^{y} & 0 & F_{\bm{Q}_1}^{z}  
\end{pmatrix}$ & 5  &  $ \begin{pmatrix}
F_{\bm{Q}_1}^{x} & iD_{\bm{Q}_1}^{z} & E_{\bm{Q}_1}^{y} \\
-iD_{\bm{Q}_1}^{z} & F_{\bm{Q}_1}^{y}  & iD_{\bm{Q}_1}^{x} \\
E_{\bm{Q}_1}^{y} & -iD_{\bm{Q}_1}^{x}  & F_{\bm{Q}_1}^{z}  
\end{pmatrix}$ & 6
\\
$P\bar{3}$ & $\begin{pmatrix}
F_{\bm{Q}_1}^{x} & E_{\bm{Q}_1}^{z} & E_{\bm{Q}_1}^{y} \\
E_{\bm{Q}_1}^{z} & F_{\bm{Q}_1}^{y}  & E_{\bm{Q}_1}^{x} \\
E_{\bm{Q}_1}^{y} & E_{\bm{Q}_1}^{x}  & F_{\bm{Q}_1}^{z}  
\end{pmatrix}$ & 6  &  $ \begin{pmatrix}
F_{\bm{Q}_1}^{x} & E_{\bm{Q}_1}^{z} & E_{\bm{Q}_1}^{y} \\
E_{\bm{Q}_1}^{z} & F_{\bm{Q}_1}^{y}  & E_{\bm{Q}_1}^{x} \\
E_{\bm{Q}_1}^{y} & E_{\bm{Q}_1}^{x}  & F_{\bm{Q}_1}^{z}    
\end{pmatrix}$ & 6
 \\ 
$P3$ & $ \begin{pmatrix}
F_{\bm{Q}_1}^{x} & E_{\bm{Q}_1}^{z}+iD_{\bm{Q}_1}^{z} & E_{\bm{Q}_1}^{y}-iD_{\bm{Q}_1}^{y} \\
E_{\bm{Q}_1}^{z}-iD_{\bm{Q}_1}^{z} & F_{\bm{Q}_1}^{y}  & E_{\bm{Q}_1}^{x}+iD_{\bm{Q}_1}^{x} \\
E_{\bm{Q}_1}^{y}+iD_{\bm{Q}_1}^{y} & E_{\bm{Q}_1}^{x}-iD_{\bm{Q}_1}^{x}  & F_{\bm{Q}_1}^{z}  
\end{pmatrix}$ & 9  &  $ \begin{pmatrix}
F_{\bm{Q}_1}^{x} & E_{\bm{Q}_1}^{z}+iD_{\bm{Q}_1}^{z} & E_{\bm{Q}_1}^{y}-iD_{\bm{Q}_1}^{y} \\
E_{\bm{Q}_1}^{z}-iD_{\bm{Q}_1}^{z} & F_{\bm{Q}_1}^{y}  & E_{\bm{Q}_1}^{x}+iD_{\bm{Q}_1}^{x} \\
E_{\bm{Q}_1}^{y}+iD_{\bm{Q}_1}^{y} & E_{\bm{Q}_1}^{x}-iD_{\bm{Q}_1}^{x}  & F_{\bm{Q}_1}^{z}  
\end{pmatrix}$ & 9
\end{tabular}
\end{ruledtabular}
\end{table*}

Once $\{\bm{Q}\}$ and the space group in the model in Eq.~(\ref{eq:ABS2}) are determined, we can write down nonzero components of $X_{\bm{q}}$ following the rules in Sec.~\ref{sec:Rule}.  
We here present the results of $X_{\bm{q}}$ in the tetragonal, hexagonal, and trigonal crystal systems, where the wave vectors lie along the high symmetry lines; $\bm{Q}_1$ is taken along the $x$ [Fig.~\ref{fig:Qset}(a)] and $[110]$ [Fig.~\ref{fig:Qset}(b)] axes in the tetragonal crystal system, while $\bm{Q}_1$ is taken along the $x$ [Fig.~\ref{fig:Qset}(c)] and $y$ [Fig.~\ref{fig:Qset}(d)] axes in the hexagonal and trigonal crystal systems. 
The specific form of $X_{\bm{Q}_1}$ is summarized in Table~\ref{tab:tetragonal} for the tetragonal crystal systems ($P4/mmm, P422, P\bar{4}2m, P\bar{4}m2, P4mm, P4/m, P4$, and $P\bar{4}$), Table~\ref{tab:hexagonal} for the hexagonal crystal systems ($P6/mmm, P622, P\bar{6}m2, P\bar{6}2m, P6mm, P6/m, P\bar{6}$, and $P6$), and Table~\ref{tab:trigonal} for the trigonal crystal systems ($P\bar{3}m1, P\bar{3}1m, P321, P312, P3m1, P31m, P\bar{3}$, and $P3$). 
In Tables~\ref{tab:tetragonal}-\ref{tab:trigonal}, $N_{\rm c}$ stands for the number of independent components of $X_{\bm{Q}_1}$. 
The results for the low-symmetric $\{\bm{Q}\}$ are shown in Appendix~\ref{ap:ModelWithLowQ}. 

From Tables~\ref{tab:tetragonal}-\ref{tab:trigonal}, one finds two features irrelevant to the details of the space group. 
First, there are at least three independent components ($N_\mathrm{c}\ge3$) in all cases. 
Among them, one component corresponds to the isotropic contribution, $F_{\bm{q}}^\mathrm{iso}=(F_{\bm{q}}^{x}+F_{\bm{q}}^{y}+F_{\bm{q}}^{z})/3$. 
Second, the antisymmetric interaction $\bm{D}_{\bm{q}}$ only appears in the absence of the spatial inversion symmetry, while the symmetric ones $\bm{E}_{\bm{q}}$ and $\bm{F}_{\bm{q}}$ appear irrespective of the inversion symmetry, as shown in the rule (a) in Sec.~\ref{sec:Rule}. 

In addition, there are three characteristics in Tables~\ref{tab:tetragonal}-\ref{tab:trigonal}. 
The first is that the interaction matrix depends on not only the space group but also the direction of $\bm{Q}_1$, which reflects the different symmetry of the wave vectors. 
In particular, $X_{\bm{Q}_1\parallel \hat{\bm{x}}}$ and $X_{\bm{Q}_1\parallel \hat{\bm{y}}}$ in $P\bar{6}m2$, $P\bar{6}2m$, $P321$, $P312$, $P3m1$, and $P31m$ crystals have a different number of independent components. 
The second is that the diagonal components of the interactions are different for all the space groups except for $\bm{Q}_1 \parallel [110]$ in $P4/mmm, P422, P\bar{4}2m, P\bar{4}m2, P4mm$, and $P4/m$ crystal systems, i.e., $F_{\bm{Q}_1}^{x}\neq F_{\bm{Q}_1}^{y} \neq F_{\bm{Q}_1}^{z}$; the difference between $F_{\bm{Q}_1}^{x}$ and $F_{\bm{Q}_1}^{z}$ ($F_{\bm{Q}_1}^{y}$ and $F_{\bm{Q}_1}^{z}$) is owing to an inequivalence between the in-plane and $z$ directions, while that between $F_{\bm{Q}_1}^{x}$ and $F_{\bm{Q}_1}^{y}$ is owing to the discrete rotational symmetry around the principal axis. 
It is noted that in the case of $\bm{Q}_1 \parallel [110]$ in $P4/mmm, P422, P\bar{4}2m, P\bar{4}m2, P4mm$, and $P4/m$ crystal systems, all the space groups allow nonzero $E^{z}_{\bm{Q}_1}$ instead of different $F_{\bm{Q}_1}^{x}$ and $F_{\bm{Q}_1}^{y}$. 
The third is that the symmetric off-diagonal components, $E^x_{\bm{Q}_1}$ and $E^y_{\bm{Q}_1}$, only appear in the trigonal crystal systems, which do not have the twofold axis along the $z$ direction and the horizontal mirror plane [see rules (c) and (d)]. 
Thus, a qualitative different multiple-$Q$ state is expected under $E^x_{\bm{Q}_1}$ and $E^y_{\bm{Q}_1}$ in the trigonal crystal systems from that in the tetragonal and hexagonal crystal systems, as discussed in Sec.~\ref{sec:Modulation}. 

The other relevant interactions at the symmetry-related wave vectors in $\{\bm{Q}\}$ are obtained by rotating the interaction matrix $X_{\bm{Q}_1}$ by the angle $\phi$, which is represented by 
\begin{align}
\label{eq:rotation}
X_{\bm{Q}_\eta} &= \Gamma(\phi) X_{\bm{Q}_1}  \Gamma^{-1}(\phi) ,
\end{align}
where  
\begin{align}
\label{eq:rep_rotation}
\Gamma(\phi)  =
 \begin{pmatrix}
 \sigma  \cos\phi & -\sigma \sin\phi & 0 \\
 \sigma \sin\phi & \sigma \cos\phi & 0 \\
 0 & 0 & 1 
\end{pmatrix}
\end{align}
with $\sigma=1$ ($\sigma=-1$) for the rotation (improper rotation) and $\eta=2,3$.  
Specifically, 
$X_{\bm{Q}_2}$ for $P4/mmm$, $P422$, $P4mm$, $P4/m$ and $P4$ ($P\bar{4}2m$, $P\bar{4}m2$, and $P\bar{4}$) are obtained by using Eqs.~(\ref{eq:rotation}) and (\ref{eq:rep_rotation}) with $\phi=\pi/2$ and $\sigma=1$ ($\sigma=-1$), and $X_{\bm{Q}_2}$ ($X_{\bm{Q}_3}$) in the hexagonal and trigonal systems are obtained with $\phi=2\pi/3$ ($\phi=4\pi/3$) and $\sigma=1$.
$X_{-\bm{Q}}$ is obtained from $X_{-\bm{Q}} = X^*_{\bm{Q}}$ by the time-reversal symmetry.

Tables~\ref{tab:tetragonal}-\ref{tab:trigonal} are useful to construct the model not only with $\{\bm{Q}\}$ shown in Fig.~\ref{fig:Qset} but also with other $\{\bm{Q}\}$.
For example,  the model with $\{\bm{Q}\}=\{\pm\bm{Q}_1\parallel \hat{\bm{x}}, \pm\bm{Q}_2\parallel \hat{\bm{y}}, \bm{Q}_1\pm\bm{Q}_2, -\bm{Q}_1\pm\bm{Q}_2\}$ in $P4/mmm$ crystal is constructed from $X_{\bm{Q}_1\parallel \hat{\bm{x}}}$ and $X_{(\bm{Q}_1+\bm{Q}_2)\parallel [110]}$, which are given in Table~\ref{tab:tetragonal}.
Then, the number of independent interactions in the model is six.
Furthermore, the general model in Eq.~(\ref{eq:ABS}) with the interactions at the two-dimensional wave vectors can be constructed from Tables~\ref{tab:tetragonal}-\ref{tab:trigonal} and Appendix~\ref{ap:ModelWithLowQ}, which will give an insight into the stability of the two-dimensional multiple-$Q$ states, such as the SkX.
Similar to the case with two-dimensional wave vectors, one can construct the spin model with the interactions at three-dimensional wave vectors based on the rules (a)-(f), which leads to a minimal effective spin model to investigate an instability toward three-dimensional multiple-$Q$ states~\cite{Okumura_PhysRevB.101.144416,hayami2021field,kato2021spin,kato2022magnetic}, such as the hedgehog lattice.

\subsection{Spin configurations under the anisotropic interactions}
\label{sec:Modulation}

\begin{figure*}[t!]
\begin{center}
\includegraphics[width=1.0\hsize]{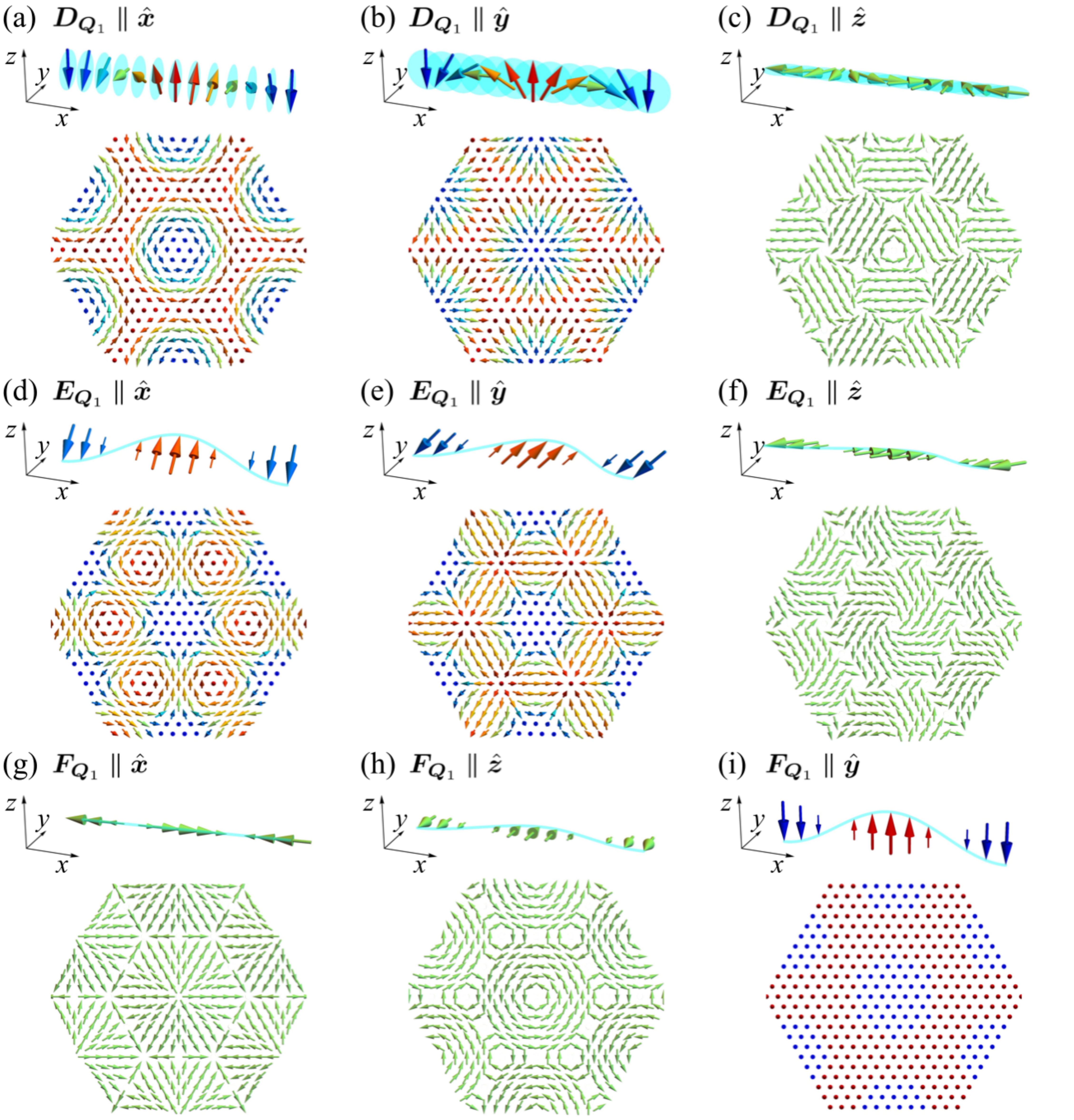} 
\caption{
\label{fig:modulation}
Spin configurations in the presence of the anisotropic interactions:
 (a) $\bm{D}_{\bm{Q}_1}\parallel \hat{\bm{x}}$, 
 (b) $\bm{D}_{\bm{Q}_1}\parallel \hat{\bm{y}}$, 
 (c) $\bm{D}_{\bm{Q}_1}\parallel \hat{\bm{z}}$, 
 (d) $\bm{E}_{\bm{Q}_1}\parallel \hat{\bm{x}}$, 
 (e) $\bm{E}_{\bm{Q}_1}\parallel \hat{\bm{y}}$, 
 (f) $\bm{E}_{\bm{Q}_1}\parallel \hat{\bm{z}}$, 
 (g) $\bm{F}_{\bm{Q}_1}\parallel \hat{\bm{x}}$, 
 (h) $\bm{F}_{\bm{Q}_1}\parallel \hat{\bm{y}}$, 
 and (i) $\bm{F}_{\bm{Q}_1}\parallel \hat{\bm{z}}$.
 Upper panel: Single-$Q$ spin structures with the ordering vector $\bm{Q}_1\parallel  \hat{\bm{x}}$ of 
 (a)-(c) spiral waves and (d-i) sinusoidal waves. 
 In (a-c), the spiral planes in the spiral wave are (a) $yz$, (b) $zx$, and (c) $xy$ planes. 
 In (d-i), the oscillating directions in the sinusoidal wave are (d)  
[011], (e) [101], (f) [110], (g) [100], (h) [010], and (i) [001] directions.
Lower panel: Triple-$Q$ structures consisting of the three single-$Q$ waves in the upper panel at $\bm{Q}_1$, $\bm{Q}_2$, and $\bm{Q}_3$ in Fig.~\ref{fig:Qset}(c).
The color of arrows represents the $z$ spin component, where red, blue, and green stand for positive, negative, and zero values.    
}
\end{center}
\end{figure*}

The symmetry argument in Secs.~\ref{sec:Rule} and \ref{sec:Model1} provides a plausible spin configuration in the presence of the anisotropic interactions. 
Here, we discuss the modulation tendency in the case of the single-$Q$ spin configuration in Sec.~\ref{sec:Single} and the triple-$Q$ spin configuration in Sec.~\ref{sec:Triple}.

\subsubsection{Single-$Q$ case}
\label{sec:Single}

We discuss the single-$Q$ spin configurations under the momentum-resolved interaction at the wave vector $\bm{Q}_1\parallel \hat{\bm{x}}$.
In the case of the isotropic interaction, the spiral state with a wave vector $\bm{q}^*=\bm{Q}_1$ has the lowest energy, as described above, where the spiral plane is arbitrary.
When additionally considering the anisotropic interactions, the spiral wave is modulated depending on the type of them. 
For example, $\bm{D}_{\bm{Q}_1}$ fixes the spiral plane perpendicular to $\bm{D}_{\bm{Q}_1}$: The proper-screw (out-of-plane cycloidal) spiral wave is favored in the space group $P622$ ($P6mm$) with nonzero $D^x_{\bm{Q}_1}$ ($D^y_{\bm{Q}_1}$), and the in-plane cycloidal spiral wave is favored in the space group $P\bar{6}m2$ with nonzero $D^z_{\bm{Q}_1}$.
The proper-screw, out-of-plane cycloidal, and in-plane cycloidal spiral waves are shown in the upper panel of Figs.~\ref{fig:modulation}(a)-\ref{fig:modulation}(c), respectively.
Meanwhile, when considering the effect of $\bm{E}_{\bm{Q}_1}$ instead of $\bm{D}_{\bm{Q}_1}$, the spiral plane by the isotropic interaction is elliptically modulated so as to have more perpendicular spin component to $\bm{E}_{\bm{Q}_1}$. 
In other words, $\bm{E}_{\bm{Q}_1}$ favors the sinusoidal wave oscillating in one direction. 
For example, $E^x_{\bm{Q}_1}$, $E^y_{\bm{Q}_1}$, and $E^z_{\bm{Q}_1}$ favor the sinusoidal wave with the spin oscillation along the [011], [101], and [110] directions in spin space, respectively, which are  shown in the upper panel of Figs.~\ref{fig:modulation}(d)-\ref{fig:modulation}(f). 
Such a sinusoidal modulation by $E^x_{\bm{Q}_1}$, $E^y_{\bm{Q}_1}$, and $E^z_{\bm{Q}_1}$ is expected in the space groups $P\bar{3}m1$, $P\bar{3}1m$, and $P6/m$, respectively.
Moreover, $\bm{F}_{\bm{Q}_1}$ also modulates the spiral wave into the sinusoidal wave. 
For example, in the case of $F^x_{\bm{Q}_1}$, $F^y_{\bm{Q}_1}$, and $F^z_{\bm{Q}_1}$, the sinusoidal wave with the spin oscillation along the [100], [010], and [001] directions in spin space is favored, respectively, as shown in the upper panel of Figs.~\ref{fig:modulation}(g)-\ref{fig:modulation}(i). 

\subsubsection{Triple-$Q$ case}
\label{sec:Triple}

Similar to the single-$Q$ case, one can expect a tendency of the multiple-$Q$ spin configuration under the anisotropic interactions. 
We here discuss the relationship between the anisotropic interactions and the triple-$Q$ spin 
configuration by superposing the three spirals with the same intensity on the triangular lattice belonging to the hexagonal and trigonal space groups, where we consider the superposition of the spin density waves at $\bm{Q}_1$, $\bm{Q}_2$, and $\bm{Q}_3$ in Fig.~\ref{fig:Qset}(c) and neglect the phase degree of freedom of the spin density wave for simplicity~\cite{hayami2021phase,Hayami_PhysRevResearch.3.043158,shimizu2022phase}.

In the case of $\bm{D}_{\bm{Q}_1}\parallel \hat{\bm{x}}$, i.e., $\bm{D}_{\bm{Q}_1}\parallel\bm{Q}_1$, the threefold rotational symmetry imposes $\bm{D}_{\bm{Q}_2}\parallel\bm{Q}_2$ and $\bm{D}_{\bm{Q}_3}\parallel\bm{Q}_3$.
Then, $\bm{D}_{\bm{Q}_1}\parallel \hat{\bm{x}}$ favors the triple-$Q$ proper-screw spiral wave expressed as the superposition of the three proper-screw spirals in the lower panel of Fig.~\ref{fig:modulation}(a), which corresponds to the Bloch SkX with $
N_\mathrm{sk}=-1$ per magnetic unit cell [see Eq.~(\ref{eq:Nsk}) for the definition of the skyrmion number $N_\mathrm{sk}$ in Sec.~\ref{sec:NumericalSimulation}].
Similarly, $\bm{D}_{\bm{Q}_1}\parallel \hat{\bm{y}}$ and $\bm{D}_{\bm{Q}_1}\parallel \hat{\bm{z}}$ favor the triple-$Q$ out-of-plane cycloidal spiral wave corresponding to the N\'{e}el SkX with $N_\mathrm{sk}=-1$ and the triple-$Q$ in-plane spiral wave, as shown in the lower panel of Figs.~\ref{fig:modulation}(b) and \ref{fig:modulation}(c), respectively.
As shown in Figs.~\ref{fig:modulation}(a) and \ref{fig:modulation}(b) [Fig.~\ref{fig:modulation}(c)], the superposition of spirals in different spiral planes (the same spiral plane) leads to the noncoplanar (coplanar) structure.

Meanwhile, $\bm{E}_{\bm{Q}_1}$ and $\bm{F}_{\bm{Q}_1}$ tend to favor triple-$Q$ sinusoidal waves, as shown in the lower panel of Figs.~\ref{fig:modulation}(d)-\ref{fig:modulation}(i). 
Among them, $\bm{E}_{\bm{Q}_1}\parallel \hat{\bm{x}}$ and $\bm{E}_{\bm{Q}_1}\parallel \hat{\bm{y}}$ tend to favor the noncoplanar triple-$Q$ sinusoidal waves since they consist of the three sinusoidal waves oscillating in different out-of-plane directions, as shown in the lower panel of Figs.~\ref{fig:modulation}(d) and \ref{fig:modulation}(e).
This noncoplanar triple-$Q$ sinusoidal states correspond to the SkXs with $N_\mathrm{sk}=+2$.  
The cases for $\bm{E}_{\bm{Q}_1}\parallel \hat{\bm{z}}$, $\bm{F}_{\bm{Q}_1}\parallel \hat{\bm{x}}$, and  $\bm{F}_{\bm{Q}_1}\parallel \hat{\bm{y}}$ favor the coplanar triple-$Q$ sinusoidal waves consisting of the three sinusoidal waves oscillating in different in-plane directions, as shown in the lower panel of Figs.~\ref{fig:modulation}(f)-\ref{fig:modulation}(h).
The remaining $\bm{F}_{\bm{Q}_1}\parallel \hat{\bm{z}}$ favors the collinear triple-$Q$ sinusoidal wave consisting of the three sinusoidal waves oscillating in the same direction [the lower panel of Fig.~\ref{fig:modulation}(i)], which is the so-called magnetic bubble.

The above intuitive analysis provides two important pieces of information about the SkXs.
The first is that the anisotropic interactions in all the hexagonal and trigonal crystal systems do not tend to favor the anti-type SkXs with $N_{\rm sk}=+1$ without the threefold rotational symmetry, since the 
anisotropic exchange interactions connected by the threefold rotation [see Eq.~(\ref{eq:rotation})] lead to the energy loss to form such SkXs free from threefold symmetry.
Meanwhile, there is no preference between the SkXs and anti-type SkXs in terms of the symmetric anisotropic exchange interactions in the tetragonal crystal systems.
The second is that there is an instability tendency toward the SkXs with $|N_\mathrm{sk}|=2$ only in the trigonal crystal systems with $E^x_{\bm{Q}_1}$ and  $E^y_{\bm{Q}_1}$ within the bilinear exchange interactions.

Such an argument in terms of the spin modulations under the anisotropic exchange interactions is consistent with the previous studies, where unbiased numerical simulations have been performed for a similar effective spin model under the space groups $P4/mmm$~\cite{Hayami_doi:10.7566/JPSJ.89.103702,Hayami_PhysRevB.103.024439}, $P4mm$~\cite{Hayami_PhysRevLett.121.137202}, $P4/m$~\cite{Hayami2022Helicity}, $P6/mmm$~\cite{hayami2020multiple,Hayami_PhysRevB.103.054422}, $P6mm$~\cite{Hayami_PhysRevB.104.094425}, and $P\bar{3}m1$~\cite{yambe2021skyrmion}.
For example, the $P\bar{3}m1$ system with nonzero $E^x_{\bm{Q}_1}$ in addition to the isotropic exchange interaction exhibits the instability toward the SkX with $|N_\mathrm{sk}|=2$ in Fig.~\ref{fig:modulation}(g).
Besides, the $P6/mmm$ system with nonzero $F^x_{\bm{Q}_1}$ ($F^y_{\bm{Q}_1}$) in addition to the isotropic exchange interaction and Zeeman coupling to an external magnetic field leads to the SkX with $N_\mathrm{sk}=-1$, whose spin configuration is similar to that in  Fig.~\ref{fig:modulation}(b) [\ref{fig:modulation}(a)]. 

\section{Origin of the anisotropic exchange interactions: case of itinerant electron models}
\label{sec:Origin1}

We discuss how to derive the momentum-resolved anisotropic exchange interaction in Eq.~(\ref{eq:ABS}) based on a microscopic Hamiltonian for itinerant magnets.
Starting from the multi-band anisotropic PAM with the SOC in Sec.~\ref{sec:PAM}, we present the important parameters for nonzero anisotropic interactions.
For that purpose, we perform the Schrieffer-Wolff transformation~\cite{Schrieffer_PhysRev.149.491} to derive the Kondo lattice model with the anisotropic exchange coupling between itinerant electron spins and localized spins in Sec.~\ref{sec:KL}.
Then, we trace out the itinerant electron degree of freedom to derive the effective spin model by supposing the weak exchange coupling in Sec.~\ref{sec:Model2}.  

\subsection{Anisotropic periodic Anderson model}
\label{sec:PAM}

We consider the multi-band anisotropic PAM incorporating the effect of the SOC~\cite{PhysRevB.55.12561,yambe2021skyrmion}, which is represented by 
\begin{align}
\label{eq:PAM}
\mathcal{H}^{\rm PAM}
&=\mathcal{H}^c+\mathcal{H}^f+\mathcal{H}^{cf},
\end{align}
where
\begin{align}
\label{eq:Hc}
\mathcal{H}^c&=\sum_{m,\bm{k},\sigma}(\varepsilon_{m\bm{k}}-\mu) c^{\dagger}_{m\bm{k}\sigma} c_{m\bm{k}\sigma},\\
\label{eq:Hf}
\mathcal{H}^f&=(E_f-\mu)\sum_{i,\sigma}n_{i\sigma}+U\sum_i n_{i\uparrow}n_{i\downarrow},\\
\mathcal{H}^{cf}&=\sum_{m,i,\bm{k},\sigma,\sigma'}\frac{e^{i\bm{k}\cdot\bm{R}_i}}{\sqrt{N}} f^{\dagger}_{i\sigma} (V^0_{m\bm{k}}\delta+\bm{V}_{m\bm{k}} \cdot\bm{\sigma} )_{\sigma\sigma'}c_{m\bm{k}\sigma'} \nonumber \\
&\hspace{4cm}+\rm{h.c.}
\end{align}
Here, $c^{\dagger}_{m\bm{k}\sigma}$ $(c_{m\bm{k}\sigma})$ is a creation (annihilation) operator of an itinerant electron with band $m$, wave vector $\bm{k}$, and spin $\sigma$, $f^{\dagger}_{i\sigma}$ $(f_{i\sigma})$ is a creation (annihilation) operator of a localized $f$ electron at position vector $\bm{R}_i$ with spin $\sigma$, and $n_{i\sigma}=f^{\dagger}_{i\sigma}f_{i\sigma}$.  
$\mathcal{H}^c$ represents the Hamiltonian of the itinerant electron with the energy dispersion $\varepsilon_{m\bm{k}}$ and  the chemical potential $\mu$.
$\mathcal{H}^f$ represents the Hamiltonian of the localized $f$ electron, where $E_f$ is the atomic energy and $U$ is the Coulomb interaction.
$\mathcal{H}_{cf}$ stands for the Hamiltonian consisting of the hybridization between the itinerant electrons and localized electrons; $V_{m\bm{k}}^0\delta_{\sigma\sigma'}$ represents the spin-independent hybridization and $\bm{V}_{m\bm{k}}\cdot\bm{\sigma}_{\sigma\sigma'}=\sum_{\alpha=x,y,z}V^\alpha_{m\bm{k}}\sigma^\alpha_{\sigma\sigma'}$ represents the spin-dependent hybridization, where $\delta_{\sigma\sigma'}$ is the Kronecker delta, $\bm{\sigma}_{\sigma\sigma'}=(\sigma^x,\sigma^y,\sigma^z)_{\sigma\sigma'}$ is a vector of the Pauli matrices, and $N$ is the number of unit cells.
The contribution of spin-dependent hybridization arises from the mixture of up- and down-spin basis functions of the itinerant and/or localized electrons due to the SOC, where the spin index $\sigma$ in the spin-orbital-coupled basis is regarded as the pseudospin.

\subsection{Anisotropic Kondo lattice model}
\label{sec:KL}

We derive a low-energy effective model when $E_f$ ($E_f+U$) is much smaller (larger) than the Fermi energy.
In this situation, the $f$ electron state at each site is occupied by a single electron ($\sum_\sigma n_{i\sigma}=1$) and the $f$ electron is approximately regarded as the localized spin.
When the hybridizations are treated as the perturbation, the low-energy effective model is derived by the Schrieffer-Wolff transformation as $e^{\mathcal{S}}\mathcal{H}^\mathrm{PAM}e^{-\mathcal{S}}$ with the generator $\mathcal{S}$; $\mathcal{S}$ satisfies $\mathcal{H}^{cf}+[\mathcal{S},\mathcal{H}^0]=0$, where $\mathcal{H}^0=\mathcal{H}^c+\mathcal{H}^f$ and $[\mathcal{S},\mathcal{H}^0]$ represents the communication relation.
Then, $\mathcal{S}$ is given by
\begin{align}
\label{eq:S}
&\mathcal{S}=\frac{1}{\sqrt{N}}\sum_{m,i,\bm{k},\sigma,\sigma'}(A_{m\bm{k}}+B_{m\bm{k}} n_{i\bar{\sigma}}) \nonumber\\&
\times\left\{ e^{i\bm{k}\cdot\bm{R}_i}f^{\dagger}_{i\sigma}(V^0_{m\bm{k}}\delta_{\sigma\sigma'}+ \bm{V}_{m\bm{k}} \cdot\bm{\sigma}_{\sigma\sigma'})c_{m\bm{k}\sigma'}-{\rm h.c.}\right\}, 
\end{align}
where $\bar{\sigma}=-\sigma$ and
\begin{align}
A_{m\bm{k}}&=\frac{1}{E_f-\varepsilon_{m\bm{k}}}, \\
B_{m\bm{k}}&=\frac{1}{\varepsilon_{m\bm{k}}-E_f} -\frac{1}{\varepsilon_{m\bm{k}}-E_f-U}.
\end{align}
Then, the low-energy effective model up to the second order of the hybridizations, $\mathcal{H}^{\rm PAM (2)}$, is approximately given by
\begin{align}
\label{eq:Heff}
\mathcal{H}^{\rm PAM (2)}
&=\mathcal{H}^0+\frac{1}{2}[\mathcal{S},\mathcal{H}^{cf}] \\
&= \mathcal{H}^c+
\sum_{m,m'}\sum_{\sigma,\sigma'} \left(
\mathcal{H}'_{m\sigma;m'\sigma}\delta_{\sigma\sigma'}+ \nonumber\right.\\
&\left. \mathcal{H}^\mathrm{ex}_{m\sigma;m'\sigma'} + \mathcal{H}^\mathrm{SOC}_{m\sigma;m'\sigma'} \right),
\end{align}
where the subscript ${m\sigma;m'\sigma'}$ represents a matrix element between itinerant electrons with $(m,\sigma)$ and $(m',\sigma')$. 
In the derivation, we drop off the constant terms such as $\mathcal{H}^f$. 
The details of $\mathcal{H}^{\rm PAM (2)}$  are given in Appendix~\ref{ap:EffHam}.

To focus on the origin of the anisotropic interactions, we further neglect the contributions from
the spin-independent term $\mathcal{H}'$ and from the different bands in $\mathcal{H}^\mathrm{ex}$ and $\mathcal{H}^\mathrm{SOC}$. 
In the end, $\mathcal{H}^{\rm PAM (2)}$ reduces to an anisotropic Kondo lattice model as 
\begin{align}
\label{eq:KLM}
\mathcal{H}^\mathrm{KLM}&= \mathcal{H}^c+\sum_m\sum_{\sigma,\sigma'} \left(\mathcal{H}^\mathrm{ex}_{m\sigma\sigma'} + \mathcal{H}^\mathrm{SOC}_{m\sigma\sigma'} \right),
\end{align}
where the subscript ${m\sigma\sigma'}$ represents a matrix element between itinerant electrons with $(m,\sigma)$ and $(m,\sigma')$. 
 
The Kondo lattice model includes two spin-dependent terms. One is the exchange interaction between itinerant electron spins and localized spins, $\mathcal{H}^\mathrm{ex}_{m\sigma\sigma'}$, which is given by
\begin{align}
\mathcal{H}^\mathrm{ex}_{m\sigma\sigma'} &= \frac{1}{\sqrt{N}}\sum_{\bm{k},\bm{q},\alpha,\beta}J^{\alpha\beta}_{m\bm{k}+\bm{q}\bm{k}}c^{\dagger}_{m\bm{k}+\bm{q}\sigma}\sigma^{\alpha}_{\sigma\sigma'}c_{m\bm{k}\sigma'}S^{\beta}_{\bm{q}}.
\end{align}
Here, $\bm{S}_{\bm{q}}$ is the Fourier transform of the localized spin $\bm{S}_i=\sum_{\sigma\sigma'}f^\dagger_{i\sigma} \bm{\sigma}_{\sigma\sigma'} f_{i\sigma'}/2$.
The exchange interaction is decomposed into isotropic, symmetric anisotropic, and antisymmetric anisotropic exchange interactions in spin space~\cite{PhysRevB.55.12561,yambe2021skyrmion,PhysRevLett.108.046601} as 
\begin{align} 
J_{m\bm{k}\bm{k}'}^{\alpha\beta}&=J^{\rm ISO}_{m\bm{k}\bm{k}'}\delta_{\alpha\beta}+[J^{\rm S}_{m\bm{k}\bm{k}'}]^{\alpha\beta}+[J^{\rm AS}_{m\bm{k}\bm{k}'}]^{\alpha\beta},
\end{align}
where
\begin{align}
\label{eq:Jiso} 
J^{\rm ISO}_{m\bm{k}\bm{k}'}&=C^{(1)}_{m\bm{k}\bm{k}'} \left(V^{0}_{m\bm{k}'}V^{0*}_{m\bm{k}}-\bm{V}_{m\bm{k}'}\cdot\bm{V}^*_{m\bm{k}}\right), \\
\label{eq:Jsa} 
[J^{\rm S}_{m\bm{k}\bm{k}'}]^{\alpha\beta}&=C^{(1)}_{m\bm{k}\bm{k}'}\left(V^{\alpha}_{m\bm{k}'} V^{\beta *}_{m\bm{k}} + V^{\alpha*}_{m\bm{k}} V^{\beta}_{m\bm{k}'}\right),  \\
\label{eq:Jdm} 
[J^{\rm AS}_{m\bm{k}\bm{k}'}]^{\alpha\beta}&=iC^{(1)}_{m\bm{k}\bm{k}'}\sum_{\gamma}\epsilon_{\alpha\beta\gamma}\left( V^{\gamma}_{m\bm{k}'} V^{0 *}_{m\bm{k}} - V^{\gamma*}_{m\bm{k}} V^{0 }_{m\bm{k}'}\right),
\end{align}
with $C^{(1)}_{m\bm{k}\bm{k}'}=(B_{m\bm{k}}+B_{m\bm{k}'})/2$ and the Levi-Civita symbol $\epsilon_{\alpha\beta\gamma}$. 
The symmetric and antisymmetric exchange interactions satisfy $[J^{\rm S
}_{m\bm{k}\bm{k}'}]^{\alpha\beta}=[J^{\rm S
}_{m\bm{k}\bm{k}'}]^{\beta\alpha}$ and $[J^{\rm AS}_{m\bm{k}\bm{k}'}]^{\alpha\beta}=-[J^{\rm AS}_{m\bm{k}\bm{k}'}]^{\beta\alpha}$, respectively.
The anisotropic exchange interactions vanish in the absence of the spin-dependent hybridizations. 
In addition, it is noted that these anisotropic interactions also vanish when $\mathcal{H}^{cf}$ includes a single component of $(V^0_{m\bm{k}}, \bm{V}_{m\bm{k}})$.

The other spin-dependent term in Eq.~(\ref{eq:KLM}) is the effective SOC for itinerant electrons, $\mathcal{H}^\mathrm{SOC}_{m\sigma\sigma'}$, which is given by
\begin{align}
\mathcal{H}^\mathrm{SOC}_{m\sigma\sigma'} &=  \sum_{\bm{k}}\bm{g}_{m\bm{k}}\cdot  c^{\dagger}_{m\bm{k}\sigma}\bm{\sigma}_{\sigma\sigma'}c_{m\bm{k}\sigma '},
\end{align}
where
\begin{align} 
g^{\alpha}_{m\bm{k}} =& C^{(2)}_{m\bm{k}}\left( V^{\alpha}_{m\bm{k}} V^{0 *}_{m\bm{k}} + V^{\alpha*}_{m\bm{k}} V^{0 }_{m\bm{k}}
-i\sum_{\beta,\gamma}\epsilon_{\alpha\beta\gamma}V^{\beta}_{m\bm{k}} V^{\gamma *}_{m\bm{k}} \right),
 \end{align}
with $C^{(2)}_{m\bm{k}}=-(A_{m\bm{k}}+B_{m\bm{k}}/2)$. 
The effective SOC is induced by the spin-dependent hybridizations; $g^{\alpha}_{m\bm{k}}$ vanishes for $\bm{V}_{m\bm{k}}=\bm{0}$.
The expression of $g^{\alpha}_{m\bm{k}}$ reduces to the antisymmetric spin-orbit interaction in the single-band system, which only appears in noncentrosymmetric crystal systems.

\subsection{Effective spin model}
\label{sec:Model2}

An effective  spin model of the anisotropic Kondo lattice model in Eq.~(\ref{eq:ABS}) is obtained by expanding the grand potential to the second order with respect to the exchange coupling~\cite{Akagi_PhysRevLett.108.096401,Hayami_PhysRevB.90.060402,Hayami_PhysRevB.95.224424}.
When taking $\bm{g}_{m\bm{k}}=0$ for simplicity, the lowest second-order contribution of the exchange energy to the grand potential is given by
\begin{align}
\Omega^{(2)}&=
- \frac{T}{2} \int^{1/T}_0 d\tau_1  \int^{1/T}_0 d\tau_2 \langle \mathcal{T}_{\tau}\mathcal{H}^\mathrm{ex}_{\tau_1} \mathcal{H}^\mathrm{ex}_{\tau_2} \rangle_{\rm con}  \nonumber\\
\label{eq:Omega}
 &=-\sum_{\bm{q}}\sum_{\alpha\beta} X^{\alpha\beta}_{\bm{q}} S^{\alpha}_{\bm{q}}S^\beta_{-\bm{q}},
 \end{align}
where $T$ is the temperature, $\tau$ is the imaginary time, $\mathcal{T}_\tau$ is the time ordered product, $\mathcal{H}^\mathrm{ex}_\tau=e^{\tau \mathcal{H}^c}\mathcal{H}^\mathrm{ex}e^{-\tau \mathcal{H}^c}$, and $\langle\cdots\rangle_{\rm con}$ represents the contributions from the connected Feynman diagrams.
$X^{\alpha\beta}_{\bm{q}}$ in the second line corresponds to the interaction matrix in Sec.~\ref{sec:Symmetry}, which is given by 
\begin{align}
 X^{\alpha\beta}_{\bm{q}} &= \frac{T}{N}
\sum_{m,\bm{k},\gamma,\omega_n}G_{m\bm{k}+\bm{q}}(i\omega_n)G_{m\bm{k}}(i\omega_n) J^{\gamma\alpha}_{m\bm{k}+\bm{q}\bm{k}} J^{\gamma\beta}_{m\bm{k}\bm{k}+\bm{q}}, \nonumber \\
\label{eq:Xq} 
&=\frac{1}{N}\sum_{m,\bm{k},\gamma}\frac{f(\varepsilon_{m\bm{k}})-f(\varepsilon_{m\bm{k}+\bm{q}})}{\varepsilon_{m\bm{k}+\bm{q}}-\varepsilon_{m\bm{k}}}J^{\gamma\alpha}_{m\bm{k}+\bm{q}\bm{k}} J^{\gamma\beta}_{m\bm{k}\bm{k}+\bm{q}}.
 \end{align}
 where $ G_{m\bm{k}}(i\omega_n)=1/(i\omega_n-\varepsilon_{m\bm{k}}+\mu)$ is the noninteracting Green's function with the Matsubara frequency $\omega_n$ and $f(\varepsilon_{m\bm{k}})$ is the Fermi distribution function. 
 It is noted that Green's function does not depend on the spin, since we neglect the effective SOC ($\bm{g}_{m\bm{k}}=0$), and then, we omit its spin dependence for notational simplicity.
 $\bm{D}_{\bm{q}}$, $\bm{E}_{\bm{q}}$, and $\bm{F}_{\bm{q}}$ in Eq.~(\ref{eq:CouplingMatrix}) are related to $ X^{\alpha\beta}_{\bm{q}}$ in Eq.~(\ref{eq:Xq}) as
\begin{align}
\label{eq:D}
D^{\alpha}_{\bm{q}}&=\frac{1}{2}\sum_{\beta,\gamma}\epsilon_{\alpha\beta\gamma}\mathrm{Im}\left[ X^{\beta\gamma}_{\bm{q}}\right],\\
E^{\alpha}_{\bm{q}}&=\frac{1}{2}\sum_{\beta,\gamma}|\epsilon_{\alpha\beta\gamma}|\mathrm{Re}\left[ X^{\beta\gamma}_{\bm{q}}\right],\\
\label{eq:F}
F^\alpha_{\bm{q}}&=X^{\alpha\alpha}_{\bm{q}}.
 \end{align}
In this way, the momentum-resolved anisotropic interactions introduced in Eq.~(\ref{eq:CouplingMatrix}) are obtained based on the itinerant electron model.
It is noted that the anisotropic interactions are also obtained from the Kondo lattice model with $\bm{g}_{m\bm{k}}\neq\bm{0}$ instead of $J^\mathrm{\rm S}_{m\bm{k}\bm{k}'}$ and $J^\mathrm{\rm AS}_{m\bm{k}\bm{k}'}$~\cite{shibuya2016magnetic,Hayami_PhysRevLett.121.137202,Okada_PhysRevB.98.224406}.

 The effective spin model in Eq.~(\ref{eq:Omega}) is justified when the energy scale of the exchange interaction is smaller than that of the bandwidth.  
In the itinerant electron model, the dominant $\bm{q}$ components in the interactions giving the largest eigenvalue of $X_{\bm{q}}$ are related to the nesting vectors of the Fermi surface, as inferred from Eq.~(\ref{eq:Xq}).
As $X_{\bm{q}}$ is calculated when $\varepsilon_{m\bm{k}}$, $\mu$, $V^0_{m\bm{k}}$, and $\bm{V}_{m\bm{k}}$ are given, one can quantitatively evaluate the contributions of the anisotropic interactions. 
For example, one can directly evaluate the anisotropic interactions in materials within the framework of the first principle calculations. 

Similar momentum-resolved spin models can be derived from other itinerant electron models.
For example, the classical Kondo lattice model in the strong exchange coupling regime (double exchange model~\cite{Zener_PhysRev.82.403,anderson1955considerations})
 is mapped onto the effective spin model with the short-range spin interactions~\cite{degennes1960effects,doi:10.1143/JPSJ.33.21,PhysRevLett.74.5144,Ishizuka_PhysRevB.92.024415}
When taking into account the Rashba- or Dresselhaus-type SOC, the short-range spin interactions become anisotropic~\cite{banerjee2013ferromagnetic,Banerjee_PhysRevX.4.031045,Kathyat_PhysRevB.102.075106,PhysRevB.104.184434,mukherjee2021antiferromagnetic,PhysRevB.103.134424}.
Furthermore, the effective spin model with the short-range spin interactions can be constructed based on the Hubbard model with the SOC~\cite{Banerjee_PhysRevX.4.031045,PhysRevLett.109.085302,chen2016exotic}. 
In these cases, the momentum-resolved effective spin model in Eq.~(\ref{eq:ABS2}) is obtained once the dominant interaction in $\bm{q}$ space (including $\bm{q}=\bm{0}$ component) are extracted.

\section{Origin of the anisotropic exchange interactions: case of localized spin models}
\label{sec:Origin2}

In the previous section, we show that the momentum-resolved anisotropic exchange interaction is obtained as the effective long(short)-range interaction for itinerant electron models.
Meanwhile, the above momentum-resolved anisotropic exchange interaction is also related to the short-range interaction in the localized spin model.
For example, a ground-state magnetic phase diagram has been constructed by considering the dominant $\bm{q}$ interactions in frustrated magnets~\cite{leonov2015multiply,Hayami_PhysRevB.103.224418} and DM-based magnets~\cite{Hayami_PhysRevB.105.014408}.
In the localized spin model, the anisotropic exchange interaction originates from the relativistic SOC and dipolar interactions, the former of which largely depends on the point-group symmetry in crystals~\cite{dzyaloshinsky1958thermodynamic,moriya1960anisotropic,kaplan1983single,PhysRevLett.69.836,Shekhtman_PhysRevB.47.174,PhysRevLett.112.077204,PhysRevLett.115.167203,PhysRevB.96.205126,Maksimov_PhysRevX.9.021017,Matsumoto_PhysRevB.101.224419,Matsumoto_PhysRevB.104.134420,PhysRevB.105.014404}. 
In such a situation, the microscopic origin of the interaction matrix $X_{\bm{q}}$ in Eq.~(\ref{eq:ABS2}) is attributed to the Fourier transform of real-space anisotropic exchange interactions. 

\section{Application to a specific hexagonal system}
\label{sec:SpecificExample}

We apply the above general expression to a specific hexagonal crystal system under the space group $P6/mmm$.
Starting from the PAM in Sec.~\ref{sec:PAM2} and mapping it onto the effective spin model in Sec.~\ref{sec:Model3}, we show the multiple-$Q$ instability by performing the simulated annealing in Sec.~\ref{sec:NumericalSimulation}.     

\subsection{Anisotropic Periodic Anderson model}
\label{sec:PAM2}

\begin{figure}[t!]
\begin{center}
\includegraphics[width=1.0\hsize]{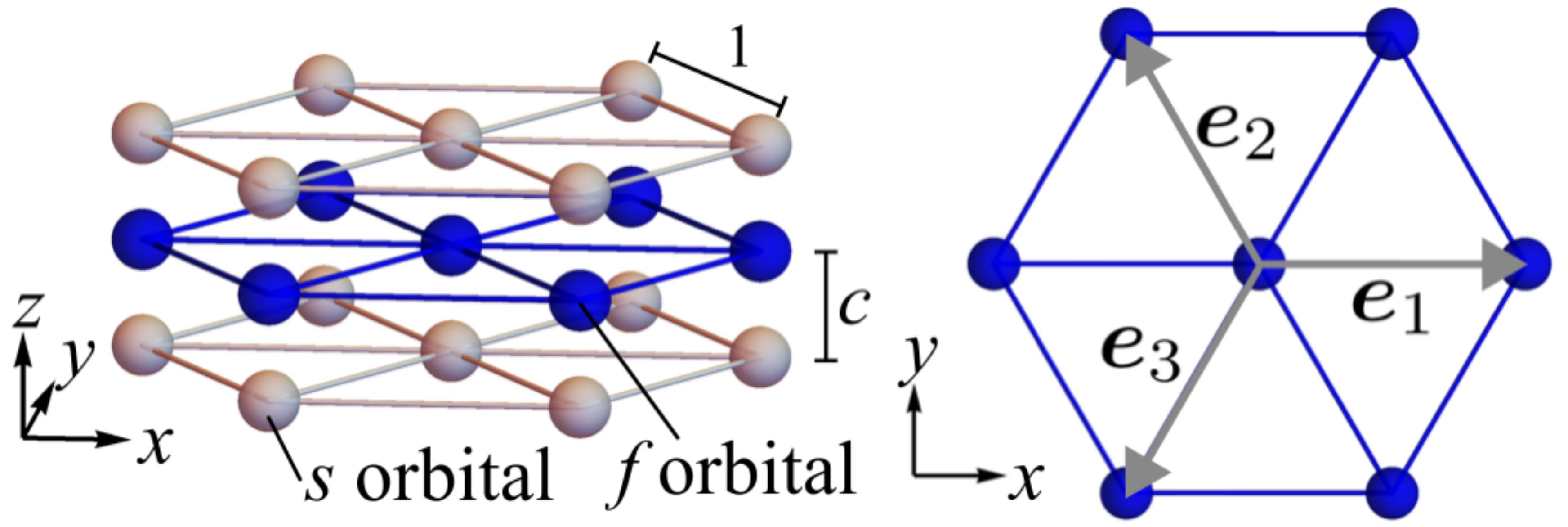} 
\caption{
\label{fig:lattice}
Left panel: $P6/mmm$ crystal lattice consisting of three layers.
The blue (gray) spheres represent magnetic (nonmagnetic) ions with the $f$ ($s$) orbital and form the triangular lattice on the $xy$ plane.
The three layers are stacked along the $z$ direction at equal intervals of $c$. 
Right panel: the triangular lattice viewed from the $z$ direction; $\bm{e}_1$, $\bm{e}_2$, and $\bm{e}_3$ are the unit vectors.  
}
\end{center}
\end{figure}

As an example, we consider the specific $P6/mmm$ crystal lattice consisting of three triangular-lattice layers separated by a distance $c$, as shown in the left panel of Fig.~\ref{fig:lattice}; the localized $f$ orbitals denoted by the blue spheres lie on the middle layer, and the itinerant $s$ orbitals denoted by the gray spheres lie on the upper and lower layers. 
We set the lattice constant of the triangular lattice as the length unit. 

The system is described by the multi-band anisotropic PAM in Eq.~(\ref{eq:PAM}) under the periodic boundary condition in the $x$ and $y$ directions.
The energy dispersion of the itinerant electron in $\mathcal{H}^c$ in upper and lower layers is given by 
\begin{align}
\label{eq:ek}
\varepsilon_{m\bm{k}} = -2\sum_{i=1,2,3}\left( 
t_1 \cos \bm{k}\cdot\bm{e}_i + t_3 \cos 2\bm{k}\cdot\bm{e}_i
\right),
\end{align}
where $\bm{k}=(k_x,k_y)$ is the two-dimensional wave vector, $\bm{e}_1=(1,0)$, $\bm{e}_2=(-1/2,\sqrt{3}/2)$, and $\bm{e}_3=(-1/2,-\sqrt{3}/2)$ are the unit vectors of the triangular lattice (the right panel of Fig.~\ref{fig:lattice}), and $m=+(-)$ represents the upper (lower) nonmagnetic layer. 
Here, we consider the hoppings between the nearest- and third-neighbor sites, $t_1$ and $t_3$, within the same layer.

Meanwhile, we suppose that the $f$ orbital with the Kramers twofold degeneracy is anisotropic in spin space by incorporating the effect of 
the SOC and the crystalline electric field (CEF) under the $P6/mmm$ symmetry in the following way.
By assuming that the magnitude of SOC is greater than that of CEF, the fourteen degenerate $f$ electron states are split into the two levels with the total angular momentum $j=7/2$ and $j=5/2$ by the SOC, and then, they are split into totally seven Kramers pairs by the CEF.
We choose one out of seven Kramers pairs, which is expressed as
\begin{align}
f^\dagger_{i\uparrow}\Ket{0} &=\alpha_\mathrm{CEF}\Ket{i,3,\frac{1}{2}} +\sqrt{1-\alpha_\mathrm{CEF}^2}\left(\sqrt{\frac{6}{7}}\Ket{i,-2,-\frac{1}{2}}  \right. \nonumber\\ 
&\left. \hspace{5mm} +\sqrt{\frac{1}{7}}\Ket{i,-3,\frac{1}{2}} \right),\\
f^\dagger_{i\downarrow}\Ket{0}  &=-\alpha_\mathrm{CEF}\Ket{i,-3,-\frac{1}{2}} \nonumber\\
&\hspace{-8mm}-\sqrt{1-\alpha_\mathrm{CEF}^2}\left(\sqrt{\frac{6}{7}}\Ket{i,2,\frac{1}{2}}+\sqrt{\frac{1}{7}}\Ket{i,3,-\frac{1}{2}} \right), 
\end{align}
where $\ket{i,l_z,s_z}$ is characterized by the site $i$ and the magnetic quantum number of the $f$ orbital ($l_z=-3,-2, \cdots 3$) and spin ($s_z=\pm1/2$) and $\alpha_\mathrm{CEF}$ $(|\alpha_\mathrm{CEF}|\le 1)$ is a constant depending on the CEF parameters. 
It is noted that the subscripts $\uparrow$ and $\downarrow$ in the left-hand side represent 
the pseudo spin to satisfy $\theta f^\dagger_{i\uparrow}\ket{0}=f^\dagger_{i\downarrow}\ket{0}$ and $\theta f^\dagger_{i\downarrow}\ket{0}=-f^\dagger_{i\uparrow}\ket{0}$ for the time-reversal operation $\theta$.
$\ket{i,\pm3,s_z}$ and $\ket{i,\pm2,s_z}$ in the right-hand side are related to the real expressions of the $f$ orbitals $\ket{3a} \propto \sqrt{10}x(x^2-3y^2)/4$, $\ket{3b} \propto \sqrt{10}y(3x^2-y^2)/4$, $\ket{\beta z} \propto \sqrt{15}z(x^2-y^2)/2$, and $\ket{xyz}\propto \sqrt{15}xyz$ as 
\begin{align}
\ket{i,\pm3,s_z} &= \mp\frac{1}{\sqrt{2}}\ket{i,3a,s_z}-\frac{i}{\sqrt{2}}\ket{i,3b,s_z}, \\
\ket{i,\pm2,s_z} &= \frac{1}{\sqrt{2}}\ket{i,\beta z,s_z}\pm\frac{i}{\sqrt{2}}\ket{i,xyz,s_z}.  
\end{align}

Then, the hybridizations $V^0_{m\bm{k}}$ and $V^\alpha_{m\bm{k}}$ ($\alpha=x,y,z$) in $\mathcal{H}^{cf}$ are given by
\begin{align}
\label{eq:V0} 
V^0_{m\bm{k}}&=\sum_{\bm{d}}T^0_{m\bm{d}}e^{i\bm{k}\cdot\bm{d}}, \\
\label{eq:V} 
V^\alpha_{m\bm{k}}&=\sum_{\bm{d}}T^\alpha_{m\bm{d}}e^{i\bm{k}\cdot\bm{d}}, 
\end{align}
where $\bm{d}$ represents the vector connecting the $s$ orbital at $\bm{R}_i+\bm{d}$ and the $f$ orbital at $\bm{R}_i$, and
 \begin{align}
T^0_{m\bm{d}}   &=\frac{-\sqrt{7}\alpha_\mathrm{CEF}+\sqrt{1-\alpha_\mathrm{CEF}^2}}{\sqrt{14}}t^{3a}_{\bm{d}}, \\
T^x_{m\bm{d}}  &=i\sqrt{1-\alpha_\mathrm{CEF}^2}\sqrt{\frac{3}{7}}t^{xyz}_{\bm{d}}, \\
T^y_{m\bm{d}}   &=i\sqrt{1-\alpha_\mathrm{CEF}^2}\sqrt{\frac{3}{7}}t^{\beta z}_{\bm{d}}, \\
T^z_{m\bm{d}}  &=i\frac{\sqrt{7}\alpha_\mathrm{CEF}+\sqrt{1-\alpha_\mathrm{CEF}^2}}{\sqrt{14}}t^{3b}_{\bm{d}}, 
 \end{align}
with $t^{3a}_{\bm{d}}=\sqrt{10}l(l^2-3m^2)(\mathrm{sf\sigma})/4$, $t^{3b}_{\bm{d}}=\sqrt{10}m(3l^2-m^2)(\mathrm{sf\sigma})/4$, $t^{\beta z}_{\bm{d}}=\sqrt{15}n(l^2-m^2)(\mathrm{sf\sigma})/2$, $t^{xyz}_{\bm{d}}=\sqrt{15}lmn(\mathrm{sf\sigma})$. 
$(l,m,n)=\bm{d}/|\bm{d}|$ and ($\mathrm{sf\sigma}$) is the Slater-Koster parameter~\cite{Takegahara_1980}.  
Hereafter, we set $(\mathrm{sf\sigma})=1$, $\bm{d}=\pm\bm{e}_1+m(0,0,c)$, $\pm\bm{e}_2+m(0,0,c)$, and $\pm\bm{e}_3+m(0,0,c)$.
Then, $V^z_{m\bm{k}}$ vanishes for any $\alpha_\mathrm{CEF}$ and $c$ due to the symmetry of $\ket{3b}$.
In addition, for $\alpha_\mathrm{CEF} = \pm 1$ or $c=0$, $V^x_{m\bm{k}}=V^y_{m\bm{k}}=0$, as $t^{xyz}_{\bm{d}}$ and $t^{\beta z}_{\bm{d}}$ are proportional to $n$.
 In this situation, the anisotropic interaction in Eqs.~(\ref{eq:D})-(\ref{eq:F}) appears for $\alpha_\mathrm{CEF} \neq \pm 1$ and $c\neq 0$.
It is noted that the nearest-neighbor hybridizations by $\bm{d}=(0,0,\pm c)$ vanish for any $\alpha_\mathrm{CEF}$ and $c$ owing to the symmetry in the present system. 

\subsection{Effective spin model}
\label{sec:Model3}

\begin{figure}[t!]
\begin{center}
\includegraphics[width=1.0\hsize]{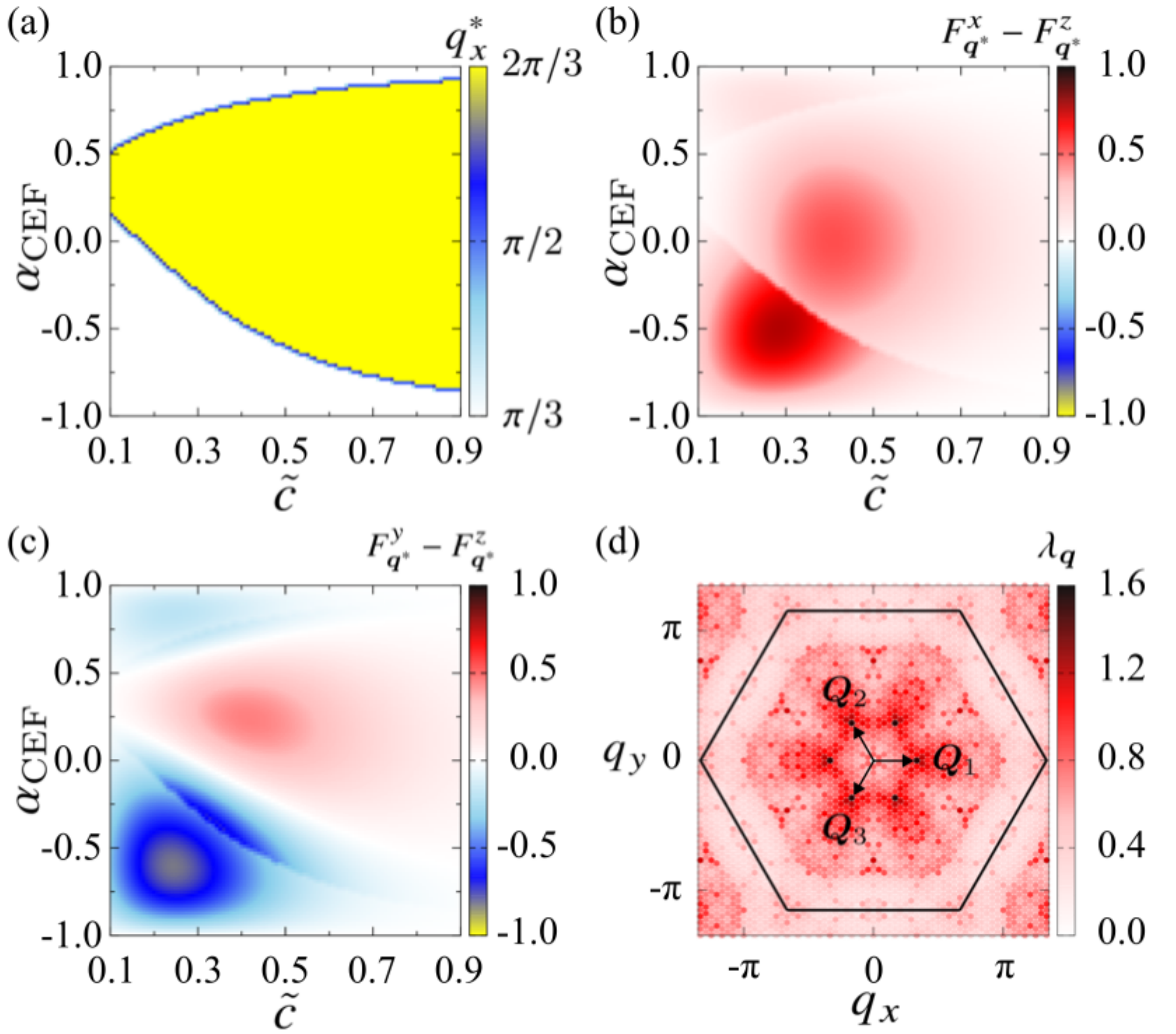} 
\caption{
\label{fig:matrix}  
$\alpha_\mathrm{CEF}$ and $\tilde{c}=c/\sqrt{3}$ dependences of 
(a) $q^*_x$ giving the largest eigenvalue of $X_{\bm{q}}$, 
(b) $F^x_{\bm{q}^*}-F^z_{\bm{q}^*}$,
and (c) $F^y_{\bm{q}^*}-F^z_{\bm{q}^*}$.
(d) Eigenvalues $\lambda_{\bm{q}}$ in momentum space at $\alpha_\mathrm{CEF}=-0.8$ and $c=0.4$, where the hexagon with a solid line shows the first Brillouin zone.
The maxima appears at $\bm{Q}_1=\bm{e}_1\pi/3$, $\bm{Q}_2=\bm{e}_2\pi/3$, and $\bm{Q}_3=\bm{e}_3\pi/3$.  
The other parameters are set as $t_1=1$, $t_3=-0.85$, $\mu=1.3$, $C^{(1)}_{m\bm{k}\bm{k}'}=1$, $T=0.02$, and $N=48^2$.
}
\end{center}
\end{figure}

Following the procedure in Sec.~\ref{sec:Origin1}, we derive the effective spin model for the present PAM. 
From the symmetry argument, there are three independent components ($\bm{F}_{\bm{q}}$) in $X_{\bm{q}}$ for the high-symmetric lines, e.g., $\bm{q}=(q_x,0)$ and $\bm{q}=(0,q_y)$, while there are four independent components ($\bm{F}_{\bm{q}}$ and $E^z_{\bm{q}}$) for a general $\bm{q}$, as shown in Tables~\ref{tab:hexagonal} and \ref{tab:hexagonal2}.
In each $\bm{q}$, the interaction matrix $X_{\bm{q}}$ is calculated when the model parameters $(t_1, t_3, \mu, U, E_f, \bm{V}_{m\bm{k}})$ are given.  
It is noted that $\bm{V}_{m\bm{k}}$ is determined by $\alpha_{\rm CEF}$ and the distance $c$, and $U$ and $E_f$ are used for $C^{(1)}_{m\bm{k}\bm{k}'}$. 
Here, we evaluate $X_{\bm{q}}$ by setting $t_1=1$, $t_3=-0.85$ and $\mu=1.3$. 
For the parameters, we neglect the wave vector dependence of $C^{(1)}_{m\bm{k}\bm{k}'}$ in Eqs.~(\ref{eq:Jiso})-(\ref{eq:Jdm}) by supposing the situation where $U$ and $|E_f|$ is larger than the bandwidth.
Besides, we set $C^{(1)}_{m\bm{k}\bm{k}'}=1$ for simplicity. 

We first calculate the optimal ordering vector $\bm{q}^*=(q^*_x,q^*_y)$ that gives the maximum eigenvalue of $X_{\bm{q}}$ while changing $\alpha_\mathrm{CEF}$ and $\tilde{c}=c/\sqrt{3}$ at a low temperature $T=0.02$ for the system size $N=48^2$. 
As shown in Fig.~\ref{fig:matrix}(a), the maximum eigenvalue of $X_{\bm{q}}$ is obtained for $
\bm{q}^*=(\pi/3,0)$ drawn by the white region, while that is for $\bm{q}^*=(q^*_x,q^*_y)$ with $q^*_x \neq \pi/3$ and $q^*_y \neq 0$ drawn by the color region. 
We also plot the anisotropic exchange interactions, $F^x_{\bm{q}^*}-F^z_{\bm{q}^*}$ and $F^y_{\bm{q}^*}-F^z_{\bm{q}^*}$, in Figs.~\ref{fig:matrix}(b) and \ref{fig:matrix}(c), respectively. 
One finds that the anisotropic interaction to satisfy $F^x_{\bm{q}^*}>F^y_{\bm{q}^*},F^z_{\bm{q}^*}$ is realized in almost the region except for $\alpha_\mathrm{CEF}=\pm1$, where only the isotropic spin interaction appears, i.e., $F^x_{\bm{q}^*}=F^y_{\bm{q}^*}=F^z_{\bm{q}^*}$. 
In other words, the magnitude of anisotropic interactions largely depends on $\alpha_\mathrm{CEF}$ and $c$. 
Especially, the reversal of the magnitude relation between $F^y_{\bm{q}^*}$ and $F^z_{\bm{q}^*}$ in Fig.~\ref{fig:matrix}(c) indicates the instability toward the different spiral or multiple-$Q$ states. 
For example, the tendency toward the out-of-plane (inplane) cycloidal spin is expected for $\alpha_{\rm CEF}=-0.5$ and $\tilde{c}=0.3$ ($\alpha_{\rm CEF}=0.3$ and $\tilde{c}=0.5$).

In the following, we fix the parameters as $\alpha_\mathrm{CEF}=-0.8$ and $\tilde{c}=0.4$, which gives the optimal ordering vectors as $\pm \bm{Q}_1=\pm \bm{e}_1\pi/3$, $\pm \bm{Q}_2=\pm \bm{e}_2\pi/3$, and $\pm \bm{Q}_3=\pm \bm{e}_3\pi/3$. 
We plot the $\bm{q}$ dependence of the largest eigenvalue of $X_{\bm{q}}$ at each $\bm{q}$ denoted as $\lambda_{\bm{q}}$ in Fig.~\ref{fig:matrix}(d).
We summarize the numerical values of $\lambda_{\bm{q}}$, $F^x_{\bm{q}}$, $F^y_{\bm{q}}$, $F^z_{\bm{q}}$, and $E^z_{\bm{q}}$ at $\bm{Q}_1$ in Table~\ref{tab:matrix}.
In addition, we show them at wave vectors given by linear combinations of $\bm{Q}_1$, $\bm{Q}_2$, and $\bm{Q}_3$ for later convenience.  

\begin{table}
\caption{\label{tab:matrix}
$\lambda_{\bm{q}}$, $F^x_{\bm{q}}$, $F^y_{\bm{q}}$, $F^z_{\bm{q}}$, and $E^z_{\bm{q}}$ at $\bm{Q}_1$, $2\bm{Q}_1$, $3\bm{Q}_1$, $2\bm{Q}_1-\bm{Q}_3$, and $\bm{Q}_1-\bm{Q}_3$ at $\alpha_\mathrm{CEF}=-0.8$ and $c=0.4$, where $\bm{Q}_1=\bm{e}_1\pi/3$, $\bm{Q}_2=\bm{e}_2\pi/3$, and $\bm{Q}_3=\bm{e}_3\pi/3$. 
The other parameters are the same as those in Fig.~\ref{fig:matrix}.
}
\begin{ruledtabular}
\begin{tabular}{llllll}
$\bm{q}$ & $\lambda_{\bm{q}}$ & $F^x_{\bm{q}}$ & $F^y_{\bm{q}}$ & $F^z_{\bm{q}}$ & $E^z_{\bm{q}}$ \\ \hline
$\bm{Q}_1$                  & 1.53 & 1.53 & 0.76 & 1.16 & 0.00  \\
$2\bm{Q}_1$                & 0.82      &  0.82 & 0.24 & 0.79 & 0.00  \\
$3\bm{Q}_1$                 & 0.80 & 0.80  & 0.45 & 0.65 & 0.00 \\
$2\bm{Q}_1-\bm{Q}_3$ & 0.64 & 0.55 & 0.30 & 0.64 & -0.09  \\ 
$\bm{Q}_1-\bm{Q}_3$   & 0.55 & 0.46 & 0.28 & 0.52 & 0.16  
\end{tabular}
\end{ruledtabular}
\end{table}     

\subsection{Multiple-$Q$ instability}
\label{sec:NumericalSimulation}

\subsubsection{Simulated annealing}

We investigate the low-temperature magnetic phases in the presence of the effective anisotropic interactions $X_{\bm{q}}$ obtained in Sec.~\ref{sec:Model3}. 
Here, we add the Zeeman term, $\mathcal{H}^\mathrm{Z}=-H\sum_iS_i^z$, to the effective anisotropic spin model in Eq.~(\ref{eq:ABS}) in order to investigate the effects of the magnetic field $H$. 
The spin configuration is obtained by using the simulated annealing combined with the standard Metropolis local updates.
We gradually reduce the temperature with a rate $T_{n+1}=\alpha T_n$, where $T_n$ is the temperature at the $n$th step. 
We set the initial temperature $T_0=1$ and the coefficient $\alpha\approx0.993116$.
A final temperature $T_\mathrm{f} = 0.001$ is reached after total $10^5$ Monte Carlo steps, where we perform $10^2$ Monte Carlo steps at each temperature  $T_n$. 
At the final temperature, we perform $10^4$ Monte Carlo steps for thermalization and measurements, respectively.
To determine the phase boundary, we set the spin configuration obtained near the phase boundary as the initial spin configuration and perform the simulated annealing starting at a low temperature ($T_0=0.05,0.01$).
We set $\lambda_{\bm{Q}_1}$ as the energy unit and $|\bm{S}_i|=1$.

We identify magnetic phases by measuring a magnetic moment, a spin scalar chirality, and the skyrmion number.
The magnetic moment with wave vector $\bm{q}$ is defined as 
\begin{align}
m^\alpha_{\bm{q}} = \sqrt{\left\langle\frac{1}{N^2}\sum_{j,k}S^\alpha_jS^\alpha_k e^{i\bm{q}\cdot(\bm{R}_j-\bm{R}_k)}\right\rangle},
\end{align}
where $\alpha=x,y,z$ and $\langle\cdots\rangle$ is the average over the Monte Carlo samples.
The in-plane and out-of-plane magnetic moments are given by $m_{\bm{q}}^\perp=\sqrt{(m_{\bm{q}}^x)^2+(m_{\bm{q}}^y)^2}$ and $m_{\bm{q}}^z$, respectively.
$\bm{m}_{\bm{q}=\bm{0}}$ corresponds to the uniform magnetization $\bm{M}$.
The spin scalar chirality of the triangle is defined as $\chi_{\bm{r}}=[ \bm{S}_j\cdot( \bm{S}_k\times \bm{S}_l)]$, where the position vector $\bm{r}$ represents the triangle center and the triangle consists of $(j,k,l)$ sites labeled in the counterclockwise order.
The uniform spin scalar chirality is given by $\chi_\mathrm{sc} =  \langle \sum_{\bm{r}} \chi_{\bm{r}}/N \rangle$.
The spin scalar chirality with wave vector $\bm{q}$ is given by 
\begin{align}
\chi_{\bm{q}} = \sqrt{\left\langle\frac{1}{N^2}\sum_{\mu}\sum_{\bm{r},\bm{r}'\in\mu}  \chi_{\bm{r}}  \chi_{\bm{r}'} e^{i\bm{q}\cdot(\bm{r}-\bm{r}')}\right\rangle},
\end{align}
where $\mu=(u,d)$ represents upward and downward triangles, respectively.
A skyrmion density $\Omega_{\bm{r}}$~\cite{BERG1981412} at the triangle $\bm{r}$ is defined as  
\begin{align}
\tan\left(\frac{\Omega_{\bm{r}}}{2}\right) = \left[\frac{\bm{S}_j\cdot( \bm{S}_k\times \bm{S}_l)}{1+\bm{S}_j\cdot\bm{S}_k+\bm{S}_k\cdot\bm{S}_l+\bm{S}_l\cdot\bm{S}_j} \right].
\end{align}
Then, the skyrmion number is given by  
\begin{align}
\label{eq:Nsk}
N_\mathrm{sk} = \frac{1}{4\pi N_\mathrm{m}}\left\langle \sum_{\bm{r}} \Omega_{\bm{r}} \right\rangle,
\end{align}
where $N_\mathrm{m}$ is the number of the magnetic unit cell.

In the following, we discuss three situations with different sets of wave vectors, $\{\bm{Q}\}$. 
First, we analyze the ground state of the effective spin model by taking into account all the $\bm{q}$ contributions in the interactions in Sec.~\ref{sec:all}.
As mentioned in Sec.~\ref{sec:Model1}, a part of interactions are important to describe the magnetic instability at low temperatures. 
Therefore, we discuss the minimum effective spin model to reproduce the results in Sec.~\ref{sec:all}. 
In Sec.~\ref{sec:1Q}, we find that it is not enough to reproduce the results in Sec.~\ref{sec:all} when  considering only the contributions from $\bm{Q}_1$-$\bm{Q}_3$. 
In Sec.~\ref{sec:3Q}, we show that the additional contribution from $3\bm{Q}_1$-$3\bm{Q}_3$ well reproduces the results in Sec.~\ref{sec:all}. 

\subsubsection{Case of the interactions at all the wave vectors}
\label{sec:all}

\begin{figure}[t!]
\begin{center}
\includegraphics[width=0.95\hsize]{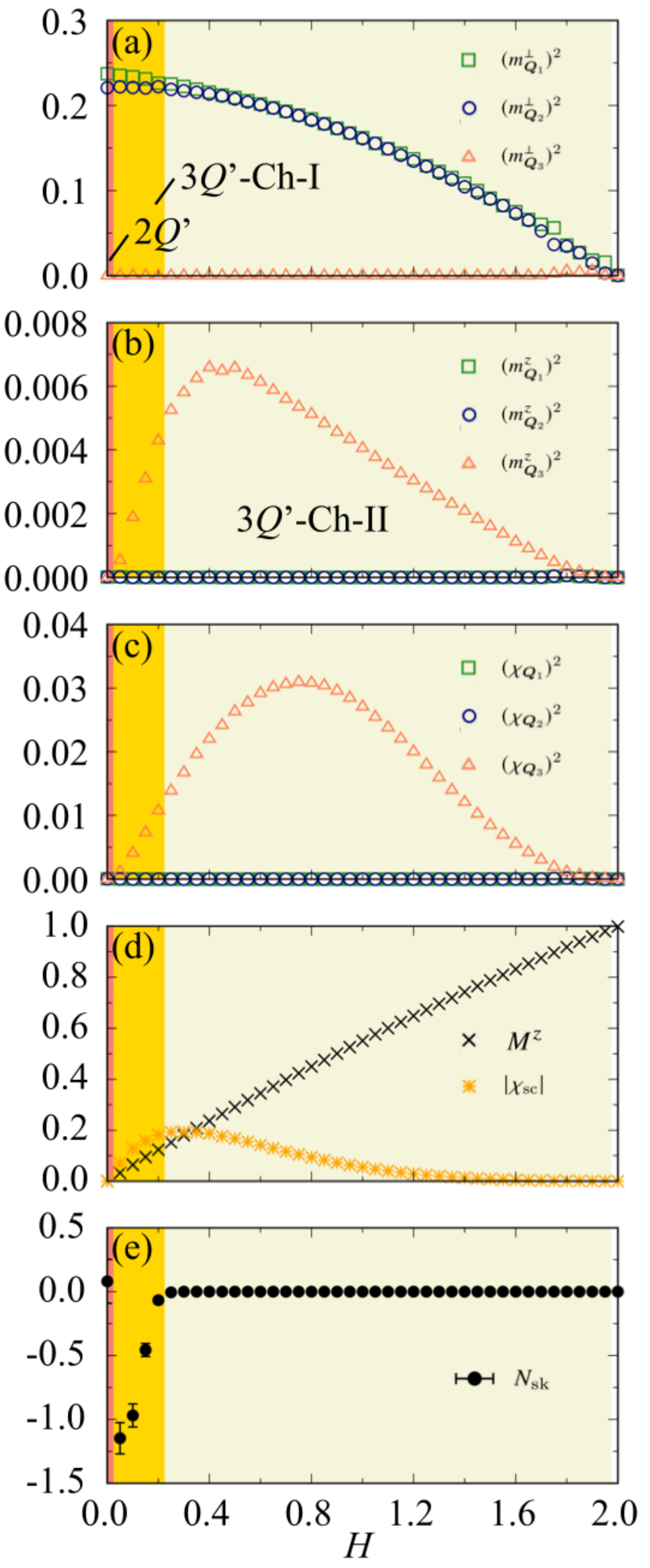} 
\caption{
\label{fig:AllQ}
$H$ dependences of 
(a) $(m^\perp_{\bm{Q}_{\eta}})^2$,
(b) $(m^z_{\bm{Q}_{\eta}})^2$,
(c) $(\chi_{\bm{Q}_{\eta}})^2$,
(d) $M^z$ and $|\chi_\mathrm{sc}|$,
and (e) $N_\mathrm{sk}$
in the model with the interactions at all the wave vectors.
We sort $m^\perp_{\bm{Q}_\eta}$, $m^z_{\bm{Q}_\eta}$, and  $\chi_{\bm{Q}_\eta}$ to satisfy $m^\perp_{\bm{Q}_1}\ge m^\perp_{\bm{Q}_2}\ge m^\perp_{\bm{Q}_3}$. 
}
\end{center}
\end{figure}

\begin{figure*}[t!]
\begin{center}
\includegraphics[width=1.0\hsize]{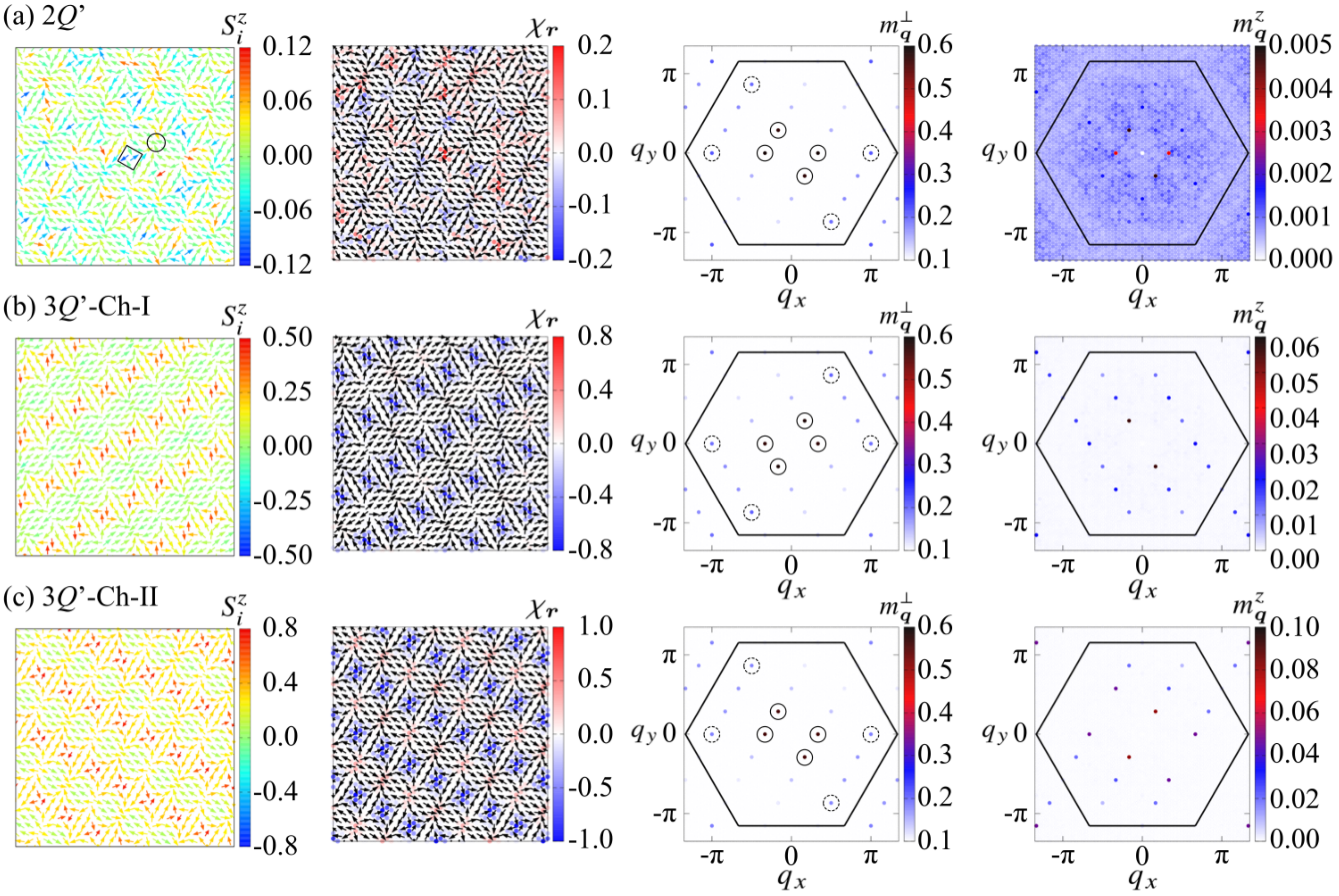} 
\caption{
\label{fig:AllQ spin} 
First column: The spin configurations averaged over 500 Monte Carlo steps of (a) the 2$Q'$ state at $H=0$, (b) 3$Q'$-Ch-I state at $H=0.15$, and (c) 3$Q'$-Ch-II state at $H=0.5$.
The arrows and contours show the $xy$ and $z$ components of the spin, respectively.
The circle (square) highlights the vortex (antivortex) structure in the $xy$ spins.   
Second column: The scalar chirality configurations of the first column. 
Third and fourth columns: The in-plane and out-of-plane magnetic moments 
in momentum space.
The solid and dashed circles in the third column highlight the $\bm{Q}_\eta$ and $3\bm{Q}_\eta$ components, respectively. 
The hexagons with a solid line show the first Brillouin zone.
The $\bm{q}=\bm{0}$ component is removed for better visibility.
}
\end{center}
\end{figure*}

In the effective spin model with the interactions at all the wave vectors $\bm{q}$ except for $\bm{q}=0$, we investigate the ground state of the effective spin model while changing the magnetic field $H$. 
We show $H$ dependences of the in-plane magnetic moment at $\bm{Q}_1$-$\bm{Q}_3$, $(m^\perp_{\bm{Q}_{\eta}})^2$, in Fig.~\ref{fig:AllQ}(a), the out-of-plane magnetic moment at $\bm{Q}_1$-$\bm{Q}_3$, $(m^z_{\bm{Q}_{\eta}})^2$, in Fig.~\ref{fig:AllQ}(b), the spin scalar chirality at $\bm{Q}_1$-$\bm{Q}_3$, $(\chi_{\bm{Q}_{\eta}})^2$, in Fig.~\ref{fig:AllQ}(c), the uniform magnetization $M^z$ and the uniform spin scalar chirality $|\chi_\mathrm{sc}|$ in Fig.~\ref{fig:AllQ}(d), and the skyrmion number $N_\mathrm{sk}$ in Fig.~\ref{fig:AllQ}(e), where we sort $m^\perp_{\bm{Q}_\eta}$, $m^z_{\bm{Q}_\eta}$, and  $\chi_{\bm{Q}_\eta}$ to satisfy $m^\perp_{\bm{Q}_1}\ge m^\perp_{\bm{Q}_2}\ge m^\perp_{\bm{Q}_3}$ for better readability. 
In addition to the fully polarized state at $H=2$, we find three types of the multiple-$Q$ states; 
all the states are characterized by $\bm{m}_{\bm{Q}_\eta}$, since $X_{\bm{q}}$ has the largest eigenvalues at $\bm{Q}_\eta$, as detailed below. 
Figure~\ref{fig:AllQ spin} shows the spin and chirality configurations in real space and the magnetic moments in momentum space for each multiple-$Q$ state.
The skyrmion density configurations in real space for each multiple-$Q$ state are shown in Fig.~\ref{fig:AllQ Nsk}.      

At $H=0$, the ground state becomes a double-$Q$ (2$Q'$) state.
In this state, the spin configuration is characterized by the double-$Q$ in-plane components $m^\perp_{\bm{Q}_1}$ and $m^\perp_{\bm{Q}_2}$ with different intensities and no out-of-plane components at $\bm{Q}_1$-$\bm{Q}_3$ ($Q'$ represents different intensities of the $\bm{Q}_1$ and $\bm{Q}_2$ components), as shown in Figs.~\ref{fig:AllQ}(a) and \ref{fig:AllQ}(b).
The real-space spin configuration is shown in the first column of Fig.~\ref{fig:AllQ spin}(a).
The in-plane spins form a periodic structure consisting of the vortex (circle) and antivortex (square), while the $z$ spins show no periodic structure.
Such a tendency is found in the presence (absence) of sharp peaks in $m^\perp_{\bm{q}}$ ($m^z_{\bm{q}}$), as shown in the third (fourth) column of Fig.~\ref{fig:AllQ spin}(a). 
In the scalar chirality sector, this state exhibits $\chi_{\bm{Q}_\eta}=0$ and $\chi_\mathrm{sc}=0$, as shown in Figs.~\ref{fig:AllQ}(c) and \ref{fig:AllQ}(d), respectively. 
In the real-space picture, the local scalar chirality is randomly distributed, as shown in the second column of Fig.~\ref{fig:AllQ spin}(a).
Accordingly, there is no skyrmion number ($N_\mathrm{sk}=0$) in Fig.~\ref{fig:AllQ}(e).
      
By applying a magnetic field, the 2$Q'$ state changes into a triple-$Q$ chiral I (3$Q'$-Ch-I) state, whose spin structure is characterized by the double-$Q$ in-plane components $m^\perp_{\bm{Q}_1}$>$m^\perp_{\bm{Q}_2}$ and the single-$Q$ $z$ component $m^z_{\bm{Q}_3}$, as shown in Figs.~\ref{fig:AllQ}(a) and \ref{fig:AllQ}(b).
The in-plane spin configuration of the 3$Q'$-Ch-I state is similar to that of the 2$Q'$ state, as shown in the first and third columns of Fig.~\ref{fig:AllQ spin}(b). 
Meanwhile, the first and fourth columns of Fig.~\ref{fig:AllQ spin}(b) show a structure of $z$ spin components due to the single-$Q$ peak of $m^z_{\bm{Q}_\eta}$, where the $z$ spins have positive (small positive or negative) values at antivortices (vortices).
The undetermined sign of the $z$ spins at vortices is owing to the small value of $m^z_{\bm{Q}_\eta}$, which results in the fluctuations of $N_{\rm sk}$ characterized by non-integer values, as shown in Fig.~\ref{fig:AllQ}(e). 
The 3$Q'$-Ch-I state shows a nonzero uniform scalar chirality [Fig.~\ref{fig:AllQ}(d)] as well as the chirality density wave along the $\bm{Q}_3$ direction [Fig.~\ref{fig:AllQ}(c)].
The nonzero uniform scalar chirality is attributed to the inequivalence between the $z$ spin component at antivortices and vortices, as found in the real-space spin and chirality configurations in Fig.~\ref{fig:AllQ spin}(b); there is a large negative chirality at antivortices with large $z$ spins and a small negative/positive chirality at vortices with small $z$ spins.

While increasing $H$, the peak structure of $m^z_{\bm{Q}_\eta}$ and $M^z$ are developed, and then, there are no fluctuations in $N_{\rm sk}$ for $H \gtrsim 0.225$. 
We call this state a triple-$Q$ chiral II (3$Q'$-Ch-II) state.
As the difference of
$m^\perp_{\bm{Q}_\eta}$, $m^z_{\bm{Q}_\eta}$, $\chi_{\bm{Q}_\eta}$, and $\chi_\mathrm{sc}$ between the 3$Q'$-Ch-I phase and the 3$Q'$-Ch-II phase seems to be slight in Figs.~\ref{fig:AllQ}(a)-\ref{fig:AllQ}(d), the similar spin and chirality configurations in real and momentum spaces appear in Figs.~\ref{fig:AllQ spin}(b) and \ref{fig:AllQ spin}(c).
By closely looking into their spin configurations, one finds that all spins have positive $z$ components in the 3$Q'$-Ch-II phase in Fig.~\ref{fig:AllQ spin}(c), which is presumably due to the development of  $m^z_{\bm{Q}_\eta}$ and $M^z$.
As a result, the positive chirality contribution appears at vortices, which leads to the suppression of the total scalar chirality, as shown in Fig.~\ref{fig:AllQ}(d). 
While further increasing $H$, the chirality contributions from the vortices and antivortices are canceled out, and then, this state turns into the fully polarized state at $H=2$.

\begin{figure}[t!]
\begin{center}
\includegraphics[width=1.0\hsize]{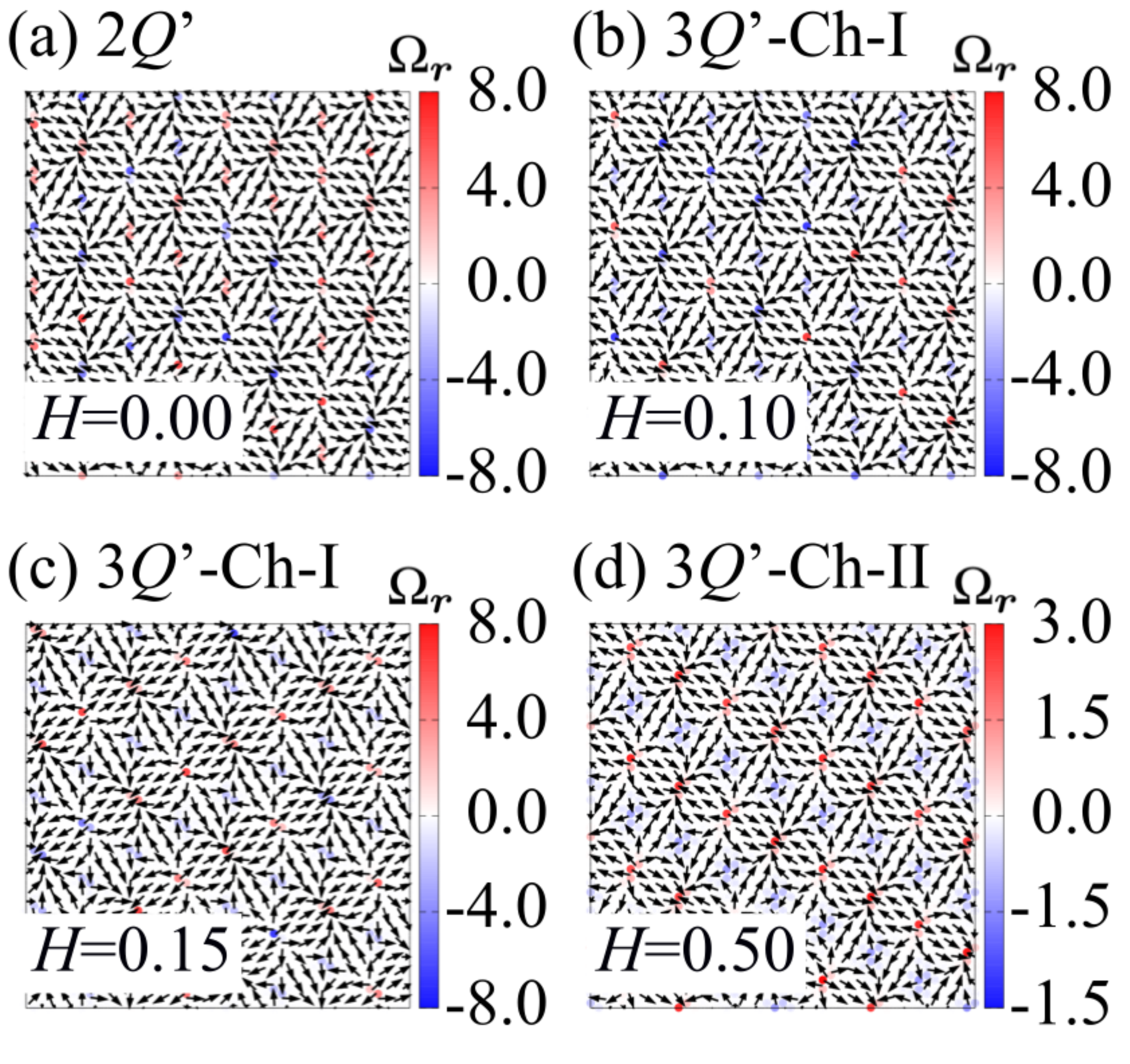} 
\caption{
\label{fig:AllQ Nsk}
Skyrmion density configurations of 
(a) the 2$Q'$ state at $H=0
$, (b) 3$Q'$-Ch-I state at $H=0.1
$, (c) 3$Q'$-Ch-I state at $H=0.15$, and (d) 3$Q'$-Ch-II state at $H=0.5$.
The skyrmion density is calculated by using the spin configuration averaged over 500 Monte Carlo steps.  
}
\end{center}
\end{figure}

We further discuss the $H$ dependence of $N_\mathrm{sk}$ in Fig.~\ref{fig:AllQ}(e), especially for the small $H$ region, where $N_\mathrm{sk}$ takes a non-integer value.
We plot the real-space skyrmion density configurations in Fig.~\ref{fig:AllQ Nsk}.
All the states have the large skyrmion density near the (anti)vortex cores.
At $H=0$, the skyrmion number becomes zero within the errorbars, where both vortices and antivortices take a random value, as shown in Fig.~\ref{fig:AllQ Nsk}(a). 
For $H>0$, $N_\mathrm{sk}$ takes a non-integer value in the 3$Q$'-Ch-I state.
In this state, the antivortices take a negative value, while the vortices take a positive or negative value at random, as shown in Figs.~\ref{fig:AllQ Nsk}(b) and \ref{fig:AllQ Nsk}(c).
This randomness is the reason why $N_{\rm sk}$ becomes the non-integer values.
Such randomness is suppressed while increasing $H$, as shown in Figs.~\ref{fig:AllQ Nsk}(b) and \ref{fig:AllQ Nsk}(c). 
In the end, the randomness vanishes in the 3$Q$'-Ch-II state, since the vortices always take a positive value, as shown in Fig.~\ref{fig:AllQ Nsk}(d). 
This result indicates that the energy scale of $F^z_{\bm{q}}$ is too small to lead to the sharp peak of $m^z_{\bm{Q}_\eta}$, which makes the skyrmion density at the vortices ambiguous. 

\subsubsection{Case of the interactions at $Q_\eta$}
\label{sec:1Q}

\begin{figure}[t!]
\begin{center}
\includegraphics[width=0.95\hsize]{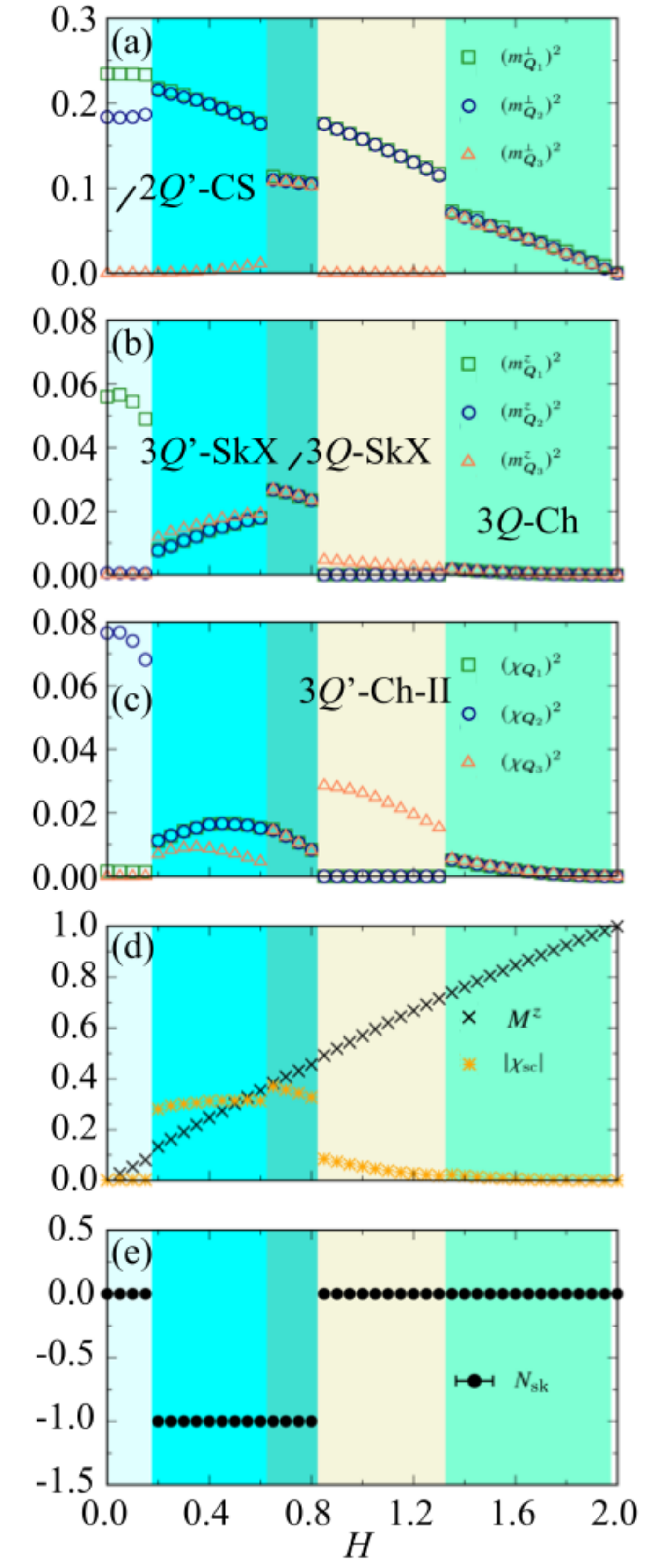} 
\caption{
\label{fig:1Q}
$H$ dependences of 
(a) $(m^\perp_{\bm{Q}_{\eta}})^2$,
(b) $(m^z_{\bm{Q}_{\eta}})^2$,
(c) $(\chi_{\bm{Q}_{\eta}})^2$,
(d) $M^z$ and $|\chi_\mathrm{sc}|$,
and (e) $N_\mathrm{sk}$
in the model with the $\bm{Q}_\eta$ channels.
We sort $m^\perp_{\bm{Q}_\eta}$, $m^z_{\bm{Q}_\eta}$, and  $\chi_{\bm{Q}_\eta}$ to satisfy $m^\perp_{\bm{Q}_1}\ge m^\perp_{\bm{Q}_2}\ge m^\perp_{\bm{Q}_3}$. 
}
\end{center}
\end{figure}

\begin{figure}[t!]
\begin{center}
\includegraphics[width=1.0\hsize]{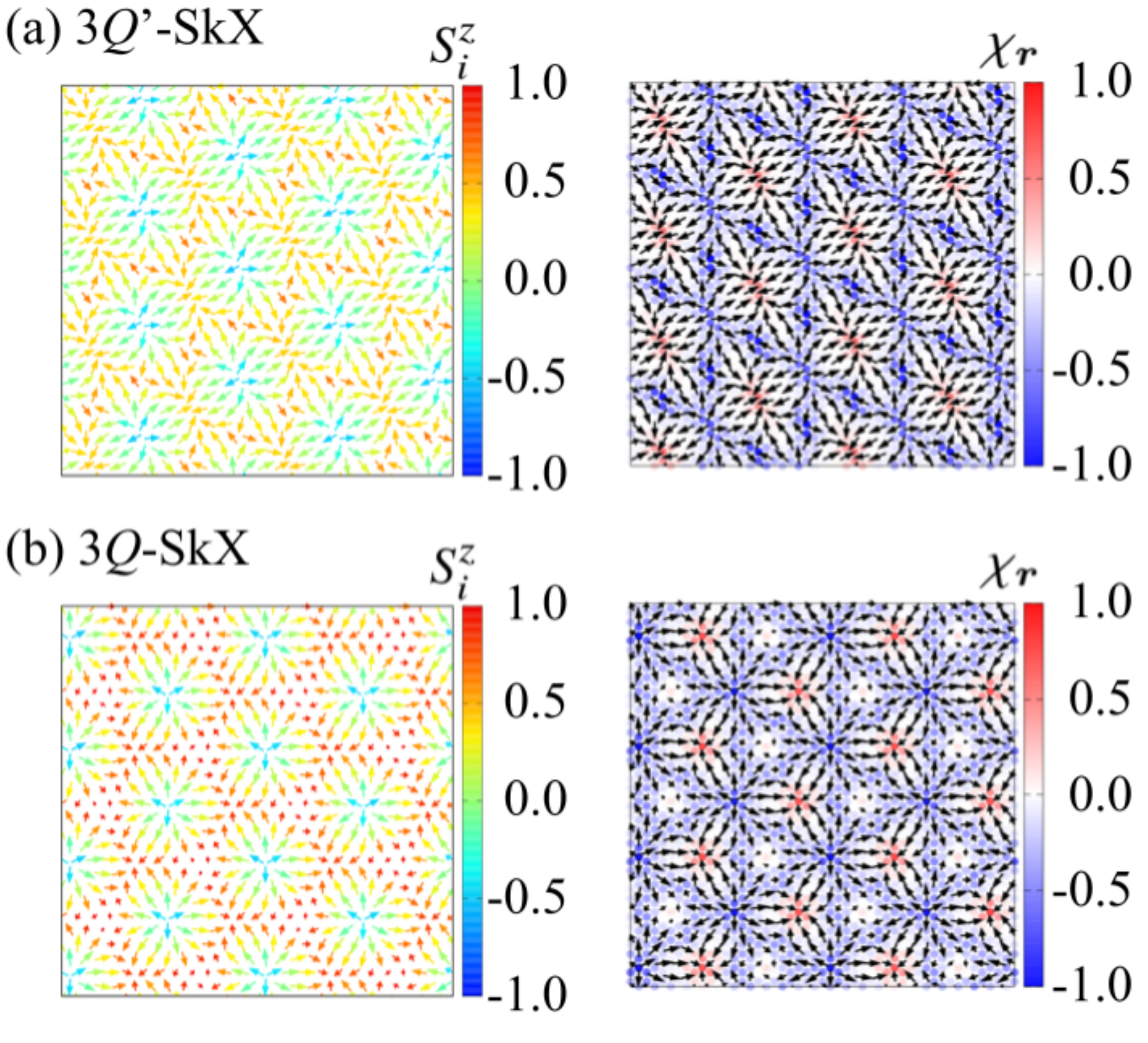} 
\caption{
\label{fig:1Q SkX}
Left panel: Snapshots of the spin configurations of (a) the 3$Q'$-SkX state at $H=0.3$ and (b) 3$Q$-SkX state at $H=0.75$.
The arrows and contours show the $xy$ and $z$ components of the spin, respectively.  
Right panel: The scalar chirality configurations corresponding to the spin configurations shown in the left panel.
}
\end{center}
\end{figure}

To identify the origin of the multiple-$Q$ states, we consider the minimum model to reproduce the results in Fig.~\ref{fig:AllQ} by dropping off the less important $\bm{q}$ component of the interactions.　
In the previous section, we find that the model shows the instability toward the multiple-$Q$ states with the scalar chirality, where there are no contributions from the interactions at almost all $\bm{q}$ channels except for $\bm{Q}_\nu$ and their higher harmonics, as discussed in Sec.~\ref{sec:Model1}. 
In this section, we only consider the contributions of the interactions at $\{\bm{Q}\}=\{\pm\bm{Q}_1,\pm\bm{Q}_2,\pm\bm{Q}_3\}$, since they give the maximum eigenvalue of $X_{\bm{q}}$.

As a result, we find that the model with the interactions at $\{\bm{Q}\}=\{\pm\bm{Q}_1,\pm\bm{Q}_2,\pm\bm{Q}_3\}$ is oversimplified in the present situation.
The $H$ dependences in Fig.~\ref{fig:1Q} show that the magnetic phases in the present model are different from those in Sec.~\ref{sec:all}; we obtain the 2$Q'$-CS, 3$Q'$-SkX, 3$Q$-SkX, and 3$Q$-Ch states that are not stabilized in the model in Sec.~\ref{sec:all}.
In particular,  the appearance of the 3$Q'$-SkX and 3$Q$-SkX with $N_\mathrm{sk}=-1$ is a characteristic of the oversimplified model, whose real-space spin and chirality configurations are shown in Fig.~\ref{fig:1Q SkX}.
In the 3$Q'$-SkX, the in-plane spin configuration is similar to that in the 3$Q'$-Ch-I state, while there is a difference in the $z$ spin configurations; the 3$Q'$-SkX in Fig~\ref{fig:1Q SkX}(a) [the 3$Q'$-Ch-I state in Fig~\ref{fig:AllQ spin}(b)] has the (no) alternating arrangement of vortices with the positive and negative $z$ spins in the $\bm{Q}_3$ direction. 
Meanwhile, The 3$Q$-SkX in Fig~\ref{fig:1Q SkX}(b) shows an entirely different structure, which is expressed as the superposition of the three cycloidal elliptical waves with the same intensity.
The 3$Q$-SkX is similar to the SkX in Fig.~\ref{fig:modulation}(b), since it is stabilized by the interplay among large $F^x_{\bm{Q}_1}$, the isotropic interaction, and the magnetic field, as discussed in Sec.~\ref{sec:Triple}. 
We show the real-space spin and chirality configurations, the $\bm{q}$-space magnetic moments, and the skyrmion density configurations for the obtained states in Appendix~\ref{ap:Phase} for reference. 

\subsubsection{Case of the interactions at $Q_\eta$ and $3Q_\eta$}
\label{sec:3Q}

\begin{figure}[t!]
\begin{center}
\includegraphics[width=0.95\hsize]{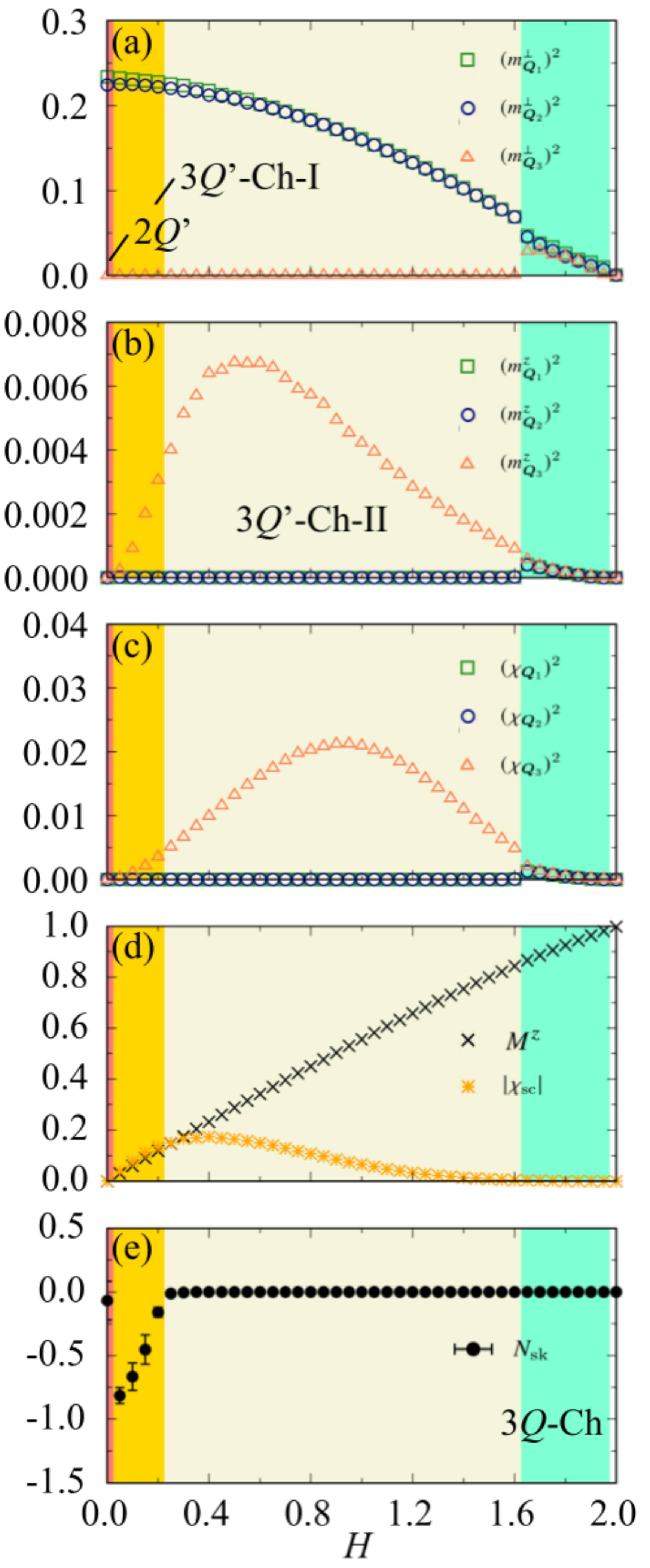} 
\caption{
\label{fig:3Q}
$H$ dependences of 
(a) $(m^\perp_{\bm{Q}_{\eta}})^2$,
(b) $(m^z_{\bm{Q}_{\eta}})^2$,
(c) $(\chi_{\bm{Q}_{\eta}})^2$,
(d) $M^z$ and $|\chi_\mathrm{sc}|$,
and (e) $N_\mathrm{sk}$
in the model with the $\bm{Q}_\eta$ and $3\bm{Q}_\eta$  channels.
We sort $m^\perp_{\bm{Q}_\eta}$, $m^z_{\bm{Q}_\eta}$, and  $\chi_{\bm{Q}_\eta}$ to satisfy $m^\perp_{\bm{Q}_1}\ge m^\perp_{\bm{Q}_2}\ge m^\perp_{\bm{Q}_3}$. 
}
\end{center}
\end{figure}

Next, we focus on the contribution from higher harmonics for the following reasons. 
By comparing the $\bm{q}$-resolved magnetic moments shown in Figs.~\ref{fig:AllQ spin} and \ref{fig:1Q spin}, we find that the discrepancy between the results in Figs.~\ref{fig:AllQ} and \ref{fig:1Q} appears in the magnetic moments at higher-harmonic wave vectors. 
Indeed, the values of $\lambda_{\bm{q}}$ and $\bm{F}_{\bm{q}}$ at $2\bm{Q}_1$, $3\bm{Q}_1$, and $2\bm{Q}_1-\bm{Q}_3$ are large enough to compete with those at $\bm{Q}_1$, as shown in Table~\ref{tab:matrix}. 
On the basis of the above discussion, we additionally take into account the interactions at the higher-harmonic wave vectors to those at $\bm{Q}_\nu$.

By performing the numerical simulations for the several models with the different $\{\bm{Q}\}$, we find that the introduction of the interactions at $\pm3\bm{Q}_1,\pm3\bm{Q}_2,\pm3\bm{Q}_3$ is enough to reproduce the results in Fig.~\ref{fig:AllQ}. 
We show the results for the model with $\{\bm{Q}\}=\{\pm\bm{Q}_1,\pm\bm{Q}_2,\pm\bm{Q}_3, \pm3\bm{Q}_1,\pm3\bm{Q}_2,\pm3\bm{Q}_3 \}$ in Fig.~\ref{fig:3Q}.
Compared to the results in Fig.~\ref{fig:AllQ}, the $H$ dependences of spin- and chirality-related quantities are reproduced except for the high field region, $H\gtrsim 1.625$.
It is noted that there is still an inconsistency in the high-field region; the 3$Q$-Ch state appears for $H\gtrsim 1.625$ corresponding to the 3$Q$-Ch state in the model with the interactions only at $\pm\bm{Q}_1,\pm\bm{Q}_2$, and $\pm\bm{Q}_3$ in Sec.~\ref{sec:1Q}, although the intensities of $\bm{Q}_1$, $\bm{Q}_2$, and $\bm{Q}_3$ components in the magnetic moments are slightly different in the present 3$Q$-Ch state.
This result indicates that the interactions at other higher harmonics like $2\bm{Q}_1- \bm{Q}_2$, which contributes to the energy in the 3$Q'$-Ch-II state [Fig.~\ref{fig:AllQ spin}(c)] might be important in the high field region. 

The reason why the contribution from the interactions at $3\bm{Q}_\nu$ is important is understood from the spiral modulation in the presence of anisotropic interactions. 
From the relation of $F^x_{\bm{Q}_{1}}>F^z_{\bm{Q}_{1}}(>F^y_{\bm{Q}_{1}})$, the spiral plane along the $\bm{Q}_{1}$ direction is elliptically modulated so as to have more $x$-spin component.
In a similar way, the multiple-$Q$ states in Sec.~\ref{sec:all} consist of a superposition of the elliptical waves along the $\bm{Q}_1$-$\bm{Q}_3$ directions. 
Such a deformation from the circular spiral plane to the elliptical spiral plane leads to the relatively large intensity at $3\bm{Q}_{\eta}$, as shown by the dashed circles in the third column in Fig.~\ref{fig:AllQ spin}. 
Thus, the interactions at $3\bm{Q}_{\eta}$ play an important role in the present situation. 
Meanwhile, it is noted that the contribution at the $2\bm{Q}_{\eta}$ channel is not important in spite of the larger value of $\lambda_{2\bm{Q}_{\eta}}$ than $\lambda_{3\bm{Q}_{\eta}}$, since the $2\bm{Q}_{\eta}$ modulation does not appear in the elliptical modulation under $\bm{F}_{\bm{Q}_{\eta}}$. 

\begin{figure}[t!]
\begin{center}
\includegraphics[width=1.0\hsize]{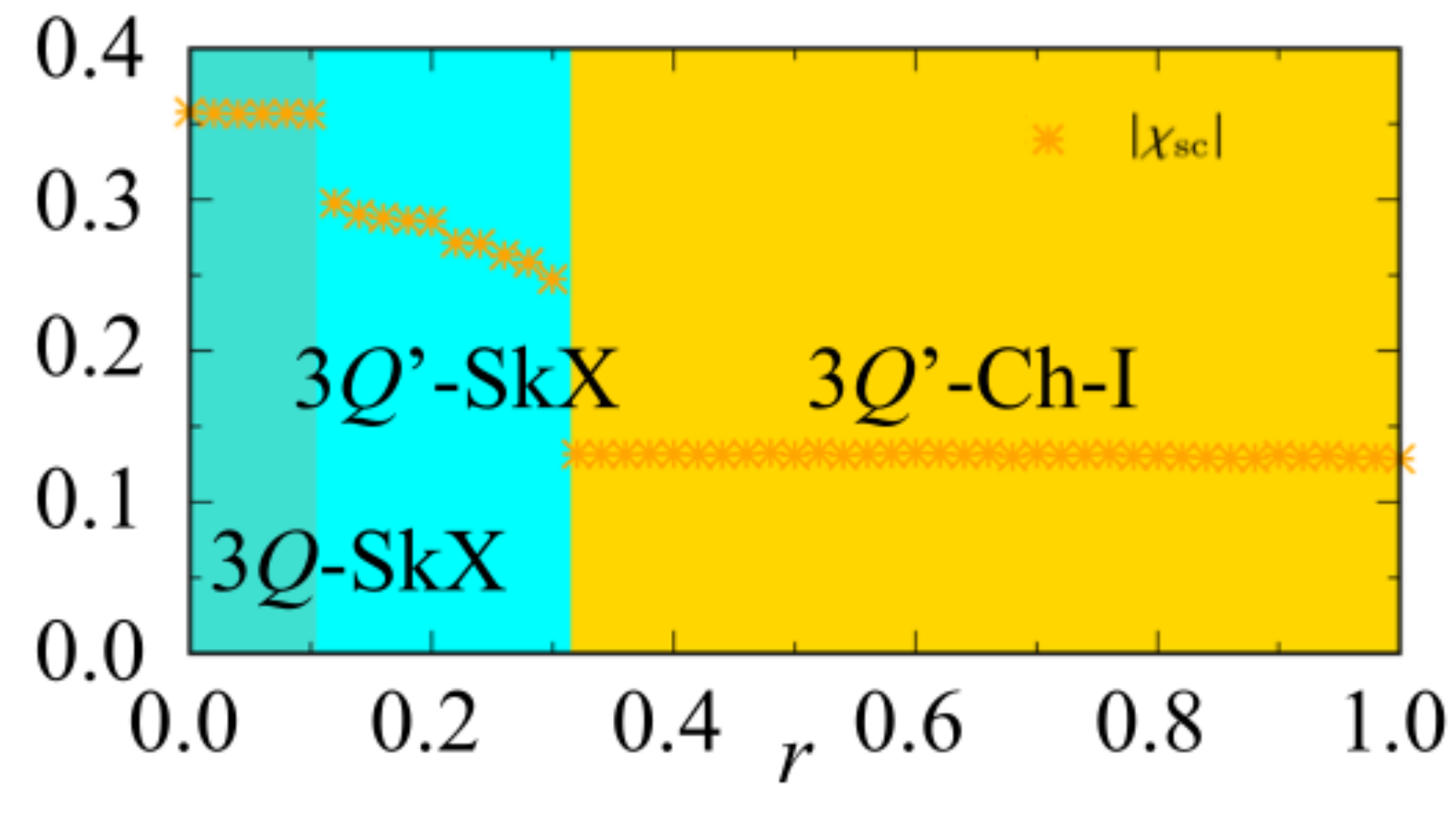} 
\caption{
\label{fig:3Q_2}
$r$ dependence of $|\chi_\mathrm{sc}|$ in the model with $X_{\bm{Q}_\eta}$ and $rX_{3\bm{Q}_\eta}$ at $H=0.7$. 
}
\end{center}
\end{figure}

Furthermore, we investigate how large contribution from the $3\bm{Q}_{\eta}$ channel requires the stabilization of the 3$Q$'-Ch-I state by multiplying the variable $0 \leq r \leq 1$ by $X_{3\bm{Q}_\eta}$.
Figure~\ref{fig:3Q_2} shows the $r$ dependence of the uniform spin scalar chirality at $H=0.7$. 
The result at $r=0$ corresponds to that in Fig.~\ref{fig:1Q}, while the result at $r=1$ corresponds to that in Fig.~\ref{fig:3Q}.
The 3$Q'$-SkX in the intermediate $r$ has similar spin and chirality textures to those in Fig.~\ref{fig:1Q SkX}(a).
The result shows that the 3$Q'$-Ch-I state appears at $r \simeq 0.31$, 
which indicates that relatively small $\lambda_{3\bm{Q}_1}\lesssim\lambda_{\bm{Q}_1}/4$ leads to the stabilization (destabilization) of the 3$Q'$-Ch-I (3$Q^{(')}$-SkX) state.

Finally, we find that the relationship of $F^x_{3\bm{Q}_1}>F^y_{3\bm{Q}_1},F^z_{3\bm{Q}_1}$ is also important.
Indeed, when we perform the simulations by setting $F^y_{3\bm{Q}_1}>F^x_{3\bm{Q}_1},F^z_{3\bm{Q}_1}$ and $F^z_{3\bm{Q}_1}>F^x_{3\bm{Q}_1},F^y_{3\bm{Q}_1}$, at the same time, we also change $3\bm{Q}_2$ and $3\bm{Q}_3$ channels to satisfy the threefold rotational symmetry, 
we could not reproduce the results in Fig.~\ref{fig:AllQ}.

\section{Summary and Perspective}
\label{sec:Summary}

To summarize, we formulated a systematic method of constructing the effective spin model with the momentum-resolved anisotropic exchange interactions based on two approaches in order to systematically understand multiple-$Q$ instabilities.
First, by performing magnetic representation analysis, we found the six symmetry rules to obtain nonzero momentum-resolved anisotropic exchange interactions.
According to the rules, one can systematically construct the effective spin model in any primitive lattices.
As a demonstration, we showed the effective spin models in the tetragonal, hexagonal, and trigonal crystal systems.  
Second, by performing perturbation analysis, we found that the spin-dependent hybridizations between itinerant electron and localized electron states are important microscopic model parameters for nonzero long-range anisotropic exchange interactions in metals.
The results beyond the symmetry argument give a way to quantitatively evaluate the contributions of the anisotropic interactions in magnetic metals within the framework of the first principle calculations.
Finally, we showed how to use the above general results by applying them to a hexagonal crystal and  how the anisotropic interactions affect multiple-$Q$ states by performing the simulated annealing for the effective model.
We found that a plethora of multiple-$Q$ states with a spin scalar chirality are stabilized by the symmetric anisotropic exchange interactions at wave vectors that give the maximum of the magnetic susceptibility as well as those at their higher harmonics.

Our results will stimulate further exploration of materials hosting SkX.
Based on the symmetry argument, one can construct the effective spin model and analyze possible SkXs stabilized by the anisotropic interactions once the crystal symmetry is provided.
Therefore, the symmetry argument provides a reference for the exploration of further SkXs in both centrosymmetric and noncentrosymmetric magnets since our results give a complete relationship between the anisotropic exchange interaction and crystal symmetry in any crystal systems. 
In particular, the symmetry rules about the symmetric anisotropic interaction makes it possible to search centrosymmetric materials hosting SkXs, which have been less studied so far compared to
noncentrosymmetric materials based on Moriya's rule.

In addition, our results will open up a possibility of exotic multiple-$Q$ states beyond the SkXs. 
As various sets of anisotropic exchange interactions emerge depending on the crystal symmetry, there are several ways to stabilize different types of multiple-$Q$ states. 
Indeed, we showed that the competition between interactions at different wave vectors leads to the emergence of the unconventional multiple-$Q$ state with a non-integer skyrmion number.
These competitions might become a source of exotic multiple-$Q$ states~\cite{Hayami_doi:10.7566/JPSJ.89.103702,doi:10.7566/JPSJ.91.023705}.

\begin{acknowledgments}
The authors thank M. Yatsushiro and T. Matsumoto for the fruitful discussions. 
This research was supported by JSPS KAKENHI Grants Numbers JP19K03752, JP19H01834, JP21H01037, and by JST PRESTO (JPMJPR20L8). 
R.Y. was supported by Forefront Physics and Mathematics Program to Drive Transformation (FoPM).
Parts of the numerical calculations were performed in the supercomputing systems in ISSP, the University of Tokyo.
\end{acknowledgments}

\appendix

\section{Magnetic representation analysis}
\label{ap:MagRep}

We show details of the magnetic representation analysis in Sec.~\ref{sec:Symmetry}.
The $\bm{q}$-resolved anisotropic spin interaction in Eq.~(\ref{eq:Interaction}) is determined so as to satisfy the crystal symmetry as well as the time-reversal symmetry.
In the following, we discuss the interaction matrix $X_{\bm{q}}$ in the gray symmorphic space group including the time-reversal operation $\theta$, space group operations, and their product.

The time-reversal symmetry connecting $\pm\bm{q}$ imposes 
\begin{align}
X^{\alpha\beta}_{\bm{q}} 
&=\theta X^{\alpha\beta}_{-\bm{q}}\theta^{-1} \nonumber \\
\label{eq:TR}
&= \left( X^{\alpha\beta}_{-\bm{q}} \right)^*, 
\end{align}
where the property of the anti-linearity of $\theta$ is used in the second line.  
From this symmetry constraint and the definition of $X_{\bm{q}}$ in Eq.~(\ref{eq:CouplingMatrix}), one obtains $\bm{D}_{\bm{q}}=-\bm{D}_{-\bm{q}}$, $\bm{E}_{\bm{q}}=\bm{E}_{-\bm{q}}$ and $\bm{F}_{\bm{q}}=\bm{F}_{-\bm{q}}$, which means that $\bm{D}_{\bm{q}}$ is antisymmetric in momentum space but $\bm{E}_{\bm{q}}$ and $\bm{F}_{\bm{q}}$ are symmetric.

We adopt point group operations in momentum space as follows.
Let us assume a crystal with a lattice vector $\bm{R}_n$ and a point group operation $P$ of the crystal.
Then, $P\bm{R}_n$ leaves the system invariant.
Meanwhile, the crystal in momentum space is characterized by a reciprocal lattice vector $\bm{G}_m$, where $P\bm{G}_m$ leaves the system invariant.
Thus, the same point group operation $P$ is present in both real and momentum spaces.
 
Since the anisotropic spin interaction, $\bm{S}_{\bm{q}}^T X_{\bm{q}}  \bm{S}_{-\bm{q}}$, is regarded as the interaction between two spins at reciprocal wave vector $\pm\bm{q}$, 
nonzero components in $X_{\bm{q}}$ are determined by the point group operation leaving the ``bond" connecting $\bm{q}$ and $-\bm{q}$.
There are two types of such operations, $P^\mathrm{I}$ and $P^\mathrm{II}$, which are given by
\begin{flushleft}
\begin{itemize}
\item[
(I)] operation $P^\mathrm{I}$ satisfying $P^\mathrm{I}\bm{q} = \bm{q}$,
\item[(II)] operation $P^\mathrm{II}$ satisfying $P^\mathrm{II}\bm{q} = -\bm{q}$.
\end{itemize}
\end{flushleft}
In terms of the magnetic space group, these point group operations form the magnetic little co-group~\cite{bradley2009mathematical}.
In other words, the anisotropic spin interaction in Eq.~(\ref{eq:Interaction}) must satisfy the magnetic little co-group symmetry rather than the point group symmetry, which is the reason why the anisotropic interaction depends on not only the crystal symmetry but also the wave vector $\bm{q}$ [see Tables.~\ref{tab:tetragonal}-\ref{tab:trigonal}].   

The symmetry constraints from point group symmetry are obtained by dividing the symmetry operations into spin and momentum space (magnetic representation~\cite{Bertaut:a05871}).
First, we rewrite the anisotropic spin interaction at $\pm\bm{q}$ as  
\begin{align}
\label{eq:interaction_q}
\bm{S}_{\bm{q}}^T X_{\bm{q}}  \bm{S}_{-\bm{q}} + \bm{S}_{-\bm{q}}^T X_{-\bm{q}}  \bm{S}_{\bm{q}}=
\tilde{\bm{S}}^T
\begin{pmatrix}
0 & X_{\bm{q}} \\
 X_{\bm{q}}^* & 0
\end{pmatrix}
\tilde{\bm{S}},
\end{align}
with
\begin{align}
\tilde{\bm{S}}=(S^{x_s}_{\bm{q}},S^{y_s}_{\bm{q}},S^{z_s}_{\bm{q}},S^{x_s}_{-\bm{q}},S^{y_s}_{-\bm{q}},S^{z_s}_{-\bm{q}})^T.
\end{align}
Here, $S^{\alpha}_{\bm{q}}$ is the classical spin (axial vector) at wave vector $\bm{q}$ in the cartesian coordinates $\alpha=(x_s,y_s,z_s)$ and $X_{\bm{q}}$ represents the $3\times 3$ interaction matrix.

By using the magnetic representation $\Gamma(P)$ for the operation $P$, the symmetry constraint is obtained from
\begin{align}
\label{eq:constraint}
\begin{pmatrix}
0 & X_{\bm{q}} \\
 X_{\bm{q}}^* & 0
\end{pmatrix} =  \Gamma(P)
\begin{pmatrix}
0 & X_{\bm{q}} \\
 X_{\bm{q}}^* & 0
\end{pmatrix} \Gamma^{-1}(P).
\end{align}
$\Gamma(P)$ is given by
\begin{align}
 \Gamma(P)=\Gamma_\mathrm{perm}(P)\otimes\Gamma_\mathrm{ax}(P),
 \end{align}
 where $2\times 2$ matrix $\Gamma_\mathrm{perm}(P)$ is the permutation representation for $\bm{q}$ and $-\bm{q}$ and  $3\times 3$ matrix $\Gamma_\mathrm{ax}(P)$ is the axial vector representation for the three spin components.
The permutation representation $\Gamma_\mathrm{perm}^\mathrm{I}$ for any type I operations is defined as 
$P^\mathrm{I}(\bm{q},-\bm{q})=(\bm{q},-\bm{q})\Gamma_\mathrm{perm}^\mathrm{I}=(\bm{q},-\bm{q})$,
while $\Gamma_\mathrm{perm}^\mathrm{II}$ for any type II operations is defined as $P^\mathrm{II}(\bm{q},-\bm{q})=(\bm{q},-\bm{q})\Gamma_\mathrm{perm}^\mathrm{II}=(-\bm{q},\bm{q})$.
Then, $\Gamma_\mathrm{perm}$ is explicitly given by 
\begin{align}
\Gamma_\mathrm{perm}^\mathrm{I}=
\begin{pmatrix}
1 & 0\\
0 & 1
\end{pmatrix},
\Gamma_\mathrm{perm}^\mathrm{II}=
\begin{pmatrix}
0 & 1\\
1 & 0
\end{pmatrix}.
\end{align}
Meanwhile, The axial vector representation is defined as $\Gamma_\mathrm{ax}(P)_{\alpha\beta}=\bra{\alpha}P\ket{\beta}$ $(\alpha,\beta=x_s, y_s, z_s)$, where $\ket{\alpha}$ is the basis in classical spin space (axial vector space).

Then, the rules (a)-(f) in Sec.~\ref{sec:Rule} are obtained from the following magnetic representations by setting $\ket{x_s}\parallel\bm{q}$: 
\begin{itemize}
\item[(A)] The representation of the space inversion center corresponding to Fig.~\ref{fig:Rule}(a) is given by 
\begin{align}
\Gamma_\mathrm{perm}^\mathrm{II}\otimes
\begin{pmatrix}
1 & 0 & 0\\
0 & 1 & 0\\
0 & 0 & 1
\end{pmatrix}.
\end{align}
\item[(B)]  The representation of the mirror plane perpendicular to $\bm{q}$ corresponding to Fig.~\ref{fig:Rule}(b)  is given by 
\begin{align}
\Gamma_\mathrm{perm}^\mathrm{II}\otimes
\begin{pmatrix}
1 & 0 & 0\\
0 & -1 & 0\\
0 & 0 & -1
\end{pmatrix}.
\end{align}
\item[(C)] The representation of the twofold axis perpendicular to $\bm{q}$ corresponding to Fig.~\ref{fig:Rule}(c) is given by 
\begin{align}
\Gamma_\mathrm{perm}^\mathrm{II}\otimes
\begin{pmatrix}
-1 & 0 & 0\\
0 & -1 & 0\\
0 & 0 & 1
\end{pmatrix},
\end{align}
where the direction of $\ket{z_s}$ is parallel to the axis.
\item[(D)]  The representation of the mirror plane including $\bm{q}$ corresponding to Fig.~\ref{fig:Rule}(d) is given by 
\begin{align}
\Gamma_\mathrm{perm}^\mathrm{I}\otimes
\begin{pmatrix}
-1 & 0 & 0\\
0 & -1 & 0\\
0 & 0 & 1
\end{pmatrix},
\end{align}
where the direction of $\ket{z_s}$ is perpendicular to the mirror plane.
\item[(E)] The representation of the twofold axis including $\bm{q}$ corresponding to Fig.~\ref{fig:Rule}(e) is given by 
\begin{align}
\Gamma_\mathrm{perm}^\mathrm{I}\otimes
\begin{pmatrix}
1 & 0 & 0\\
0 & -1 & 0\\
0 & 0 & -1
\end{pmatrix}.
\end{align}
\item[(F)] The representation of the $n$-fold ($n=3,4,6$) axis including $\bm{q}$ corresponding to Fig.~\ref{fig:Rule}(f) is given by 
\begin{align}
\Gamma_\mathrm{perm}^\mathrm{I}\otimes
\begin{pmatrix}
1 & 0 & 0\\
0 & \cos\phi & -\sin\phi\\
0 & \sin\phi & \cos\phi
\end{pmatrix},
\end{align}
with $\phi=2\pi/n$.
\end{itemize}
Since the operation $P$ in the rules (a)-(c) [(d)-(f)] is the type II (I), the rules (a)-(c) [(d)-(f)] are obtained from $X_{\bm{q}}=\Gamma_\mathrm{ax}(P)X_{\bm{q}}^*\Gamma_\mathrm{ax}^{-1}(P)$ [$X_{\bm{q}}=\Gamma_\mathrm{ax}(P)X_{\bm{q}}\Gamma_\mathrm{ax}^{-1}(P)$].
Thus,  the rules (a)-(c) [(d) and (e)] are imposed by the point group operation (not) combined with the time-reversal operation, which results in the different (same) nonzero components of $\bm{E}_{\bm{q}}$ and $\bm{D}_{\bm{q}}$.
In the magnetic representations (A)-(E), the axial vector representations do not have the off-diagonal components, which results in no constraint on $\bm{F}_{\bm{q}}$.
  
In Sec.~\ref{sec:Model1}, we use the axial vector representation by setting $\ket{x_s}\parallel \ket{x}$, $\ket{y_s}\parallel \ket{y}$, and $\ket{z_s}\parallel \ket{z}$, where $(\ket{x},\ket{y},\ket{z})$ is the basis set of the crystal lattice shown in Fig.~\ref{fig:Qset}(a).
Then, the axial vector representation has the off-diagonal components depending on the symmetry of the space group and the wave vector, which results in different constraints on the interactions. Equation~(\ref{eq:rotation}) is obtained by using the permutation representation for $(\pm\bm{Q}_1,\pm\bm{Q}_2)$ or $(\pm\bm{Q}_1,\pm\bm{Q}_2, \pm\bm{Q}_3)$ space. 
    

\section{Effective spin model with the interactions at low symmetric wave vectors in tetragonal, hexagonal, and trigonal crystal systems}
\label{ap:ModelWithLowQ}

\begin{figure}[t!]
\begin{center}
\includegraphics[width=1.0\hsize]{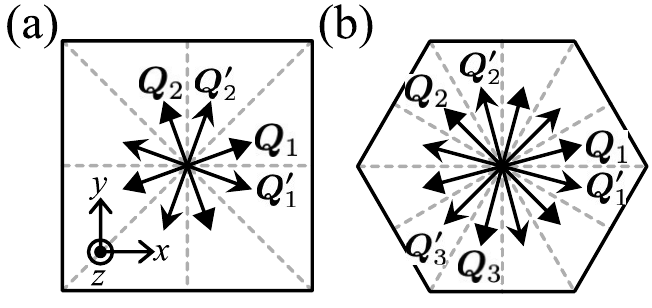} 
\caption{
\label{fig:LowQset}
A set of low symmetric wave vectors along the low symmetric lines inside the first Brillouin zone in (a) tetragonal crystal systems and (b) hexagonal and trigonal crystal systems.
The dashed lines represent the high symmetric lines.
In (a), $\bm{Q}_1$ and $\bm{Q}_2$ ($\bm{Q}'_1$ and $\bm{Q}'_2$) are connected by the fourfold rotation around the $z$ axis, while in (b), $\bm{Q}_1$, $\bm{Q}_2$, and $\bm{Q}_3$ ($\bm{Q}'_1$, $\bm{Q}'_2$, and $\bm{Q}'_3$) are connected by the threefold rotation.
$\bm{Q}_1$ and $\bm{Q}'_1$ are connected by the twofold rotation around the $x$ axis, the mirror reflection on the $xz$ plane, the time-reversal operation after the twofold rotation around the $y$ axis, or the time-reversal operation after the mirror reflection on the $yz$ plane.
The wave vectors in $\{\bm{Q}\}$ lie on the $xy$ plane.   
}
\end{center}
\end{figure}

We here present the model in Eq.~(\ref{eq:ABS2}) with the low symmetric wave vectors in the tetragonal, hexagonal, and trigonal crystal systems.
Figure~\ref{fig:LowQset}(a)[(b)] shows the schematic pictures of the low symmetric wave vectors for the tetragonal (hexagonal and trigonal) crystal systems. 
In the $P4/mmm$, $P422$, $P\bar{4}2m$, $P\bar{4}m2$, and $P4mm$ ($P4/m, P4$, and $P\bar{4}$) crystals, there are four (two) equivalent wave vectors $\bm{Q}_1$, $\bm{Q}_2$, $\bm{Q}'_1$, and $\bm{Q}'_2$ ($\bm{Q}_1$ and $\bm{Q}_2$) connected by the crystal symmetry; the effective spin model in Eq.~(\ref{eq:ABS2}) is described to have the interactions at $\{\bm{Q}\}=\{\pm\bm{Q}_1,\pm\bm{Q}_2,\pm\bm{Q}'_1,\pm\bm{Q}'_2\}$ ($\{\bm{Q}\}=\{\pm\bm{Q}_1,\pm\bm{Q}_2\}$ or $\{\pm\bm{Q}'_1,\pm\bm{Q}'_2\}$).
Meanwhile, in the $P6/mmm$, $P622$, $P\bar{6}m2$, $P\bar{6}2m$, $P6mm$, $P\bar{3}m1$, $P\bar{3}1m$, $P321$, $P312$, and $P3m1$ ($P6/m, P\bar{6}, P6, P\bar{3}$, and $P3$) crystals, there are six (three) equivalent wave vectors $\bm{Q}_1$, $\bm{Q}_2$, $\bm{Q}_3$, $\bm{Q}'_1$, $\bm{Q}'_2$, and $\bm{Q}'_3$ ($\bm{Q}_1$, $\bm{Q}_2$, and $\bm{Q}_3$) connected by the crystal symmetry. 
In this case, the dominant exchange interactions in the effective spin model in Eq.~(\ref{eq:ABS2}) are described by ones at 
$\{\bm{Q}\}=\{\pm\bm{Q}_1,\pm\bm{Q}_2,\pm\bm{Q}_3,\pm\bm{Q}'_1,\pm\bm{Q}'_2,\pm\bm{Q}'_3\}$ ($\{\bm{Q}\}=\{\pm\bm{Q}_1,\pm\bm{Q}_2,\pm\bm{Q}_3\}$ or $\{\pm\bm{Q}'_1,\pm\bm{Q}'_2,\pm\bm{Q}'_3\}$).

Tables~\ref{tab:tetragonal2}-\ref{tab:trigonal2} show the results of $X_{\bm{Q}_1}$ and $X_{\bm{Q}_1'}$ in the tetragonal, hexagonal, trigonal crystal systems, respectively.
In addition, the number of independent components ($N_\mathrm{c}$) of the interaction matrix is shown.
In all cases, $X_{\bm{Q}_1}$ has at least four independent components ($N_\mathrm{c} \ge 4$).
In the $P4/mmm$, $P422$, $P\bar{4}2m$, $P\bar{4}m2$, $P4mm$, $P6/mmm$, $P622$, $P\bar{6}m2$, $P\bar{6}2m$, $P6mm$, $P\bar{3}m1$, $P\bar{3}1m$, $P321$, $P312$, and $P3m1$ crystals, $N_\mathrm{c}$ of $X_{\bm{Q}_1'}$ is zero since the components of $X_{\bm{Q}_1'}$ are related to those of $X_{\bm{Q}_1}$.
For example, nonzero components of $X_{\bm{Q}_1'}$ is obtained from those of $X_{\bm{Q}_1}$ by using the twofold rotation about the $x$ axis, the mirror reflection on the $xz$ plane, the time-reversal operation after the twofold rotation about the $y$ axis, or the time-reversal operation after the mirror reflection on the $yz$ plane depending on the space group.
The other relevant interactions at the symmetry-related wave vectors in $\{\bm{Q}\}$ are obtained by using  Eq.~(\ref{eq:rotation}) in a similar way.

\begin{table*}
\caption{\label{tab:tetragonal2}
Interaction matrices $X_{\bm{Q}_1}$ and $X_{\bm{Q}_1'}$ and the number of independent components $N_\mathrm{c}$ in the tetragonal crystal systems for the low symmetric wave vectors $\bm{Q}_1$ and $\bm{Q}_1'$ shown in Fig.~\ref{fig:LowQset}(a).
The spin coordinates $x_\mathrm{s}$, $y_\mathrm{s}$, and $z_\mathrm{s}$ are taken along the $x$, $y$, and $z$ directions in Fig.~(\ref{fig:LowQset})(a), respectively.
}
\begin{ruledtabular}
\begin{tabular}{ccccc}
 &\multicolumn{2}{c}{$\bm{Q}_1$}  & \multicolumn{2}{c}{$\bm{Q}'_1$}  \\
\cline{2-3} \cline{4-5}
space group $\mathbf{H}$&  $X_{\bm{Q}_1}$ & $N_\mathrm{c}$ & $X_{\bm{Q}'_1}$ & $N_\mathrm{c}$ \\ \hline
$P4/mmm$ & $\begin{pmatrix}
F_{\bm{Q}_1}^{x} & E_{\bm{Q}_1}^{z} & 0 \\
E_{\bm{Q}_1}^{z} & F_{\bm{Q}_1}^{y}  & 0 \\
0& 0  & F_{\bm{Q}_1}^{z}  
\end{pmatrix}$ & 4 &$\begin{pmatrix}
F_{\bm{Q}_1}^{x} & -E_{\bm{Q}_1}^{z} & 0 \\
-E_{\bm{Q}_1}^{z} & F_{\bm{Q}_1}^{y}  & 0 \\
0& 0  & F_{\bm{Q}_1}^{z}   
\end{pmatrix}$  & 0\\ 
$P422$ & $\begin{pmatrix}
F_{\bm{Q}_1}^{x} & E_{\bm{Q}_1}^{z} & -iD_{\bm{Q}_1}^{y} \\
E_{\bm{Q}_1}^{z} & F_{\bm{Q}_1}^{y}  & iD_{\bm{Q}_1}^{x} \\
iD_{\bm{Q}_1}^{y} &-iD_{\bm{Q}_1}^{x}  & F_{\bm{Q}_1}^{z} 
\end{pmatrix}$ & 6 & $\begin{pmatrix}
F_{\bm{Q}_1}^{x} & -E_{\bm{Q}_1}^{z} & iD_{\bm{Q}_1}^{y} \\
-E_{\bm{Q}_1}^{z} & F_{\bm{Q}_1}^{y}  & iD_{\bm{Q}_1}^{x} \\
-iD_{\bm{Q}_1}^{y} &-iD_{\bm{Q}_1}^{x}  & F_{\bm{Q}_1}^{z}  
\end{pmatrix}$ & 0 \\
 $P\bar{4}2m$ & $\begin{pmatrix}
F_{\bm{Q}_1}^{x} & E_{\bm{Q}_1}^{z} & -iD_{\bm{Q}_1}^{y} \\
E_{\bm{Q}_1}^{z} & F_{\bm{Q}_1}^{y}  & iD_{\bm{Q}_1}^{x} \\
iD_{\bm{Q}_1}^{y} &-iD_{\bm{Q}_1}^{x}  & F_{\bm{Q}_1}^{z} 
\end{pmatrix}$ & 6 & $\begin{pmatrix}
F_{\bm{Q}_1}^{x} & -E_{\bm{Q}_1}^{z} & iD_{\bm{Q}_1}^{y} \\
-E_{\bm{Q}_1}^{z} & F_{\bm{Q}_1}^{y}  & iD_{\bm{Q}_1}^{x} \\
-iD_{\bm{Q}_1}^{y} &-iD_{\bm{Q}_1}^{x}  & F_{\bm{Q}_1}^{z}  
\end{pmatrix}$ & 0\\
 $P\bar{4}m2$ & $\begin{pmatrix}
F_{\bm{Q}_1}^{x} & E_{\bm{Q}_1}^{z} & -iD_{\bm{Q}_1}^{y} \\
E_{\bm{Q}_1}^{z} & F_{\bm{Q}_1}^{y}  & iD_{\bm{Q}_1}^{x} \\
iD_{\bm{Q}_1}^{y} &-iD_{\bm{Q}_1}^{x}  & F_{\bm{Q}_1}^{z} 
\end{pmatrix}$ & 6 & $\begin{pmatrix}
F_{\bm{Q}_1}^{x} & -E_{\bm{Q}_1}^{z} & -iD_{\bm{Q}_1}^{y} \\
-E_{\bm{Q}_1}^{z} & F_{\bm{Q}_1}^{y}  & -iD_{\bm{Q}_1}^{x} \\
iD_{\bm{Q}_1}^{y} &iD_{\bm{Q}_1}^{x}  & F_{\bm{Q}_1}^{z}  
\end{pmatrix}$ &  0\\
$P4mm$ & $\begin{pmatrix}
F_{\bm{Q}_1}^{x} & E_{\bm{Q}_1}^{z} & -iD_{\bm{Q}_1}^{y} \\
E_{\bm{Q}_1}^{z} & F_{\bm{Q}_1}^{y}  & iD_{\bm{Q}_1}^{x} \\
iD_{\bm{Q}_1}^{y} &-iD_{\bm{Q}_1}^{x}  & F_{\bm{Q}_1}^{z} 
\end{pmatrix}$ & 6 &  $\begin{pmatrix}
F_{\bm{Q}_1}^{x} & -E_{\bm{Q}_1}^{z} & -iD_{\bm{Q}_1}^{y} \\
-E_{\bm{Q}_1}^{z} & F_{\bm{Q}_1}^{y}  & -iD_{\bm{Q}_1}^{x} \\
iD_{\bm{Q}_1}^{y} &iD_{\bm{Q}_1}^{x}  & F_{\bm{Q}_1}^{z}  
\end{pmatrix}$ & 0\\ 
$P4/m$ & $\begin{pmatrix}
F_{\bm{Q}_1}^{x} & E_{\bm{Q}_1}^{z} & 0 \\
E_{\bm{Q}_1}^{z} & F_{\bm{Q}_1}^{y}  &0  \\
0& 0 & F_{\bm{Q}_1}^{z}  
\end{pmatrix}$  & 4 &   
 $\begin{pmatrix}
F_{\bm{Q}_1'}^{x} & E_{\bm{Q}_1'}^{z} & 0 \\
E_{\bm{Q}_1'}^{z} & F_{\bm{Q}_1'}^{y}  &0  \\
0& 0 & F_{\bm{Q}_1'}^{z}  
\end{pmatrix}$  & 4\\
$P4$ & $\begin{pmatrix}
F_{\bm{Q}_1}^{x} & E_{\bm{Q}_1}^{z} & -iD_{\bm{Q}_1}^{y} \\
E_{\bm{Q}_1}^{z} & F_{\bm{Q}_1}^{y}  & iD_{\bm{Q}_1}^{x} \\
iD_{\bm{Q}_1}^{y} &-iD_{\bm{Q}_1}^{x}  & F_{\bm{Q}_1}^{z}  
\end{pmatrix}$ & 6 &  $\begin{pmatrix}
F_{\bm{Q}_1'}^{x} & E_{\bm{Q}_1'}^{z} & -iD_{\bm{Q}_1'}^{y} \\
E_{\bm{Q}_1'}^{z} & F_{\bm{Q}_1'}^{y}  & iD_{\bm{Q}_1'}^{x} \\
iD_{\bm{Q}_1'}^{y} &-iD_{\bm{Q}_1'}^{x}  & F_{\bm{Q}_1'}^{z} 
\end{pmatrix}$ & 6\\
$P\bar{4}$ & $\begin{pmatrix}
F_{\bm{Q}_1}^{x} & E_{\bm{Q}_1}^{z} & -iD_{\bm{Q}_1}^{y} \\
E_{\bm{Q}_1}^{z} & F_{\bm{Q}_1}^{y}  & iD_{\bm{Q}_1}^{x} \\
iD_{\bm{Q}_1}^{y} &-iD_{\bm{Q}_1}^{x}  & F_{\bm{Q}_1}^{z} 
\end{pmatrix}$ & 6 & $\begin{pmatrix}
F_{\bm{Q}_1'}^{x} & E_{\bm{Q}_1'}^{z} & -iD_{\bm{Q}_1'}^{y} \\
E_{\bm{Q}_1'}^{z} & F_{\bm{Q}_1'}^{y}  & iD_{\bm{Q}_1'}^{x} \\
iD_{\bm{Q}_1'}^{y} &-iD_{\bm{Q}_1'}^{x}  & F_{\bm{Q}_1'}^{z}  
\end{pmatrix}$ & 6\\
        \hline
  \end{tabular}
\end{ruledtabular}
\end{table*}

\begin{table*}
\caption{\label{tab:hexagonal2}
Interaction matrices $X_{\bm{Q}_1}$ and $X_{\bm{Q}_1'}$ and the number of independent components $N_\mathrm{c}$ in the hexagonal crystal systems for the low symmetric wave vectors $\bm{Q}_1$ and $\bm{Q}_1'$ shown in Fig.~\ref{fig:LowQset}(b).
The spin coordinates $x_\mathrm{s}$, $y_\mathrm{s}$, and $z_\mathrm{s}$ are taken along the $x$, $y$, and $z$ directions in Fig.~(\ref{fig:LowQset})(a), respectively. 
}
\begin{ruledtabular}
\begin{tabular}{ccccc}
  &\multicolumn{2}{c}{$\bm{Q}_1$}  & \multicolumn{2}{c}{$\bm{Q}'_1$}  \\
\cline{2-3} \cline{4-5}
space group $\mathbf{H}$&  $X_{\bm{Q}_1}$ & $N_\mathrm{c}$ & $X_{\bm{Q}_1}$ & $N_\mathrm{c}$ \\ \hline
$P6/mmm$ & $\begin{pmatrix}
F_{\bm{Q}_1}^{x} & E_{\bm{Q}_1}^{z} & 0 \\
E_{\bm{Q}_1}^{z} & F_{\bm{Q}_1}^{y}  & 0 \\
0& 0  & F_{\bm{Q}_1}^{z}  
\end{pmatrix}$ & 4 &$\begin{pmatrix}
F_{\bm{Q}_1}^{y} & -E_{\bm{Q}_1}^{z} & 0 \\
-E_{\bm{Q}_1}^{z} & F_{\bm{Q}_1}^{x}  &0  \\
0& 0 & F_{\bm{Q}_1}^{z}  
\end{pmatrix}$  & 0\\ 
$P622$ & $\begin{pmatrix}
F_{\bm{Q}_1}^{x} & E_{\bm{Q}_1}^{z} & -iD_{\bm{Q}_1}^{y} \\
E_{\bm{Q}_1}^{z} & F_{\bm{Q}_1}^{y}  & iD_{\bm{Q}_1}^{x} \\
iD_{\bm{Q}_1}^{y} &-iD_{\bm{Q}_1}^{x}  & F_{\bm{Q}_1}^{z} 
\end{pmatrix}$ & 6 & $\begin{pmatrix}
F_{\bm{Q}_1}^{x} & -E_{\bm{Q}_1}^{z} & iD_{\bm{Q}_1}^{y} \\
-E_{\bm{Q}_1}^{z} & F_{\bm{Q}_1}^{y}  & iD_{\bm{Q}_1}^{x} \\
-iD_{\bm{Q}_1}^{y} &-iD_{\bm{Q}_1}^{x}  & F_{\bm{Q}_1}^{z}  
\end{pmatrix}$ & 0 \\
 $P\bar{6}m2$ & $\begin{pmatrix}
F_{\bm{Q}_1}^{x} & E_{\bm{Q}_1}^{z}+iD_{\bm{Q}_1}^{z} &0  \\
E_{\bm{Q}_1}^{z}-iD_{\bm{Q}_1}^{z}  & F_{\bm{Q}_1}^{y}  & 0 \\
0 & 0& F_{\bm{Q}_1}^{z}  
\end{pmatrix}$ & 5 & $\begin{pmatrix}
F_{\bm{Q}_1}^{x} & -E_{\bm{Q}_1}^{z}-iD_{\bm{Q}_1}^{z} &0  \\
-E_{\bm{Q}_1}^{z}+iD_{\bm{Q}_1}^{z}  & F_{\bm{Q}_1}^{y}  & 0 \\
0 & 0& F_{\bm{Q}_1}^{z}  
\end{pmatrix}$ & 0\\
 $P\bar{6}2m$ & $\begin{pmatrix}
F_{\bm{Q}_1}^{x} & E_{\bm{Q}_1}^{z}+iD_{\bm{Q}_1}^{z} &0  \\
E_{\bm{Q}_1}^{z}-iD_{\bm{Q}_1}^{z}  & F_{\bm{Q}_1}^{y}  & 0 \\
0 & 0& F_{\bm{Q}_1}^{z}  
\end{pmatrix}$ & 5 & $\begin{pmatrix}
F_{\bm{Q}_1}^{x} & -E_{\bm{Q}_1}^{z}-iD_{\bm{Q}_1}^{z} &0  \\
-E_{\bm{Q}_1}^{z}+iD_{\bm{Q}_1}^{z}  & F_{\bm{Q}_1}^{y}  & 0 \\
0 & 0& F_{\bm{Q}_1}^{z}    
\end{pmatrix}$ &  0\\
$P6mm$ & $\begin{pmatrix}
F_{\bm{Q}_1}^{x} & E_{\bm{Q}_1}^{z} & -iD_{\bm{Q}_1}^{y} \\
E_{\bm{Q}_1}^{z} & F_{\bm{Q}_1}^{y}  & iD_{\bm{Q}_1}^{x} \\
iD_{\bm{Q}_1}^{y} &-iD_{\bm{Q}_1}^{x}  & F_{\bm{Q}_1}^{z} 
\end{pmatrix}$ & 6 &  $\begin{pmatrix}
F_{\bm{Q}_1}^{x} & -E_{\bm{Q}_1}^{z} & -iD_{\bm{Q}_1}^{y} \\
-E_{\bm{Q}_1}^{z} & F_{\bm{Q}_1}^{y}  & -iD_{\bm{Q}_1}^{x} \\
iD_{\bm{Q}_1}^{y} &iD_{\bm{Q}_1}^{x}  & F_{\bm{Q}_1}^{z}   
\end{pmatrix}$ & 0\\ 
$P6/m$ & $\begin{pmatrix}
F_{\bm{Q}_1}^{x} & E_{\bm{Q}_1}^{z} & 0 \\
E_{\bm{Q}_1}^{z} & F_{\bm{Q}_1}^{y}  &0  \\
0& 0 & F_{\bm{Q}_1}^{z}  
\end{pmatrix}$  & 4 &   
 $\begin{pmatrix}
F_{\bm{Q}_1'}^{x} & E_{\bm{Q}_1'}^{z} & 0 \\
E_{\bm{Q}_1'}^{z} & F_{\bm{Q}_1'}^{y}  &0  \\
0& 0 & F_{\bm{Q}_1'}^{z}  
\end{pmatrix}$  & 4\\
$P\bar{6}$ & $\begin{pmatrix}
F_{\bm{Q}_1}^{x} & E_{\bm{Q}_1}^{z}+iD_{\bm{Q}_1}^{z} &0  \\
E_{\bm{Q}_1}^{z}-iD_{\bm{Q}_1}^{z}  & F_{\bm{Q}_1}^{y}  & 0 \\
0 & 0& F_{\bm{Q}_1}^{z}  
\end{pmatrix}$ & 5 & $\begin{pmatrix}
F_{\bm{Q}_1'}^{x} & E_{\bm{Q}_1'}^{z}+iD_{\bm{Q}_1'}^{z} &0  \\
E_{\bm{Q}_1'}^{z}-iD_{\bm{Q}_1'}^{z}  & F_{\bm{Q}_1'}^{y}  & 0 \\
0 & 0& F_{\bm{Q}_1'}^{z}  
\end{pmatrix}$ & 5\\
$P6$ & $\begin{pmatrix}
F_{\bm{Q}_1}^{x} & E_{\bm{Q}_1}^{z} & -iD_{\bm{Q}_1}^{y} \\
E_{\bm{Q}_1}^{z} & F_{\bm{Q}_1}^{y}  & iD_{\bm{Q}_1}^{x} \\
iD_{\bm{Q}_1}^{y} &-iD_{\bm{Q}_1}^{x}  & F_{\bm{Q}_1}^{z}  
\end{pmatrix}$ & 6 &  $\begin{pmatrix}
F_{\bm{Q}_1'}^{x} & E_{\bm{Q}_1'}^{z} & -iD_{\bm{Q}_1'}^{y} \\
E_{\bm{Q}_1'}^{z} & F_{\bm{Q}_1'}^{y}  & iD_{\bm{Q}_1'}^{x} \\
iD_{\bm{Q}_1'}^{y} &-iD_{\bm{Q}_1'}^{x}  & F_{\bm{Q}_1'}^{z} 
\end{pmatrix}$ & 6\\
        \hline
  \end{tabular}
\end{ruledtabular}
\end{table*}

\begin{table*}
\caption{\label{tab:trigonal2}
Interaction matrices $X_{\bm{Q}_1}$ and $X_{\bm{Q}_1'}$ and the number of independent components $N_\mathrm{c}$ in the trigonal crystal systems for the low symmetric wave vectors $\bm{Q}_1$ and $\bm{Q}_1'$ shown in Fig.~\ref{fig:LowQset}(b).
The spin coordinates $x_\mathrm{s}$, $y_\mathrm{s}$, and $z_\mathrm{s}$ are taken along the $x$, $y$, and $z$ directions in Fig.~(\ref{fig:LowQset})(a), respectively. 
}
\begin{ruledtabular}
\begin{tabular}{cccccccc}
 &\multicolumn{2}{c}{$\bm{Q}_1$}  & \multicolumn{2}{c}{$\bm{Q}'_1$}  \\
\cline{2-3} \cline{4-5}
space group $\mathbf{H}$&  $X_{\bm{Q}_1}$ & $N_\mathrm{c}$ & $X_{\bm{Q}_1}$ & $N_\mathrm{c}$ \\ \hline
$P\bar{3}m1$ & $\begin{pmatrix}
F_{\bm{Q}_1}^{x} & E_{\bm{Q}_1}^{z} & E_{\bm{Q}_1}^{y} \\
E_{\bm{Q}_1}^{z} & F_{\bm{Q}_1}^{y}  & E_{\bm{Q}_1}^{x} \\
E_{\bm{Q}_1}^{y} & E_{\bm{Q}_1}^{x}  & F_{\bm{Q}_1}^{z}   
\end{pmatrix}$ & 6  &  $ \begin{pmatrix}
F_{\bm{Q}_1}^{x} & -E_{\bm{Q}_1}^{z} & -E_{\bm{Q}_1}^{y} \\
-E_{\bm{Q}_1}^{z} & F_{\bm{Q}_1}^{y}  & E_{\bm{Q}_1}^{x} \\
-E_{\bm{Q}_1}^{y} & E_{\bm{Q}_1}^{x}  & F_{\bm{Q}_1}^{z}  
\end{pmatrix}$
& 0 
 \\
 $P\bar{3}1m$ & $\begin{pmatrix}
F_{\bm{Q}_1}^{x} & E_{\bm{Q}_1}^{z} & E_{\bm{Q}_1}^{y} \\
E_{\bm{Q}_1}^{z} & F_{\bm{Q}_1}^{y}  & E_{\bm{Q}_1}^{x} \\
E_{\bm{Q}_1}^{y} & E_{\bm{Q}_1}^{x}  & F_{\bm{Q}_1}^{z}   
\end{pmatrix}$ & 6  &  $ \begin{pmatrix}
F_{\bm{Q}_1}^{x} & -E_{\bm{Q}_1}^{z} & E_{\bm{Q}_1}^{y} \\
-E_{\bm{Q}_1}^{z} & F_{\bm{Q}_1}^{y}  & -E_{\bm{Q}_1}^{x} \\
E_{\bm{Q}_1}^{y} & -E_{\bm{Q}_1}^{x}  & F_{\bm{Q}_1}^{z}    
\end{pmatrix}$
& 0 \\ 
 $P321$ & 
$ \begin{pmatrix}
F_{\bm{Q}_1}^{x} & E_{\bm{Q}_1}^{z}+iD_{\bm{Q}_1}^{z} & E_{\bm{Q}_1}^{y}-iD_{\bm{Q}_1}^{y} \\
E_{\bm{Q}_1}^{z}-iD_{\bm{Q}_1}^{z} & F_{\bm{Q}_1}^{y}  & E_{\bm{Q}_1}^{x}+iD_{\bm{Q}_1}^{x} \\
E_{\bm{Q}_1}^{y}+iD_{\bm{Q}_1}^{y} & E_{\bm{Q}_1}^{x}-iD_{\bm{Q}_1}^{x}  & F_{\bm{Q}_1}^{z} 
\end{pmatrix}$ & 9  &  $ \begin{pmatrix}
F_{\bm{Q}_1}^{x} & -E_{\bm{Q}_1}^{z}-iD_{\bm{Q}_1}^{z} & -E_{\bm{Q}_1}^{y}+iD_{\bm{Q}_1}^{y} \\
-E_{\bm{Q}_1}^{z}+iD_{\bm{Q}_1}^{z} & F_{\bm{Q}_1}^{y}  & E_{\bm{Q}_1}^{x}+iD_{\bm{Q}_1}^{x} \\
-E_{\bm{Q}_1}^{y}-iD_{\bm{Q}_1}^{y} & E_{\bm{Q}_1}^{x}-iD_{\bm{Q}_1}^{x}  & F_{\bm{Q}_1}^{z}
\end{pmatrix}$
 & 0 
\\ 
 $P312$ & $\begin{pmatrix}
F_{\bm{Q}_1}^{x} & E_{\bm{Q}_1}^{z}+iD_{\bm{Q}_1}^{z} & E_{\bm{Q}_1}^{y}-iD_{\bm{Q}_1}^{y} \\
E_{\bm{Q}_1}^{z}-iD_{\bm{Q}_1}^{z} & F_{\bm{Q}_1}^{y}  & E_{\bm{Q}_1}^{x}+iD_{\bm{Q}_1}^{x} \\
E_{\bm{Q}_1}^{y}+iD_{\bm{Q}_1}^{y} & E_{\bm{Q}_1}^{x}-iD_{\bm{Q}_1}^{x}  & F_{\bm{Q}_1}^{z}  
\end{pmatrix}$ & 9  & $ \begin{pmatrix}
F_{\bm{Q}_1}^{x} & -E_{\bm{Q}_1}^{z}+iD_{\bm{Q}_1}^{z} & E_{\bm{Q}_1}^{y}+iD_{\bm{Q}_1}^{y} \\
-E_{\bm{Q}_1}^{z}-iD_{\bm{Q}_1}^{z} & F_{\bm{Q}_1}^{y}  & -E_{\bm{Q}_1}^{x}+iD_{\bm{Q}_1}^{x} \\
E_{\bm{Q}_1}^{y}-iD_{\bm{Q}_1}^{y} & -E_{\bm{Q}_1}^{x}-iD_{\bm{Q}_1}^{x}  & F_{\bm{Q}_1}^{z}
\end{pmatrix}$
 & 0 
\\
 $P3m1$ & $\begin{pmatrix}
F_{\bm{Q}_1}^{x} & E_{\bm{Q}_1}^{z}+iD_{\bm{Q}_1}^{z} & E_{\bm{Q}_1}^{y}-iD_{\bm{Q}_1}^{y} \\
E_{\bm{Q}_1}^{z}-iD_{\bm{Q}_1}^{z} & F_{\bm{Q}_1}^{y}  & E_{\bm{Q}_1}^{x}+iD_{\bm{Q}_1}^{x} \\
E_{\bm{Q}_1}^{y}+iD_{\bm{Q}_1}^{y} & E_{\bm{Q}_1}^{x}-iD_{\bm{Q}_1}^{x}  & F_{\bm{Q}_1}^{z}   
\end{pmatrix}$ & 9  & $ \begin{pmatrix}
F_{\bm{Q}_1}^{x} & -E_{\bm{Q}_1}^{z}+iD_{\bm{Q}_1}^{z} & -E_{\bm{Q}_1}^{y}-iD_{\bm{Q}_1}^{y} \\
-E_{\bm{Q}_1}^{z}-iD_{\bm{Q}_1}^{z} & F_{\bm{Q}_1}^{y}  & E_{\bm{Q}_1}^{x}-iD_{\bm{Q}_1}^{x} \\
-E_{\bm{Q}_1}^{y}+iD_{\bm{Q}_1}^{y} & E_{\bm{Q}_1}^{x}+iD_{\bm{Q}_1}^{x}  & F_{\bm{Q}_1}^{z}  
\end{pmatrix}$
& 0
\\
$P31m$ & $\begin{pmatrix}
F_{\bm{Q}_1}^{x} & E_{\bm{Q}_1}^{z}+iD_{\bm{Q}_1}^{z} & E_{\bm{Q}_1}^{y}-iD_{\bm{Q}_1}^{y} \\
E_{\bm{Q}_1}^{z}-iD_{\bm{Q}_1}^{z} & F_{\bm{Q}_1}^{y}  & E_{\bm{Q}_1}^{x}+iD_{\bm{Q}_1}^{x} \\
E_{\bm{Q}_1}^{y}+iD_{\bm{Q}_1}^{y} & E_{\bm{Q}_1}^{x}-iD_{\bm{Q}_1}^{x}  & F_{\bm{Q}_1}^{z}   
\end{pmatrix}$ & 9  &  $ \begin{pmatrix}
F_{\bm{Q}_1}^{x} & -E_{\bm{Q}_1}^{z}-iD_{\bm{Q}_1}^{z} & E_{\bm{Q}_1}^{y}-iD_{\bm{Q}_1}^{y} \\
-E_{\bm{Q}_1}^{z}+iD_{\bm{Q}_1}^{z} & F_{\bm{Q}_1}^{y}  & -E_{\bm{Q}_1}^{x}-iD_{\bm{Q}_1}^{x} \\
E_{\bm{Q}_1}^{y}+iD_{\bm{Q}_1}^{y} & -E_{\bm{Q}_1}^{x}+iD_{\bm{Q}_1}^{x}  & F_{\bm{Q}_1}^{z}  
\end{pmatrix}$
& 0
\\
$P\bar{3}$ & $\begin{pmatrix}
F_{\bm{Q}_1}^{x} & E_{\bm{Q}_1}^{z} & E_{\bm{Q}_1}^{y} \\
E_{\bm{Q}_1}^{z} & F_{\bm{Q}_1}^{y}  & E_{\bm{Q}_1}^{x} \\
E_{\bm{Q}_1}^{y} & E_{\bm{Q}_1}^{x}  & F_{\bm{Q}_1}^{z}  
\end{pmatrix}$ & 6  &  $ \begin{pmatrix}
F_{\bm{Q}_1'}^{x} & E_{\bm{Q}_1'}^{z} & E_{\bm{Q}_1'}^{y} \\
E_{\bm{Q}_1'}^{z} & F_{\bm{Q}_1'}^{y}  & E_{\bm{Q}_1'}^{x} \\
E_{\bm{Q}_1'}^{y} & E_{\bm{Q}_1'}^{x}  & F_{\bm{Q}_1'}^{z}    
\end{pmatrix}$
& 6 \\
$P3$ & $ \begin{pmatrix}
F_{\bm{Q}_1}^{x} & E_{\bm{Q}_1}^{z}+iD_{\bm{Q}_1}^{z} & E_{\bm{Q}_1}^{y}-iD_{\bm{Q}_1}^{y} \\
E_{\bm{Q}_1}^{z}-iD_{\bm{Q}_1}^{z} & F_{\bm{Q}_1}^{y}  & E_{\bm{Q}_1}^{x}+iD_{\bm{Q}_1}^{x} \\
E_{\bm{Q}_1}^{y}+iD_{\bm{Q}_1}^{y} & E_{\bm{Q}_1}^{x}-iD_{\bm{Q}_1}^{x}  & F_{\bm{Q}_1}^{z}  
\end{pmatrix}$ & 9  &  $ \begin{pmatrix}
F_{\bm{Q}_1'}^{x} & E_{\bm{Q}_1'}^{z}+iD_{\bm{Q}_1'}^{z} & E_{\bm{Q}_1'}^{y}-iD_{\bm{Q}_1'}^{y} \\
E_{\bm{Q}_1'}^{z}-iD_{\bm{Q}_1'}^{z} & F_{\bm{Q}_1'}^{y}  & E_{\bm{Q}_1'}^{x}+iD_{\bm{Q}_1'}^{x} \\
E_{\bm{Q}_1'}^{y}+iD_{\bm{Q}_1'}^{y} & E_{\bm{Q}_1'}^{x}-iD_{\bm{Q}_1'}^{x}  & F_{\bm{Q}_1'}^{z}  
\end{pmatrix}$  
& 9 
 \end{tabular}
\end{ruledtabular}
\end{table*}

\section{Effective Hamiltonian of the anisotropic periodic Anderson model}
\label{ap:EffHam}
We show the details of the low-energy effective model in Eq.~(\ref{eq:Heff}) of the multi-band anisotropic periodic Anderson model.
The spin-dependent term, $\mathcal{H}'_{m\sigma;m'\sigma}$, is given by 
\begin{align}
\mathcal{H}'_{m\sigma;m'\sigma} &=  \sum_{\bm{k},\bm{q},\alpha} \tilde{\varepsilon}^\alpha_{m\bm{k}+\bm{q}m'\bm{k}}S^{\alpha}_{\bm{q}}  c^{\dagger}_{m\bm{k}+\bm{q}\sigma}c_{m'\bm{k}\sigma} \nonumber \\ &
\label{eq:H'}
+ \sum_{\bm{k}} \tilde{\varepsilon}_{mm'\bm{k}} c^{\dagger}_{m\bm{k}\sigma}c_{m'\bm{k}\sigma}
\end{align}
where 
\begin{align}
\tilde{\varepsilon}^{\alpha}_{m\bm{k}m'\bm{k}'} &= C^{(1)}_{m\bm{k}m'\bm{k}'}\left[ V^{\alpha}_{m'\bm{k}'} V^{0 *}_{m\bm{k}}+ V^{\alpha*}_{m\bm{k}} V^{0 }_{m'\bm{k}'} \right. \nonumber\\
 &\left. \qquad -i\sum_{\beta,\gamma}\epsilon_{\alpha\beta\gamma}V^{\beta}_{m'\bm{k}'} V^{\gamma *}_{m\bm{k}} \right] \\
\varepsilon'_{mm'\bm{k}} &=  C^{(2)}_{mm'\bm{k}} (V^{0}_{m'\bm{k}}V^{0*}_{m\bm{k}}+\bm{V}_{m'\bm{k}}\cdot\bm{V}^*_{m\bm{k}}
),
 \end{align}
 with
 \begin{align} 
 C^{(1)}_{m\bm{k}m'\bm{k}'} &= \frac{1}{2}(B_{m\bm{k}}+B_{m'\bm{k}'}), \\
 C^{(2)}_{mm'\bm{k}} &= -\frac{1}{2}\left(A_{m\bm{k}}+\frac{B_{m\bm{k}}}{2}+A_{m'\bm{k}}+\frac{B_{m'\bm{k}}}{2} \right).
\end{align}
The first term with $m\neq m'$ ($m=m'$) in Eq.~(\ref{eq:H'}) hybridizes different bands (the same band) at different  wave vectors, while the second term with $m\neq m'$ ($m=m'$) hybridizes different bands (the same band) at the same  wave vectors.
However, these terms keep the degeneracy in terms of the itinerant electron spin $\sigma$, so they cannot be the origin of the anisotropic exchange interactions. 

The spin-dependent terms, $\mathcal{H}^\mathrm{ex}_{m\sigma;m'\sigma'}$ and  $\mathcal{H}^\mathrm{SOC}_{m\sigma;m'\sigma'}$, include the hybridization of the different bands, which is neglected in the main text, although they also become the origin of the anisotropic exchange interactions.
When considering the hybridization, the expression of the anisotropic exchange interactions in Sec.~\ref{sec:Model2} becomes more complex.  
In the following, we show the details of the spin-dependent terms.   
The exchange interaction, $\mathcal{H}^\mathrm{ex}_{m\sigma;m'\sigma'}$, is given by 
\begin{align}
\mathcal{H}^\mathrm{ex}_{m\sigma;m'\sigma'} &= \frac{1}{\sqrt{N}}\sum_{\bm{k},\bm{q},\alpha,\beta}J^{\alpha\beta}_{m\bm{k}+\bm{q}m'\bm{k}}c^{\dagger}_{m\bm{k}+\bm{q}\sigma}\sigma^{\alpha}_{\sigma\sigma'}c_{m'\bm{k}\sigma'}S^{\beta}_{\bm{q}}.
\end{align}
where
\begin{align} 
J_{m\bm{k}m'\bm{k}'}^{\alpha\beta}&=J^{\rm ISO}_{m\bm{k}m'\bm{k}'}\delta_{\alpha\beta}+[J^{\rm S}_{m\bm{k}m'\bm{k}'}]^{\alpha\beta}+[J^{\rm AS}_{m\bm{k}m'\bm{k}'}]^{\alpha\beta},
\end{align}
with
\begin{align}
J^{\rm ISO}_{m\bm{k}m'\bm{k}'}&=C^{(1)}_{m\bm{k}m'\bm{k}'}(V^{0}_{m'\bm{k}'}V^{0*}_{m\bm{k}}- \bm{V}_{m'\bm{k}'}\cdot\bm{V}^{*}_{m\bm{k}}), \\
[J^{\rm S}_{m\bm{k}m'\bm{k}'}]^{\alpha\beta}&=C^{(1)}_{m\bm{k}m'\bm{k}'}\left(V^{\alpha}_{m'\bm{k}'} V^{\beta *}_{m\bm{k}} + V^{\alpha*}_{m\bm{k}} V^{\beta}_{m'\bm{k}'}\right),  \\
[J^{\rm AS}_{m\bm{k}m'\bm{k}'}]^{\alpha\beta}&=C^{(1)}_{m\bm{k}m'\bm{k}'}i\sum_{\gamma}\epsilon_{\alpha\beta\gamma}\left( V^{\gamma}_{m'\bm{k}'} V^{0 *}_{m\bm{k}} - V^{\gamma*}_{m\bm{k}} V^{0 }_{m'\bm{k}'}\right).
\end{align}
The effective SOC, $\mathcal{H}^\mathrm{SOC}_{m\sigma;m'\sigma'}$, is given by
\begin{align}
\mathcal{H}^\mathrm{SOC}_{m\sigma;m'\sigma'} &=  \sum_{\bm{k}}\bm{g}_{mm'\bm{k}}\cdot  c^{\dagger}_{m\bm{k}\sigma}\bm{\sigma}_{\sigma\sigma'}c_{m'\bm{k}\sigma '},
\end{align}
where
\begin{align} 
g^{\alpha}_{mm'\bm{k}} &= C^{(2)}_{mm'\bm{k}}\left[ V^{\alpha}_{m'\bm{k}} V^{0 *}_{m\bm{k}} + V^{\alpha*}_{m\bm{k}} V^{0 }_{m'\bm{k}}
 -i\sum_{\beta,\gamma}\epsilon_{\alpha\beta\gamma}V^{\beta}_{m'\bm{k}} V^{\gamma *}_{m\bm{k}} \right].
 \end{align}

\section{Magnetic phases in the case of the interactions at $Q_\eta$}
\label{ap:Phase}

\begin{figure*}[t!]
\begin{center}
\includegraphics[width=1.0\hsize]{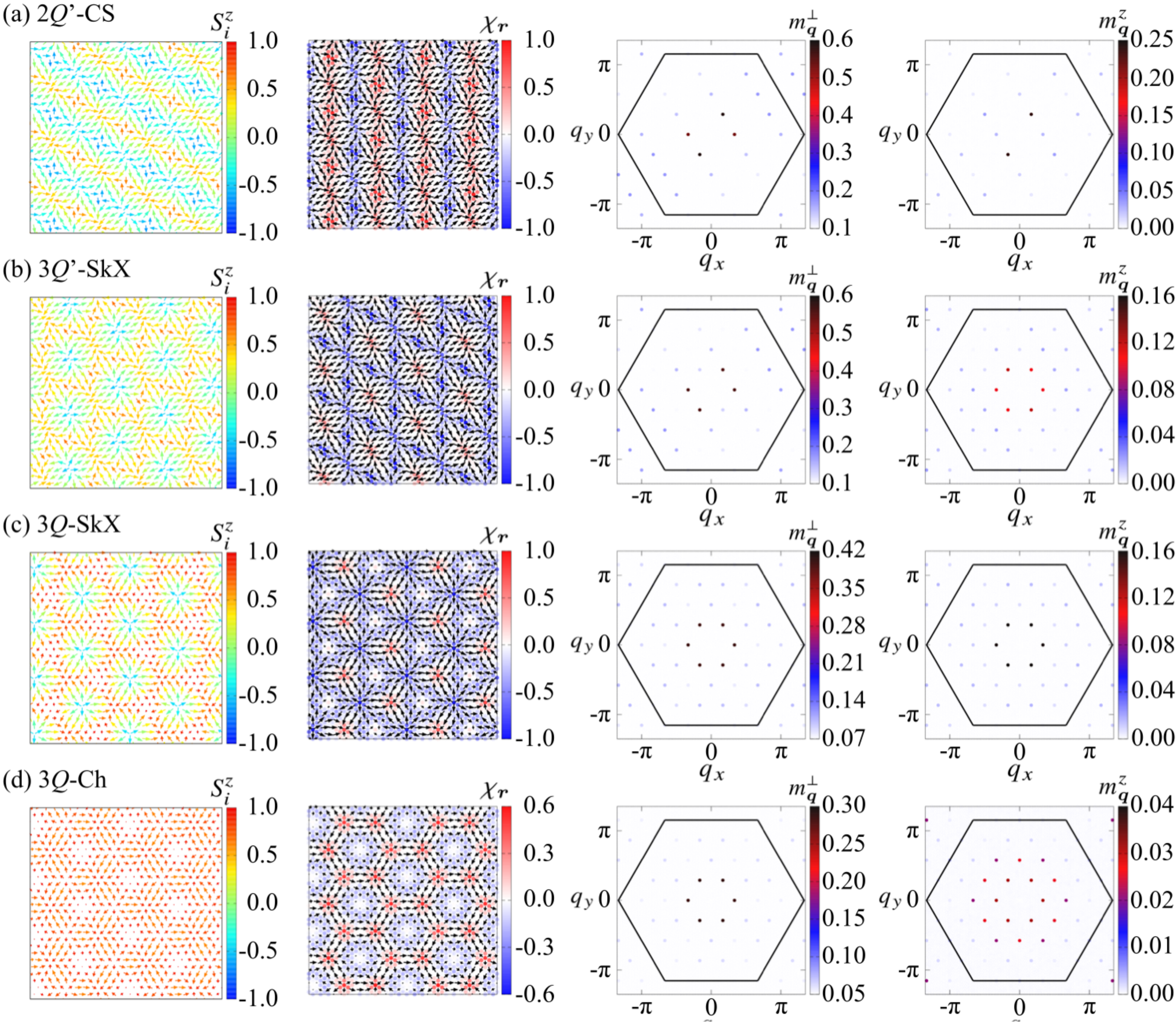} 
\caption{
\label{fig:1Q spin} 
First column: Snapshots of the spin configurations of (a) the 2$Q'$-CS state at $H=0$, (b) 3$Q'$-SkX state at $H=0.3$, (c) 3$Q$-SkX state at $H=0.75$, and (d) 3$Q$-Ch state at $H=1.5$.
The arrows and contours show the $xy$ and $z$ components of the spin, respectively.  
Second column: The scalar chirality configurations of the spin configurations shown in the first column.
Third and fourth columns: The in-plane and out-of-plane magnetic moments
in momentum space, respectively.
The hexagons with a solid line show the first Brillouin zone.
The $\bm{q}=\bm{0}$ component is removed for better visibility.
The first and second columns in (b) and (c) are the same as Figs.~\ref{fig:1Q SkX}(a) and (b), respectively.
}
\end{center}
\end{figure*}

\begin{figure}[t!]
\begin{center}
\includegraphics[width=1.0\hsize]{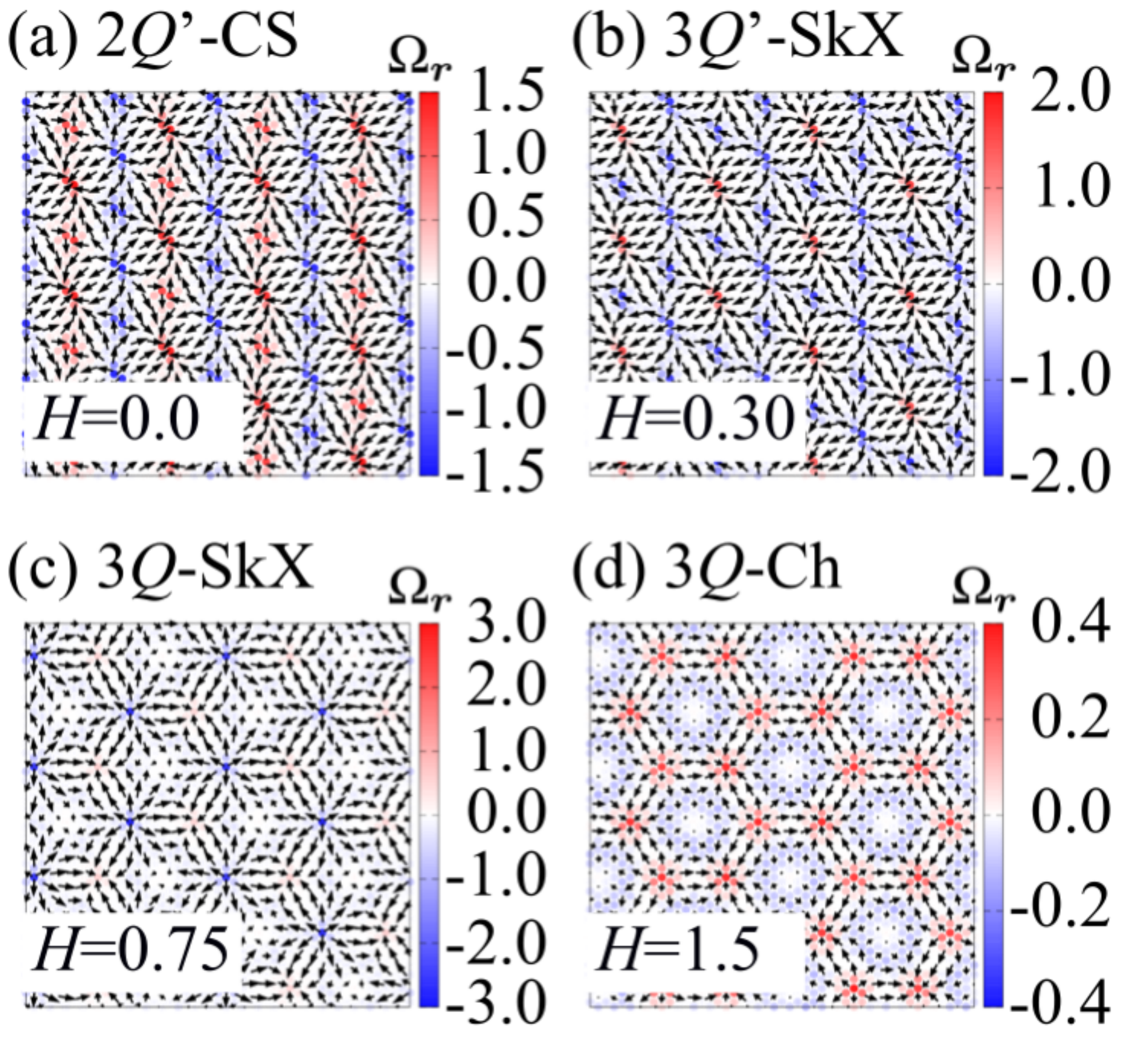} 
\caption{
\label{fig:1Q Nsk}
Skyrmion density configurations of (a) the 2$Q'$ state at $H=0$, (b) 3$Q'$-Ch-I state at $H=0.1$, (c) 3$Q'$-Ch-I state at $H=0.15$, and (d) 3$Q'$-Ch-II state at $H=0.5$. 
}
\end{center}
\end{figure}

We show the details of the multiple-$Q$ states in the model with the interactions at $\{\bm{Q}\}=\{\pm\bm{Q}_1,\pm\bm{Q}_2,\pm\bm{Q}_3\}$ in Sec.~\ref{sec:1Q}.
As shown in Fig.~\ref{fig:1Q}, we find the 2$Q'$-CS state, 3$Q'$-SkX, 3$Q$-SkX, and 3$Q$-Ch state in addition to the 3$Q'$-Ch-II state and the fully polarized state.
Here, CS represents a chiral stripe characterized by a single peak of $\chi_{\bm{q}}$~\cite{Solenov_PhysRevLett.108.096403,Ozawa_doi:10.7566/JPSJ.85.103703,yambe2020double} and 3$Q$ stands for the same intensity of $\bm{Q}_1$, $\bm{Q}_2$, and $\bm{Q}_3$ components in the magnetic moments.  
Figure~\ref{fig:1Q spin} shows the real-space spin and chirality configurations and the $\bm{q}$-space magnetic moments for the 2$Q'$-CS state, 3$Q'$-SkX, 3$Q$-SkX, and 3$Q$-Ch state.
Their skyrmion density configurations are shown in Fig.~\ref{fig:1Q Nsk}.

\bibliography{main.bbl}
\end{document}